\documentclass[11pt]{article}
\pdfoutput=1
\usepackage[utf8]{inputenc}
\usepackage{dsfont}
\usepackage{amsfonts}
\usepackage{amsmath}
\allowdisplaybreaks[4]        
\usepackage{amssymb}
\usepackage{euscript}     
\usepackage{braket}
\usepackage{starfont}
\usepackage{color,soul}         
\usepackage{tensor}        
\usepackage{amsthm}
\usepackage{graphicx}
\usepackage{slashed}
\usepackage{leftidx}
\usepackage{subfigure}
\usepackage{bbm}
\usepackage[normalem]{ulem}
\definecolor{outerspace}{rgb}{0.25, 0.29, 0.3}
\definecolor{scarlet}{rgb}{1.0, 0.13, 0.0}
\usepackage[header,title,page,titletoc]{appendix}  
\definecolor{princetonorange}{rgb}{1.0, 0.56, 0.0}
\definecolor{WildStrawberry}{rgb}{1.0, 0.26, 0.64}
\definecolor{rossocorsa}{rgb}{0.83, 0.0, 0.0}
\definecolor{navyblue}{rgb}{0.0, 0.0, 0.5}
\usepackage[numbers,sort&compress]{natbib}  
\usepackage{float}
\usepackage[paper=letterpaper,margin=1in]{geometry}
\newcommand{\labell}[1]{\label{#1}} %{\label{#1}\qquad\mt{#1}} %

\newcommand{\req}[1]{(\ref{#1})} %{Eq.\thinspace(\ref{#1})}  

\newcommand{\bea}{\begin{eqnarray}}

\newcommand{\eea}{\end{eqnarray}}
\newcommand{\ba}{\begin{eqnarray}}
\newcommand{\ea}{\end{eqnarray}}
\newcommand{\nn}{\nonumber \\}

\newcommand{\be}{\begin{equation}}
\newcommand{\ee}{\end{equation} }
\newcommand{\beqa}{\begin{eqnarray}}
\newcommand{\eeqa}{\end{eqnarray}}
\newcommand{\beqar}{\begin{eqnarray*}}
\newcommand{\eeqar}{\end{eqnarray*}}

\renewcommand{\req}[1]{eq.~(\ref{#1})}

\newcommand{\ssc}{\scriptscriptstyle}
\newcommand{\eg}{{\it e.g.,}\ }
\newcommand{\ie}{{\it i.e.,}\ }

\newcommand{\rh}{r_{\rm h}}
\newcommand{\rps}{r_{\rm ps}}
\newcommand{\rpr}{r_{\rm pr}}
\newcommand{\ats}{a_2^\star}
\newcommand{\gzs}{g_0^\star}

\newcommand{\E}{\mathcal{E}}

\newcommand{\fin}{f_\infty}

\usepackage{hyperref}
\hypersetup{
    colorlinks,
    citecolor=rossocorsa,
    filecolor=navyblue,
    linkcolor=navyblue,
    urlcolor=navyblue
}

\begin{document} 

\begin{titlepage}

\begin{center}

\phantom{ }
\vspace{3cm}

{\bf \Large{Slowly rotating black holes in Einsteinian cubic gravity}}
\vskip 0.5cm
 Connor Adair,$^{\text{\Poseidon}}$ Pablo Bueno,${}^{\text{\Zeus}}$ Pablo A. Cano,${}^{\text{\Kronos}}$ Robie A. Hennigar${}^{\text{\Apollon}}$ and Robert B. Mann$^{\text{\Poseidon}}$
\vskip 0.05in
\small{$^{\text{\Poseidon}}$  \textit{Department of Physics and Astronomy, University of Waterloo}}
\vskip -.4cm
\small{\textit{Waterloo, Ontario, Canada, N2L 3G1}}

\small{${}^{\text{\Zeus}}$ \textit{Instituto Balseiro, Centro At\'omico Bariloche}}
\vskip -.4cm
\small{\textit{ 8400-S.C. de Bariloche, R\'io Negro, Argentina}}

\small{${}^{\text{\Kronos}}$ \textit{Instituut voor Theoretische Fysica, KU Leuven}}
\vskip -.4cm
\small{\textit{Celestijnenlaan 200D, B-3001 Leuven, Belgium}}

\small{${}^{\text{\Apollon}}$ \textit{Department of Mathematics and Statistics, Memorial University of Newfoundland}}
\vskip -.4cm
\small{\textit{  St. John's, Newfoundland and Labrador, A1C 5S7, Canada}}

%\vskip -.4cm
%\small{\textit{}}
\begin{abstract}
We construct slowly rotating black-hole solutions of Einsteinian cubic gravity (ECG) in four dimensions with flat and AdS asymptotes. At leading order in the rotation parameter, the only modification with respect to the static case is the appearance of a non-vanishing $g_{t\phi}$ component. Similarly to the static case, the order of the equation determining such component can be reduced twice, giving rise to a second-order differential equation which can be easily solved numerically as a function of the ECG coupling.  We study how various physical properties of the solutions are modified with respect to the Einstein gravity case, including its angular velocity, photon sphere, photon rings, shadow, and innermost stable circular orbits (in the case of timelike geodesics).

%We also determine the general form of slowly rotating black holes for general Lovelock theories.

\end{abstract}
\end{center}
\end{titlepage}

\setcounter{tocdepth}{2}

{\parskip = .2\baselineskip \tableofcontents}

\section{Introduction}\labell{Introduction}
Constructing black hole solutions of higher-curvature modifications of Einstein gravity is a challenging task. Restricting the discussion to pure-metric diffeomorphism-covariant theories, even the simplest case of static and spherically symmetric configurations is very difficult to tackle in general, typically involving fourth-order coupled  differential equations. 

Nonetheless, there are exceptions to this. The prototypical example corresponds to Lovelock theories \cite{Lovelock1,Lovelock2}, for which (analytic) generalizations of the usual Schwarzschild solution were constructed long ago \cite{Wheeler:1985nh,Wheeler:1985qd,Boulware:1985wk,Myers:1988ze} and 
more recently \cite{Cai:1998vy,Aros:2000ij,Cai:2001dz,Dehghani:2005zm,Dehghani:2009zzb,deBoer:2009gx,Camanho:2011rj,Garraffo:2008hu}.   Naturally, such solutions are nontrivial only for $D\geq 5$, as all higher-curvature Lovelock densities are either topological or trivial in $D=4$. Also for $D\geq 5$, there exists another class of theories which admit analytic static black holes. These are the so-called Quasi-topological gravities \cite{Quasi2,Quasi,Oliva:2011xu,Oliva:2012zs,Dehghani:2011vu,Cisterna:2017umf}. The static black hole solutions of Quasi-topological gravities are characterized by a single function which, similarly to what happens for Lovelock theories, is determined by an algebraic equation. Recently, it has been realized that both Lovelock and Quasi-topological theories can be thought of as particular cases of a more general type of theories, the so-called Generalized Quasi-topological (GQT) gravities \cite{Hennigar:2017ego,PabloPablo3,Ahmed:2017jod,PabloPablo4,Bueno:2019ycr}. These are characterized by admitting single-function ($g_{tt}g_{rr}=-1$) non-hairy generalizations of the Schwarzschild black hole as well as by possessing second-order linearized equations around maximally symmetric backgrounds. A detailed list of their properties 
have been given \cite{Bueno:2019ycr,Bueno:2019ltp}, and the thermodynamics of the cubic and quartic cases have been studied in detail \cite{PabloPablo4,Hennigar:2017umz,Mir:2019rik,Mir:2019ecg}; see
 \cite{Li:2017ncu,Li:2017txk,Li:2018drw} for related references. In GQT theories, the equation that determines the black hole metric function $f(r)\equiv -g_{tt}$ can be automatically integrated once \cite{PabloPablo3}, and we can distinguish two cases. The first group corresponds to theories for which the resulting equation is algebraic, whereas the second includes theories for which it is a second-order differential equation. Lovelock and Quasi-topological gravities comprise the first group.

%\comment{blah blah} 

The first representative of the second group to be identified was Einsteinian cubic gravity (ECG). This theory was originally constructed as the most-general higher-curvature modification of Einstein gravity which, up to cubic order, only propagates the usual transverse and traceless graviton on maximally symmetric backgrounds in general dimensions \cite{PabloPablo}. Soon after, it was realized that --- analogously to Lovelock and Quasi-topological theories in $D\geq 5$ --- four-dimensional ECG admits non-hairy single-function generalizations of the Schwarzschild black hole \cite{PabloPablo2,Hennigar:2016gkm}. The action of the theory is given by
\be\label{ECG-act} 
I_{\rm \ssc ECG} = \frac{1}{16 \pi G} \int d^4x \sqrt{-g} \left[R - \frac{\mu L^4}{8}{\cal P} \right] \, ,
\ee
where $\mu$ is a dimensionless coupling, $L$ some length scale and ${\cal P}$ is given by~\cite{PabloPablo}
%, while ${\cal C}$ is a density identified in~\cite{Hennigar:2017ego}:
\begin{equation}\label{ecg-den}
{\cal P} = 12 R_{a\ b}^{\ c \ d}R_{c\ d}^{\ e \ f}R_{e\ f}^{\ a \ b}+R_{ab}^{cd}R_{cd}^{ef}R_{ef}^{ab}-12R_{abcd}R^{ac}R^{bd}+8R_{a}^{b}R_{b}^{c}R_{c}^{a} \, .
\end{equation}
The construction of ECG opened the door to the definition and classification of GQT gravities, which are now known to exist at general orders and in arbitrary dimensions \cite{Bueno:2019ycr}. The static black hole solutions of ECG have been by now studied in numerous papers and in several contexts \cite{Feng:2017tev,Hennigar:2018hza,HoloECG,Poshteh:2018wqy,Emond:2019crr,Jiang:2019kks,Mehdizadeh:2019qvc,Frassino:2020zuv, KordZangeneh:2020qeg,Bueno:2020odt}.\footnote{Cosmological applications of the theory have been also explored \cite{Arciniega:2018fxj,Cisterna:2018tgx,Arciniega:2018tnn,Erices:2019mkd,Marciu:2020ysf,Quiros:2020uhr,Arciniega:2020pcy,Pookkillath:2020iqq,Marciu:2020ski}.}

Besides ECG and its higher-order cousins, there exist a few other static black hole solutions of higher-cuvature modifications of Einstein gravity in $D=4$. An important set of such solutions (for which $g_{tt}g_{rr}\neq 1$) was presented in \cite{Lu:2015psa,Lu:2015cqa} for quadratic modifications of Einstein gravity. More solutions are available for the so-called ``non-polynomial'' gravities --- see \eg \cite{Deser:2007za,Colleaux:2019ckh} and refs. therein.

All we have said so far concerns static black holes. When rotation is included, things become considerably more involved. For instance, no analytic generalization of the higher-dimensional Kerr (or Myers-Perry \cite{Myers:1986un}) solution has been constructed so far for Lovelock theories in the case of arbitrary rotation. In fact, it has been argued that the Kerr-Schild ansatz does not help in that case \cite{Anabalon:2009kq} (at least in the Einstein-Gauss-Bonnet case).
 Only for certain values of the Lovelock couplings \cite{Anabalon:2009kq,Cvetic:2016sow} or in the slowly rotating limit \cite{Kim:2007iw,Yue:2011et,Camanho:2015ysa} explicit solutions have been constructed, though   full-fledged solutions have been numerically obtained  \cite{Brihaye:2008kh,Brihaye:2010wx}. Additional relevant approaches to higher-curvature modifications of the Kerr solution in  $D=4$ include the construction of solutions at leading order in the effective gravitational action couplings \cite{Cardoso:2018ptl,Cano:2019ore,Reall:2019sah}. 

As far as ECG is concerned, near-horizon rotating solutions have been constructed in \cite{Cano:2019ozf} and perturbative solutions in the gravitational coupling have been obtained in \cite{Burger:2019wkq}. In the present paper we construct the slowly rotating black holes of the theory. The metric takes the form
\begin{equation}
ds^2 = -  f(r)dt^2 + \frac{dr^2}{f(r)} + 2 a r^2 p(r)\sin^2\theta dt d\phi + r^2 \left[ {d\theta^2}  + \sin^2\theta  d\phi^2 \right]  \; .
\label{ansatz}
\end{equation}
With respect to the static case, fully characterized by the function $f(r)$, the only modification is the appearance of a nonvanishing $g_{t\phi}$ component. Similarly to what happens for general GQT gravities in the static case --- as well as for Taub-NUT/Bolt solutions \cite{NewTaub2} --- we observe that the order of the equation that determines the function $p(r)$  can be reduced twice, leaving us with a second-order differential equation that can be easily handled and solved numerically. This allows us to explicitly perform numerous computations of physically relevant quantities of the solution. 

Besides theoretical considerations, the birth of the gravitational wave-astronomy era \cite{Abbott:2016blz} should eventually allow for a detailed scrutiny of the validity of the Kerr solution as the putative description of astrophysical black holes. This further motivates the study of rotating black hole solutions and their physical properties in modified theories of gravity.

The structure of our paper is as follows. In Section \ref{flat} we construct the asymptotically flat slowly rotating black hole of four-dimensional ECG. As anticipated above, we show that the only new equation to be solved is a second-order differential equation. We analyze the asymptotic and near-horizon solutions, and then we construct the full solutions numerically. For a given value of the ECG coupling, each solution is fully characterized by its mass and angular momentum. In Section \ref{props} we study geodesics in the new slowly rotating black hole background and how they differ from their slowly rotating Kerr black hole counterparts. In the case of null geodesics, we study the photon sphere, the photon rings (and the Lyapunov exponents controlling their instability) and the black hole shadow as seen by an asymptotic observer. For timelike geodesics, we compute how the innermost stable circular orbit is modified.  We also compute the horizon angular velocity here.  In Section \ref{adss} we repeat the analysis of Section \ref{flat} in the presence of a negative cosmological constant and construct the corresponding asymptotically AdS solutions. We also construct a new analytic rotating solution in the critical limit of the theory. Section \ref{fincom} contains some final comments regarding possible future studies. Appendix \ref{love} contains a review of the slowly rotating black holes (with a single axis of rotation) of Einstein gravity and general Lovelock theories in arbitrary dimensions.

%\comment{Recently, new class of theories admitting static black hole solutions $g_{tt}g_{rr}=-1$, GQTs, refs. This includes Lovelock theories as particular cases, as well as the Quasi-topological class, previously identified, but which only exists in $D\geq 5$, refs. Integrability of eom, reduced order of eqs.} \comment{The first example of the broader GQT class to be identified is the so-called Einsteinan cubic gravity, define it. Black holes of this theory thoroughly studied refs.} 

%\comment{Footnote about this non-polynomial gravity theories?}

\section{Asymptotically flat solutions}\label{flat}
%Let us consider the action of four-dimensional Einsteinian cubic gravity (ECG). This reads\footnote{An interesting modification of four-dimensional ECG was found in\cite{Arciniega:2018fxj}, which replaced $\mathcal{P}$ by  $\mathcal{P} - 8 {\cal C}$ with
% \begin{equation}\mathcal{C} = R_{abcd}R^{abc}\,_{e}R^{de}-\frac{1}{4}R_{abcd}R^{abcd}R-2R_{abcd}R^{ac}R^{bd}+\frac{1}{2}R_{ab}R^{ab}R\, .
% \end{equation}
%While $\mathcal{C}$ makes no contribution to the equations of motion when considered on static and spherically symmetric backgrounds, it does affect them for other backgrounds and, in particular, the equations of  $\mathcal{P} - 8 {\cal C}$ turn out to be of second-order for the scale factor when evaluated on a Friedmann–Lema\^itre–Robertson–Walker ansatz, which is not the case for $\mathcal{P}$ alone ---see also \cite{Arciniega:2018tnn,}. In the slowly rotating case considered here, ${\cal C}$ makes no contribution to the equations at leading order in the rotation parameter, so all results presented in the paper also hold for the $\mathcal{P} - 8 {\cal C}$ combination.
 %}
%We starting by considering asymptotically flat rotating black holes in four-dimensions.  The action is
%\be\label{ECG-act} 
%I_{\rm \ssc ECG} = \frac{1}{16 \pi G} \int d^4x \sqrt{-g} \left[R - \frac{\mu L^4}{8}{\cal P} \right] \, ,
%\ee
%where $\mu$ is a dimensionless coupling, $L$ some length scale and ${\cal P}$ is given by~\cite{PabloPablo}
%, while ${\cal C}$ is a density identified in~\cite{Hennigar:2017ego}:
The equations of motion of ECG are given by \cite{Hennigar:2017ego,PabloPablo}
%\begin{equation}\label{ecg-den}
%{\cal P} = 12 R_{a\ b}^{\ c \ d}R_{c\ d}^{\ e \ f}R_{e\ f}^{\ a \ b}+R_{ab}^{cd}R_{cd}^{ef}R_{ef}^{ab}-12R_{abcd}R^{ac}R^{bd}+8R_{a}^{b}R_{b}^{c}R_{c}^{a} \, .
%\end{equation}
%Considered in $D$-dimensions, the ECG action also contains the quadratic and cubic Lovelock densities, and it turns out to be the most general higher-curvature Lagrangian involving up to cubic order terms which possesses second-order linearized equations on maximally symmetric backgrounds while constructed in a dimension-independent fashion~\cite{PabloPablo} ---\eg the relative coefficients in \req{ecg-den} are the same for any $D$.  When restricted to $D=4$, ECG has the additional interesting property of admitting non-hairy generalizations of the Schwarzschild-AdS black hole solution satisfying $g_{tt} g_{rr}=-1$ \cite{Hennigar:2016gkm,PabloPablo2} making it the simplest example of a theory of the so-called ``Generalized quasi-topological gravity'' class, recently identified \cite{Hennigar:2017ego,PabloPablo3,Ahmed:2017jod,PabloPablo4,Bueno:2019ycr,Bueno:2019ltp}\comment{more}.\footnote{\comment{More precisely, GQTGs are defined by ...}}
%
%We note that ${\cal C}$ makes no contribution to the field equations for a spherically symmetric black hole solution, nor does it contribute in the slowly rotating case. However, it was pointed out in~\cite{Arciniega:2018fxj} --- see also~\cite{Arciniega:2018tnn} --- that this term is essential in the cosmological context. 
%
%The equations of motion of ECG can be written as
\be \label{eqsECG} 
 P_{a cde}R_{b}{}^{cde} - \frac{1}{2} g_{ab} {\cal L}_{\ssc \rm ECG}- 2 \nabla^c \nabla^d P_{acdb} = 0\, , \quad P_{abcd} \equiv  \,  \frac{\partial {\cal L}_{\ssc \rm ECG}}{\partial R^{abcd}}\, ,
\ee
where
\begin{align}\label{P_thing}
P_{abcd} &= \frac{1}{16\pi G}\Big[ g_{a[c}g_{b]d} - \frac{3\mu L^4}{4} \big[  \,  R _{ad} R _{bc} -  R_{ac} R _{bd} +  g_{bd} R _{a}{}^{e} \
R _{ce} -  g_{ad} R _{b}{}^{e} R_{ce}  -  g_{bc} R _{a}{}^{e} R_{de} \\ \notag
&+  g_{ac} R _{b}{}^{e} R_{de} 
-  g_{bd} R ^{ef} R_{aecf} +  g_{bc} R ^{ef} R _{aedf} + \
 g_{ad} R ^{ef} R _{becf} - 3 R_{a}{}^{e}{}_{d}{}^{f} R _{becf} \\ \notag
&  - g_{ac} R ^{ef} R _{bedf} + 3 R_{a}{}^{e}{}_{c}{}^{f} R _{bedf} + \tfrac{1}{2} R_{ab}{}^{ef} R _{cdef} \big]\Big] \, .
\end{align}
We shall write the ansatz  \eqref{ansatz} for the slowly rotating black hole in the following form
\be 
ds^2 = -N(r)^2 f(r)dt^2 + \frac{dr^2}{f(r)} + 2 a r^2 p(r)(1-x^2)dt d\phi + r^2 \left[\frac{dx^2}{1-x^2} + (1- x^2) d\phi^2 \right]  \, ,
\ee
%\left(d\theta^2+\sin^2 \theta d\phi^2 \right) 
%\left[\frac{dx^2}{1-x^2} + (1- x^2) d\phi^2 \right]
where $x \equiv \cos\theta$. Note that this includes three-independent functions of the radial coordinate.
We wish to solve  the field equations \eqref{eqsECG}  for this metric ansatz, working at linear order in $a$. At this order,  only $\E_{t\phi}$ becomes modified. The $\E_{tt}$ and $\E_{rr}$ are identical to the static case, for which $\E_{t}{}^{t} = \E_{r}{}^{r}$. It follows that this relation also holds in the slowly rotating case, % that holds for static solutions also does here.  
which implies that we can set $N(r) = 1$ without loss of generality.    

This $\E_{t\phi}$ component of the field equations is a complicated, fourth-order, differential equation in the function $p(r)$. However the following combination of the components of the field equations
\be 
\frac{r^4}{f(r)} \left[\E_{t}{}^{\phi} - \frac{a r p(r) }{2} \frac{d \E_{r}{}^{r}}{dr} \right] 
\ee
remarkably admits a trivial first integral.  
The resulting third-order equation takes the form
\begin{align}
C &= r^4 p'  + \mu  \bigg[ -\frac{3}{2} \left( \frac{r f'}{2} + 1 - f \right) r^2 f p''' - \frac{3}{2} \left(\frac{r^2 f}{2} f'' + \frac{r^2 f'^2}{2} + \frac{r(2-f) f'}{2} + 2 f(1-f) \right) p'' 
	\nn
	&+ \frac{15}{2} \left( - \frac{3r^2 }{10} \left(-\frac{r f'}{3} + 1 + \frac{2 f}{3} \right)f'' - \frac{r^2 f'^2}{2} + \frac{r(1 + 7 f) f'}{5} + (1-f)\left(1 + \frac{6}{5} f \right) \right) p' 
\bigg] \, ,
\end{align}
with $C$ a constant of integration, which we show below to be proportional to the mass of the solution through $C=6M$. We note that a second remarkable thing has occurred upon integration: while the equation for $p(r)$ is now a third-order equation, there is no term proportional to $p(r)$ itself. This means that we can further simplify the problem by defining $p(r) = \int^r g(r') dr'$ which gives a second-order differential equation for the unknown function $g(r)$.

We therefore must solve the following two equations to determine the two unknown metric functions:
\begin{align}
2 G M &=  r(1-f) + \frac{\mu L^4}{4 r^2} \left[6 r \left(\frac{r f'}{2} + 1- f \right) f'' f-  \left(r^2 f'^2 + 3 r f' + 6 f (1-f) \right)f' \right] \, , \label{eqnf}
\\
C &= r^4 g  -  \frac{3 \mu L^4}{2}  \bigg[  \left( \frac{r f'}{2} + 1 - f \right) r^2 f g'' +  \left(\frac{r^2 f}{2} f'' + \frac{r^2 f'^2}{2} + \frac{r(2-f) f'}{2} + 2 f(1-f) \right) g'
	\nn
	&\, -5 \left( - \frac{3r^2 }{10} \left(-\frac{r f'}{3} + 1 + \frac{2 f}{3} \right)f'' - \frac{r^2 f'^2}{2} + \frac{r(1 + 7 f) f'}{5} + (1-f)\left(1 + \frac{6}{5} f \right) \right) g  \label{eqng}
\bigg] \, .
\end{align}
The first of these equations is the usual (integrated) ${\cal E}_r{}^r$ component of the field equations for a static black hole ansatz that has been studied in the context of four-dimensional spherically symmetric solutions~\cite{Hennigar:2016gkm, PabloPablo2}, while the second equation will account for the new physics due to the rotation. 
We would like to stress that this double order-reduction phenomenon that allows us to reduce the problem to two second-order differential equations for $f(r)$ and $g(r)$ --- which, besides, are linear in $f''(r)$ and $g''(r)$ respectively --- is highly nontrivial. It would be interesting to explore whether this extends to general GQTGs in $D=4$ (and also in higher dimensions, possibly with several independent rotation parameters turned on).

In the weak coupling limit, we can solve the equations above by assuming a perturbative expansion in $\mu$ of the functions $f$ and $g$. This process leads to the following solution up to order $\mu^2$:

\begin{align}\label{eq:fpert}
f(r)=&1-\frac{2 M}{r}+\mu  L^4 \left[-\frac{27 M^2}{r^6}+\frac{46 M^3}{r^7}\right]-\left(\mu 
   L^4\right)^2 \left[\frac{6804 M^3}{r^{11}}-\frac{27702 M^4}{r^{12}}+\frac{28014
   M^5}{r^{13}}\right]\, ,\\
r^2p(r)=&-\frac{2 M}{r}+\frac{46 \mu  L^4 M^3}{r^7}+\left(\mu  L^4\right)^2 \left[\frac{4860 L^8
   M^3}{13 r^{11}}+\frac{12393 L^8 M^4}{r^{12}}-\frac{28014 L^8 M^5}{r^{13}}\right]\, .
   \label{eq:ppert}
\end{align}
However, this solution is only valid for rather small values of $\mu L^4/M^4$, and adding more terms in the expansion does not increase the precision.\footnote{In fact, an analysis of the series in $\mu$ suggests that its radius of convergence is zero. Nevertheless the perturbative expansion with a few terms does provide a reasonable approximation for small enough coupling.}
Thus, we need to perform a non-perturbative analysis in order to understand the solutions at large coupling. Let us then study the solutions to the equations \eqref{eqnf} and \eqref{eqng} in more detail. 

\subsection{Asymptotic solution \label{sec:asympsol}}

The solution of the equation for $f(r)$ has been quite thoroughly studied in other work --- see, \eg~\cite{Hennigar:2017ego, PabloPablo2, PabloPablo4, Hennigar:2018hza, HoloECG} --- so here we will be relatively concise with the analysis. 

Asymptotically, the solution $f(r)$ consists of a particular and a homogeneous part. The particular solution can be obtained by means of a power series in $r$. Taking as an ansatz
\be 
f_{1/r}(r) =  \sum_{n=0} \frac{b_n}{r^n} \, ,
\ee
plugging this into the equations of motion and solving order-by-order in the large-$r$ limit we obtain the following:
\begin{align} 
f_{1/r} =  1  - \frac{2 M}{r}  - \frac{27  M^2 \mu L^4 }{r^6} + {\cal O}\left( r^{-7} \right)  \, .
\end{align}
To obtain the homogeneous part of the solution we plug $f(r) = f_{1/r} + f_{\rm h}(r)$ back into \eqref{eqnf} and work to linear order in $f_{\rm h}(r)$. This yields 
\be 
-\frac{9 M L^4 \mu}{2 } f_{\rm h}''(r) + r^3 f_{\rm h}(r) = 0\, , 
\ee
where we have kept only the leading terms in the large-$r$ limit.\footnote{This means, for example, that we are neglecting the term proportional to $f'_{\rm h}$, which falls off faster than the other terms. Working to higher-order and including this term does not change the conclusions.} While the equation for the homogeneous solution admits a solution in terms of special functions, it is more useful to simply present the asymptotic behavour in the $r \to \infty$ limit, which is governed by 
\be\label{f-flatasymp} 
f_{\rm h}(r) \sim A \sqrt{r} I_{1/5} \left(\frac{2 \sqrt{2} r^{5/2}}{15 \sqrt{M \mu L^4} } \right) + B  \sqrt{r} K_{1/5} \left(-\frac{2 \sqrt{2} r^{5/2}}{15 \sqrt{M \mu L^4} } \right) \, ,
\ee
where $A$ and $B$ are constants and $K_\nu (x)$ and $I_\nu (x)$ are Bessel functions.  We see that the requirement of a well-behaved positive mass solutions fixes the parameter $A = 0$ and demands $\mu > 0$. The second part of the solution, which decays super-exponentially, can be safely neglected when writing the asymptotic solution.

We now turn to an analysis of \req{eqng}. Here the general solution likewise consists of a particular solution and a homogeneous solution. The particular solution is obtained as a large-$r$ power series with the first few terms reading
\be 
g_{1/r}(r) =  \frac{C}{r^4} - \frac{69 C M \mu L^4  }{ r^{10} } + {\cal O}\left(r^{-12} \right) \, . 
\ee
The homogeneous equation can be obtained by substituting $g(r) = g_{1/r}(r) + g_{\rm h}(r)$ back into~\eqref{eqng} and keeping only the most important terms in the large-$r$ limit. This yields the following equation:
\be 
-\frac{9 M  \mu  L^4}{2}  g_{\rm h}''(r)  + r^3 g_{\rm h}(r) = 0 \, .
\ee
The equation is of exactly the same form as that presented above in \eqref{f-flatasymp}  for $f_{\rm h}$, and so the basic conclusion is the same:  we must fix one integration constant to zero (the coefficient of the growing mode) and restrict the coupling to positive values in order to have well-behaved positive mass solutions.

%Again, considering only the asymptotic behavior of the homogeneous solution, we see that it is governed by
%\be 
%g_{\rm h}(r) \sim A_g  r^{-3/4} \exp \left(\frac{2 \sqrt{2} r^{5/2} }{15 \sqrt{M \mu L^4}} \right) + B_g r^{3/4} \exp \left(- \frac{2 \sqrt{2}  r^{5/2} }{15 \sqrt{M \mu L^4 }}  \right) \, ,
%\ee
%The conclusion here is basically the same as it was for $f(r)$:

The equation that determines $p(r)$ was originally fourth-order and, as a result, the solution should be characterized by four parameters. We see that three of these parameters arise as integration constants in the solution for $g(r)$: the constant $C$ and two more constants of integration in the homogeneous solution. One of these latter parameters --- the coefficient of the growing mode --- is fixed to zero by requiring well-behaved asymptotics.  We will see in the following subsections that a free parameter in the near-horizon solution is fixed by requiring that the near horizon solution connects smoothly to the asymptotic solution. Let us now discuss the interpretation of the integration constant $C$ and the final parameter that arises when integrating $g(r)$ to obtain $p(r)$. 

In the large-$r$ limit, the behavior of $p(r)$ is accurately described by integrating just the particular solution $g_{1/r}(r)$. This gives
\be 
p(r) = -\frac{\Omega_{\infty}}{a} - \frac{C}{3 r^3}  +  \frac{23 C \mu M^2 L^4 }{3  r^9} + {\cal O} \left(r^{-11} \right) \, ,
\ee  
where $\Omega_{\infty}$ is a constant of integration. It is easy to see that $\Omega_\infty$ is related to the asymptotic angular velocity of the spacetime. We have
\be 
\Omega = - \frac{g_{t\phi}}{g_{\phi\phi}} \to - a \lim_{r\to\infty} p(r)  = \Omega_{\infty} \, .
\ee
By a suitable choice of the Killing coordinates $t$ and $\phi$ it is always possible to set $\Omega_{\infty} = 0$, and so henceforth we will make this choice. This amounts to starting the integration for $p(r)$ at $r = \infty$ and working inward.

 To interpret the integration constant $C$, we will require that our solution asymptotically approaches the slow rotation limit of the Kerr solution
\begin{align} 
ds^2 \to& - \left(1 - \frac{2 M}{ r} \right) dt^2 - \frac{4 M a (1- x^2)}{r} dt d\phi + \frac{dr^2}{\left(1 - \frac{2 M}{ r}\right)}  
+r^2 \left[\frac{dx^2}{1-x^2} + (1- x^2) d\phi^2 \right] \, .
\end{align}
Comparing this with our asymptotic expansion for $p(r)$ reveals that
\be 
C = 6 M \, .
\ee
This ensures that the spin parameter $a = J/M$ represents the angular momentum per unit mass of our solution.

\subsection{Near horizon solution \label{sec:nhsol}}

We now wish to consider the equations of motion expanded about the horizon of a black hole. Within the first-order slow rotation approximation there is no displacement of the horizon and it remains located at $f(\rh) = 0$. Let us first discuss the near horizon solution for the function $f(r)$. In this case, since we wish to study black hole solutions, we will assume that $f(r)$ goes to zero linearly as $r \to \rh$:
\be 
f(r) = 4 \pi T (r-\rh) + \sum_{n=2} a_n (r-\rh)^n \, .
\ee  
Plugging this ansatz into the first equation of~\eqref{eqnf} and expanding order by order in $(r-\rh)$, we find that the first two relationships can be expressed as
\begin{align}
2 G M &=  \rh  - \frac{3 (4 \pi T)^2 \mu L^4}{4 \rh} - \frac{(4 \pi T)^3 \mu L^4}{4}  \, ,
	\nn
0 &= -1 + 4 \pi T \rh  + \frac{3  (4 \pi T)^2 \mu L^4}{4 \rh^2}  \, .
\end{align}
These two conditions suffice to determine the horizon radius and temperature of the black hole for a given choice of the mass and coupling. At the next order, $a_3$ appears linearly and can be solved for in terms of the parameter $a_2$, along with the mass and horizon radius. The general pattern is that $a_n$ for $n > 2$ can be solved for in terms of the preceding parameters, which can themselves be expressed in terms of $a_2$, the mass and the coupling.

The parameter $a_2$ is not fixed by the near-horizon equations of motion. However, as we will see explicitly when considering the numerical solution, the value of $a_2$ \textit{is} fixed by requiring the solution joins smoothly on to the asymptotic solution~\cite{PabloPablo4, Hennigar:2018hza}, we will refer to this special value as $\ats$. An alternative prescription to determine $\ats$ involves fixing the $\mu$-derivatives of $a_2$ by the requirement that the near horizon solution admits a smooth $\mu \to 0$ limit~\cite{Hennigar:2018hza, HoloECG}. The basic idea is to remove any terms proportional to inverse powers of $\mu$ in the series expansion of the term $a_n$ for small $\mu$. This, in fact, fixes all the derivatives of $a_2$ with the first few of these reading:
\be 
a_2^{(0)} = - \frac{1}{\rh^{2}} \, , \quad a_2^{(1)} = \frac{81 L^4}{4 \rh^6} \, , \quad a_2^{(2)} = - \frac{3807 L^8}{2 \rh^{10}} \, , \quad a_2^{(3)} = \frac{17827209 L^{12}}{32 \rh^{14}} \, ,
\ee 
where $a_2^{(n)} := \left[ (d/d\mu)^n a_2(\mu) \right]|_{\mu = 0}$.  The coefficients grow faster than $n!$ and a convergence analysis reveals that the power series expansion has vanishing radius of convergence. Thus, regarded as a function of the coupling, $a_2$ is a smooth but non-analytic function. However, as was discussed in~\cite{Hennigar:2018hza}, using a Pad{\'e} approximant to represent $a_2(\mu)$ yields a result that matches quite well the numerically determined value of $\ats$.

We do not expect that the function $g(r)$ should vanish at the horizon, but we do expect that it is a smooth function there.  Therefore, we take the following series ansatz
\be 
g(r) = \sum_{n=0} g_n (r-\rh)^n \, .
\ee
Substituting this ansatz, along with the near horizon expansion for $f(r)$, into \req{eqng} we can solve order by order for the parameters $g_n$. In this case, we find that all of the parameters $g_n$ for $n > 0$ can be obtained explicitly in terms of the mass, the coupling, and the parameters $a_2$ and $g_0$. The first few relationships read:
\begin{align}
6 M &= g_0 \left(\frac{3}{4} \mu  L^4 \left(2 \rh \left[(4 \pi T)-3 a_2 \rh \right]+(4 \pi T) \rh^2 \left[2 a_2 \rh-5 (4 \pi T) \right]+10 \right)+\rh^4\right)
	\nn
	&-\frac{3}{4} (4 \pi T) g_1 \mu  L^4 \rh \left[(4 \pi T) \rh+2 \right] \, ,
\\
0 &= g_1 \left(\frac{3}{4} \mu  L^4 \left(-2  \rh \left[5 a_2 \rh+3 (4 \pi T) \right] -(4 \pi T) \rh^2 \left[4 a_2 \rh+7 (4 \pi T) \right]+10 \right)+\rh^4\right)
	\nn
	&+g_0 \bigg(\frac{3}{2} \mu  L^4 \left( \left[2 (4 \pi T)-\rh
   (9 a_3 \rh+4 a_2) \right]+\rh \left[2 a_2^2 \rh^2+3 (4 \pi T) \rh (a_3 \rh-3 a_2)+2 (4 \pi T)^2\right]\right)
   \nn
   &+4 \rh^3\bigg) -3 (4 \pi T) g_2 \mu  L^4 \rh^2 \left[(4 \pi T) \rh+2 \right] \, .
\end{align}
The situation for $g_0$ is quite similar to the situation for $a_2$. While it appears as a free parameter in the near horizon expansion, a numerical analysis reveals that the solution for $g(r)$ only joins smoothly onto the asymptotic solution if the parameter $g_0$ takes on a special value, which we will denote as $\gzs$. Again, if we demand a smooth $\mu \to 0$ limit of the near horizon solution this fixes all of the derivatives of $\gzs$, with the first few reading
\be 
g_0^{(0)} = \frac{3}{\rh^3} \, , \quad g_0^{(1)} = - \frac{219 L^4}{4 \rh^7} \, , \quad g_0^{(2)} = \frac{3681 L^4}{\rh^{11}} \, , \quad g_0^{(3)} = - \frac{22862007 L^{12}}{32 \rh^{15} } \, , 
\ee
where $g_0^{(n)} \equiv \left[ (d/d\mu)^n g_0(\mu) \right]|_{\mu = 0}$.  Once again, due to the growth of the coefficients, a power series for $g_0$ based on this expansion does not converge --- it is a smooth but non-analytic function of the coupling $\mu$.

\subsection{Numerical solution}
Let us now construct the full solution numerically. 
We start again by discussing the construction of the numerical solution for $f(r)$. This has been discussed  previously \eg~\cite{PabloPablo4}, but we include the details here for completeness, since the situation for $g(r)$ is very similar. 

The basic idea here is to use the near horizon expansion constructed in Section~\ref{sec:nhsol} to generate initial data to be used in a numerical solver that integrates from the horizon toward infinity (or toward the origin). For $r$ close to $\rh$ we can, to a good approximation, keep only the first two terms in the near horizon expansion,
\be 
f(r) \approx 4 \pi T (r-\rh) + a_2(r-\rh)^2 \, .
\ee
We write $r = \rh + \epsilon M$ where $\epsilon$ is a small quantity that we take  to be $\pm 10^{-5}$. The plus sign is chosen for constructing the exterior solution, while the minus sign is chosen to construct the interior solution. 

%Additionally, we will fix the mass $M$ of the solution, along with the coupling and shall set $L = M$ in the analysis.

\begin{figure}[t]
\centering 
\includegraphics[scale=0.45]{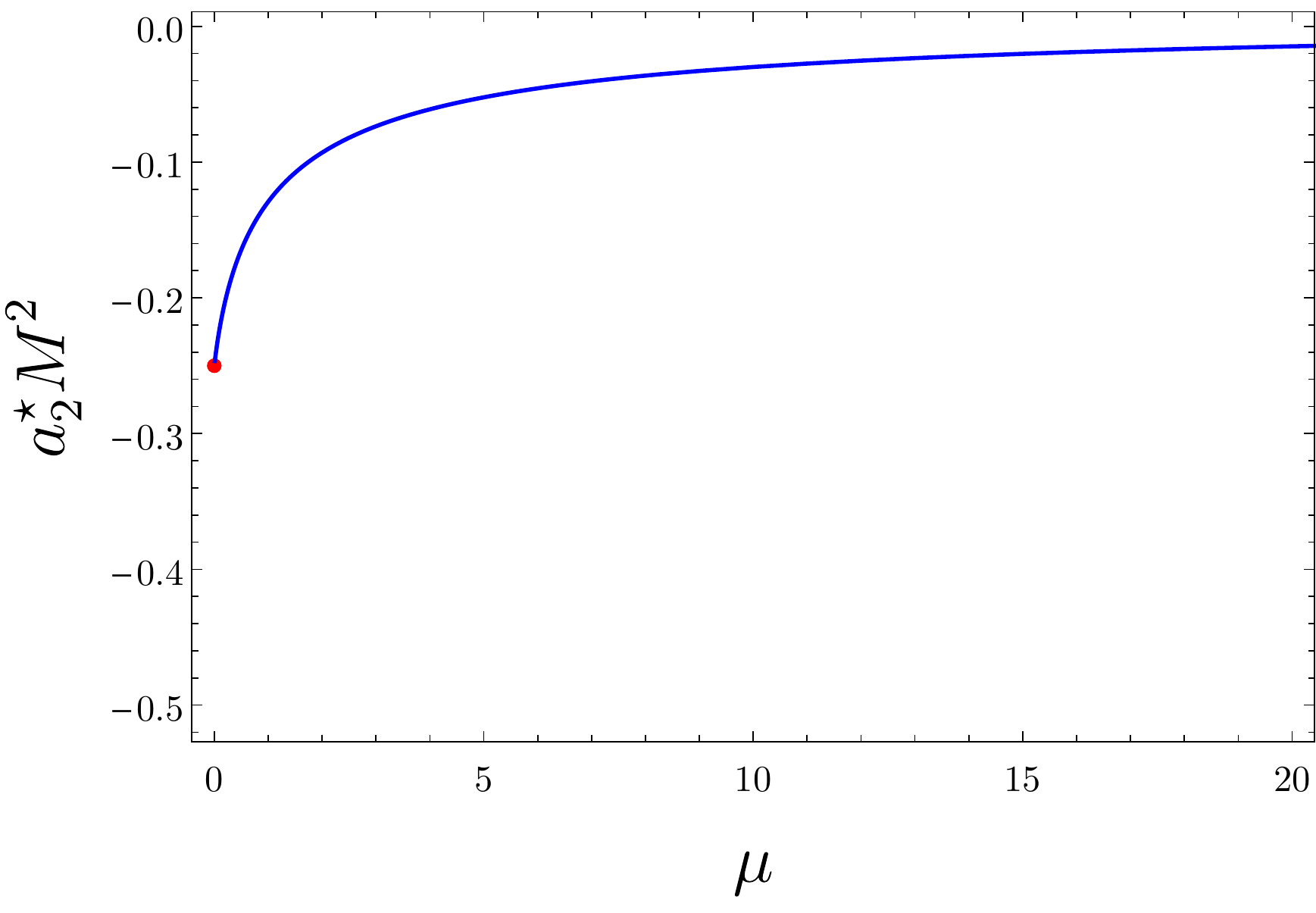}
\quad
\includegraphics[scale=0.45]{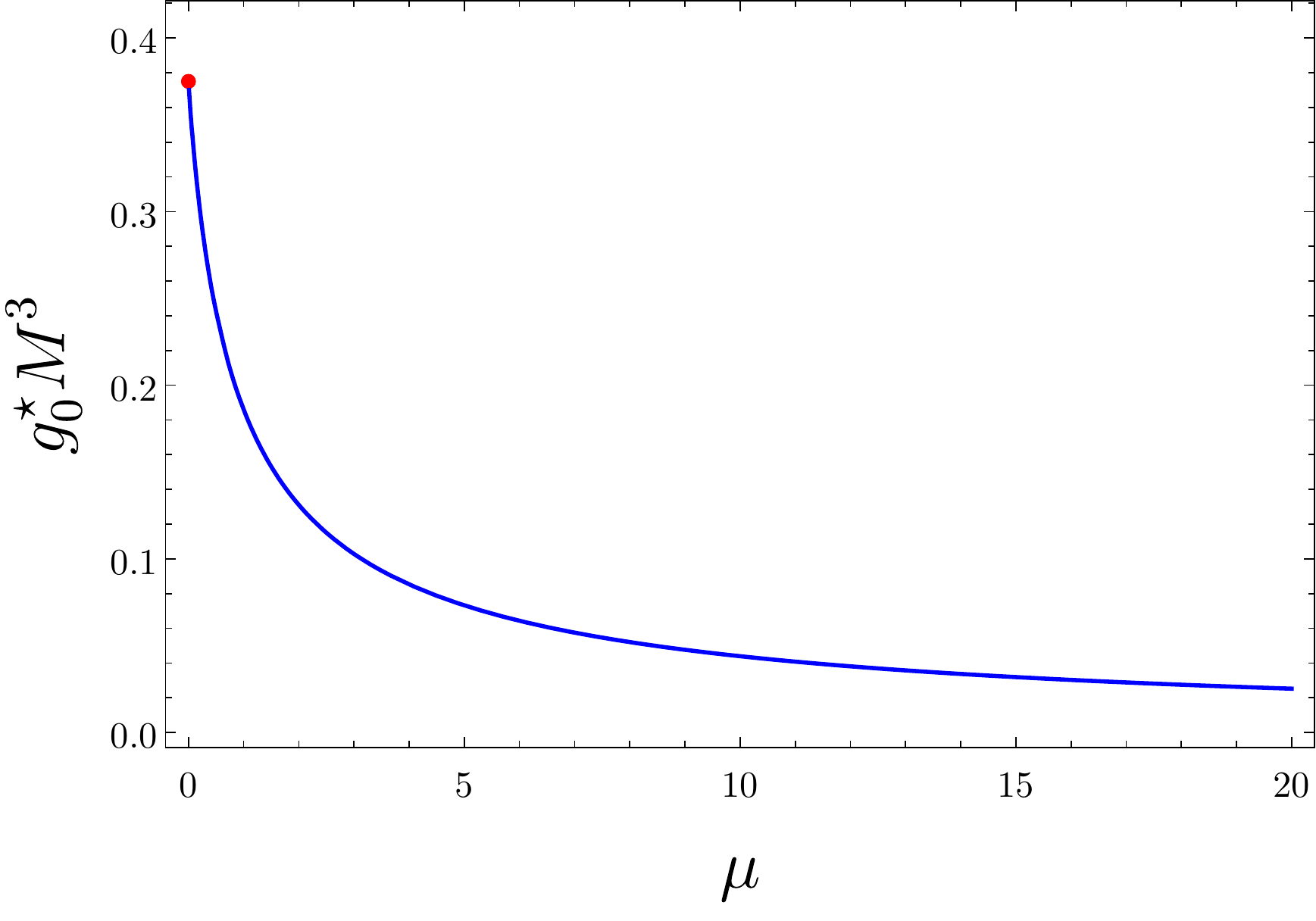}
\caption{Here we show an interpolation of the numerically determined value of $\ats$ (left) and $\gzs$ (right) as functions of the higher-curvature coupling $\mu$. The red dots indicate the values for the Einstein gravity solution, which read: $a_2 = -1/(4 M^2)$ and $g_0 = 3/(8 M^3)$.  We have set $L = M$.}
\label{solParams}
\end{figure}

Of course, \textit{a priori} $a_2$ is a free parameter in the near horizon solution.  However, not all choices of $a_2$ will lead to a sensible solution at large distances. In fact, we find that $a_2$ must take on a uniquely determined value to  join  the numerical solution smoothly to the asymptotic expansion and avoid exciting the exponentially growing mode in~\eqref{f-flatasymp} --- we refer to this value as $\ats$. The shooting method is used to determine this special value of $\ats$, using the following idea. For a given choice of $M$ and the coupling $\mu$, a value $r_{\rm max}$ is chosen by the requirement that the asymptotic solution is a good approximation for $r > r_{\rm max}$. Then for a choice of $a_2$ the field equation is solved numerically, giving a value $f_{\rm numeric}(r_{\rm max}; a_2)$ that can be compared with the asymptotic expansion. The scheme is successful when the numerically determined result $f_{\rm numeric}(r_{\rm max}; a_2)$ agrees with the asymptotic expansion at $r_{\rm max}$ --- the result for $\ats$ is shown in the left plot of Figure~\ref{solParams}.\footnote{An additional step that one can perform is to compute the integrated residual $\int_{r_{\rm max}-\delta}^{r_{\rm max}} |f_{\rm numeric}(r) - f_{\rm asymp}(r)| dr$ where $\delta$ is some small positive quantity, and determine $\ats$ through its minimization. This eliminates the possiblity that perhaps the numerical solution simply passes through the asymptotic solution at $r_{\rm max}$.} By scanning the parameter space (see \eg~\cite{Hennigar:2018hza, NewTaub2}) it can be confirmed that there is only one value of $\ats$ for which this happens. In practice, $\ats$ must be computed with very high precision in order to numerically solve the field equations at large distances. Inevitably, for any chosen precision, the numerical method will fail for large enough $r$, though this point of failure can be pushed to larger distances by increasing the precision of $\ats$. The idea then is to use the numerical scheme to compute the solution up to $r_{\rm max}$, and then use the asymptotic solution to continue it toward infinity.

Once the value of $\ats$ is determined through the procedure just described, the solution for $r < \rh$ can be constructed. In this case the idea is to choose $\epsilon$ to be some small negative number to construct initial data just inside the horizon. The solution can then be constructed numerically for all $r < \rh$, and no issues arise.

The basic idea for constructing the numerical solution for $g(r)$ is the same as for $f(r)$.  Since the equation for $g(r)$ depends on $f(r)$ we first construct a numerical solution for $f(r)$ as just described.  We then use the near horizon expansion for $g(r)$ to establish initial data for the numerical routine. The near horizon solution for $g(r)$ does not fix the parameter $g_0$, and so we once again have a one-parameter family of initial data. However, just as was the case for $a_2$ in the solution for $f(r)$, we find that $g_0$ must take on a uniquely fixed value $\gzs$ so that the numerical solution connects smoothly with the asymptotic expansion for $g(r)$, and avoids exciting the growing mode. The specific value for $\gzs$ is determined in a way completely analogous to the value of $\ats$: using the shooting method, terminating the procedure when the numerically constructed solution $g_{\rm numeric}(r;g_0)$ agrees with the asymptotic solution at a large value of $r$ where the asymptotic solution is a good approximation --- the result is shown in the right plot of Figure~\ref{solParams}.

\begin{figure}
\centering
\includegraphics[scale=0.45]{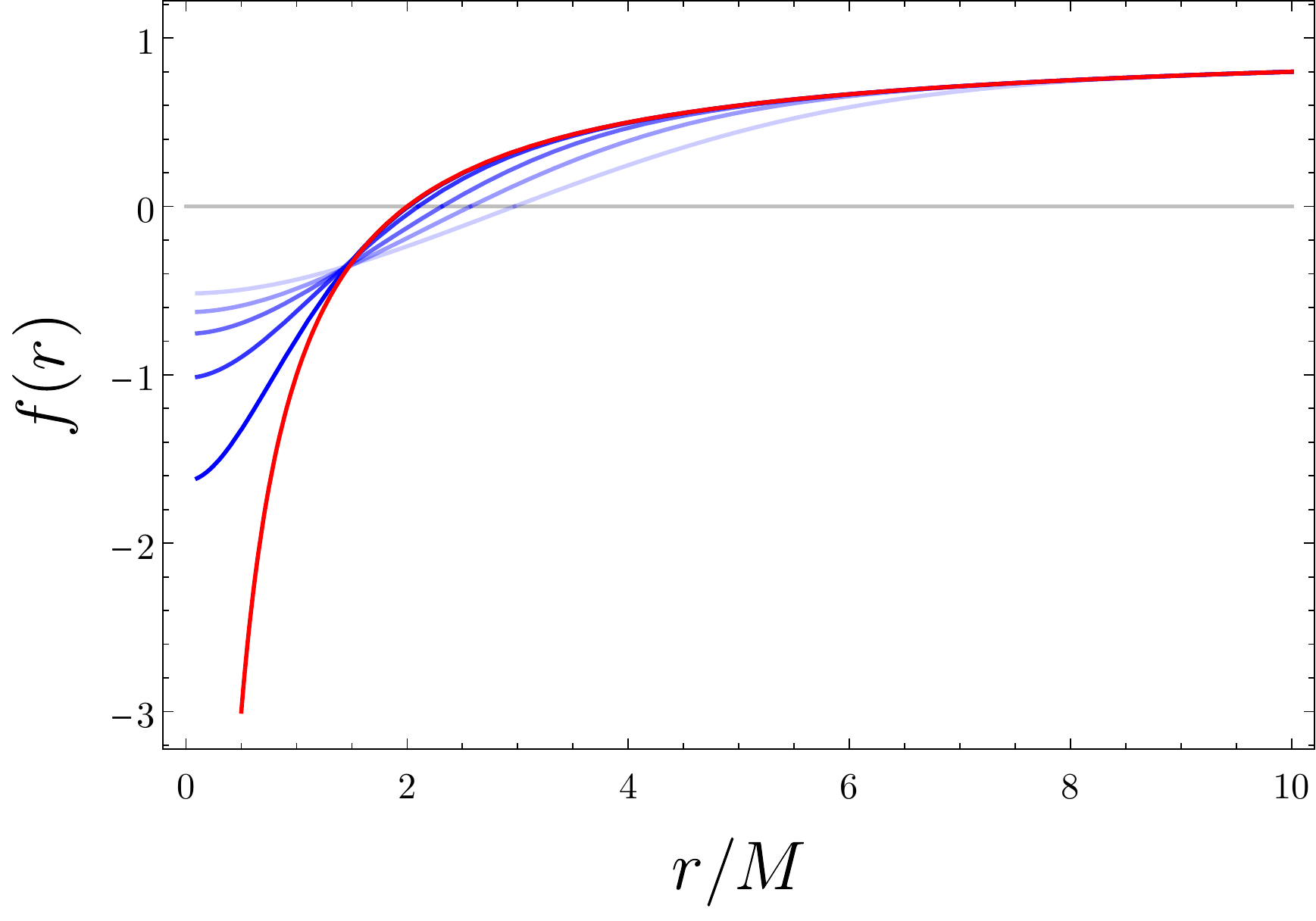}
\quad 
\includegraphics[scale=0.45]{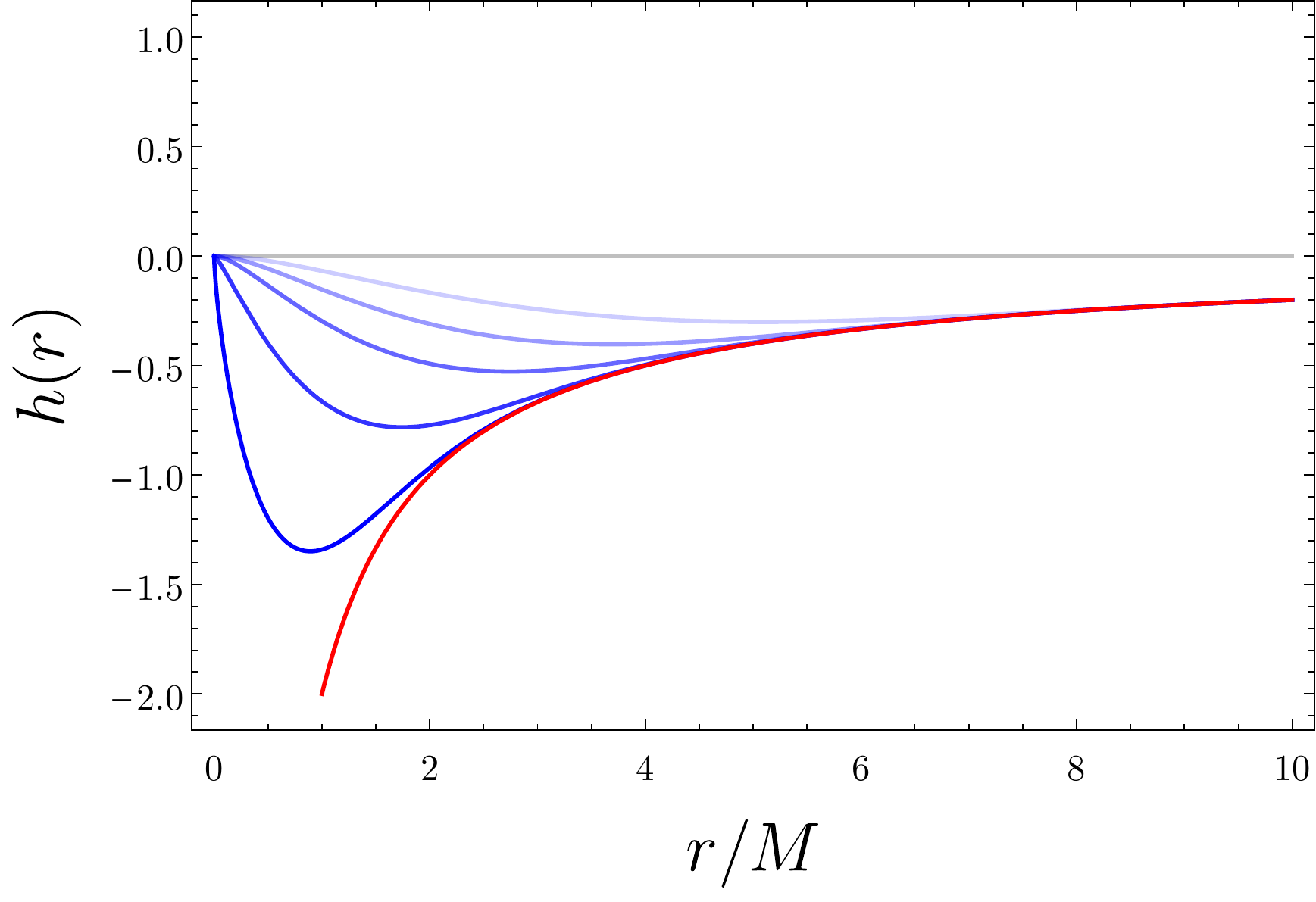}
\caption{Here we show numerical profiles for the metric functions $f(r)$ (left) and $h(r) \equiv r^2 p(r)$ (right). In each case, the blue curves correspond to $\mu = 0.1, 1, 5, 15, 50$ in order of decreasing opacity, while the red curves illustrate the profile in Einstein gravity, given by $f(r) = 1 - 2 M/r$ and $h(r) = -2 M/r$.}
\label{fgplots}
\end{figure}

Once the solution for $g(r)$ is constructed, it can then be numerically integrated to determine $p(r)$. We start the numerical integration at $r = \infty$ so that $p(r) \to 0$ as $r\to\infty$, ensuring that the solution is written in a frame that does not rotate at infinity.  In Figure~\ref{fgplots} we show the numerically constructed profiles for the metric functions for different choices of the coupling. On the left we show $f(r)$, where we see that the effect of the higher-curvature corrections is to push the horizon outward and ensure that $f(r) \to \text{constant}$ as $r\to 0$. On the right we show $h(r) = r^2 p(r)$, which is the combination appearing in $g_{t\phi}$, neglecting the spin parameter $a$ and the angular piece $(1-x^2)$. Here the curves all approach $h(r) \to 0$ as $r\to 0$, rather than decaying as $r^{-1}$ as is the case in Einstein gravity.

The fact that $h(r)$ is bounded in all its domain has some interesting consequences.  For the Kerr solution, the slowly rotating approximation only works for large enough $r$. If $\chi=a/M\ll 1$ this approximation is valid up to the level of the horizon, but it always breaks down when $r\rightarrow 0$, because $g_{t\phi}$ diverges and rotation becomes important. However, in the case of ECG we can see that $g_{t\phi}=a h(r) \sin^2\theta$ is bounded for all values of $r$ (more importantly, $g_{t\phi}/g_{\phi\phi}$ is bounded). This means that if $\chi$ is small enough so that the maximum value of $g_{t\phi}/g_{\phi\phi}$ is also sufficiently small, the slowly rotating approximation could be valid for all $r$. More interestingly, we observe that when $M\to 0$ and we keep $\chi$ constant, the quantity $a h(r)=\chi M h(r)$ goes to zero everywhere. This is telling us that, in the regime where $M\ll L\mu^{1/4}$, the slowly rotating approximation is probably valid even for large $\chi$, since the effect of rotation is almost negligible. This seems to indicate that the maximum value for the angular momentum in these black holes could be larger than in general relativity. This is, there would be black holes with $a>M$ (and even with $a\gg M$ if $M$ is small). But in order to check this one would need at least to compute the solution at order $\mathcal{O}(a^2)$, to ensure that these terms are indeed irrelevant when $M\ll L\mu^{1/4}$.

\begin{figure}
\centering
\includegraphics[scale=0.5]{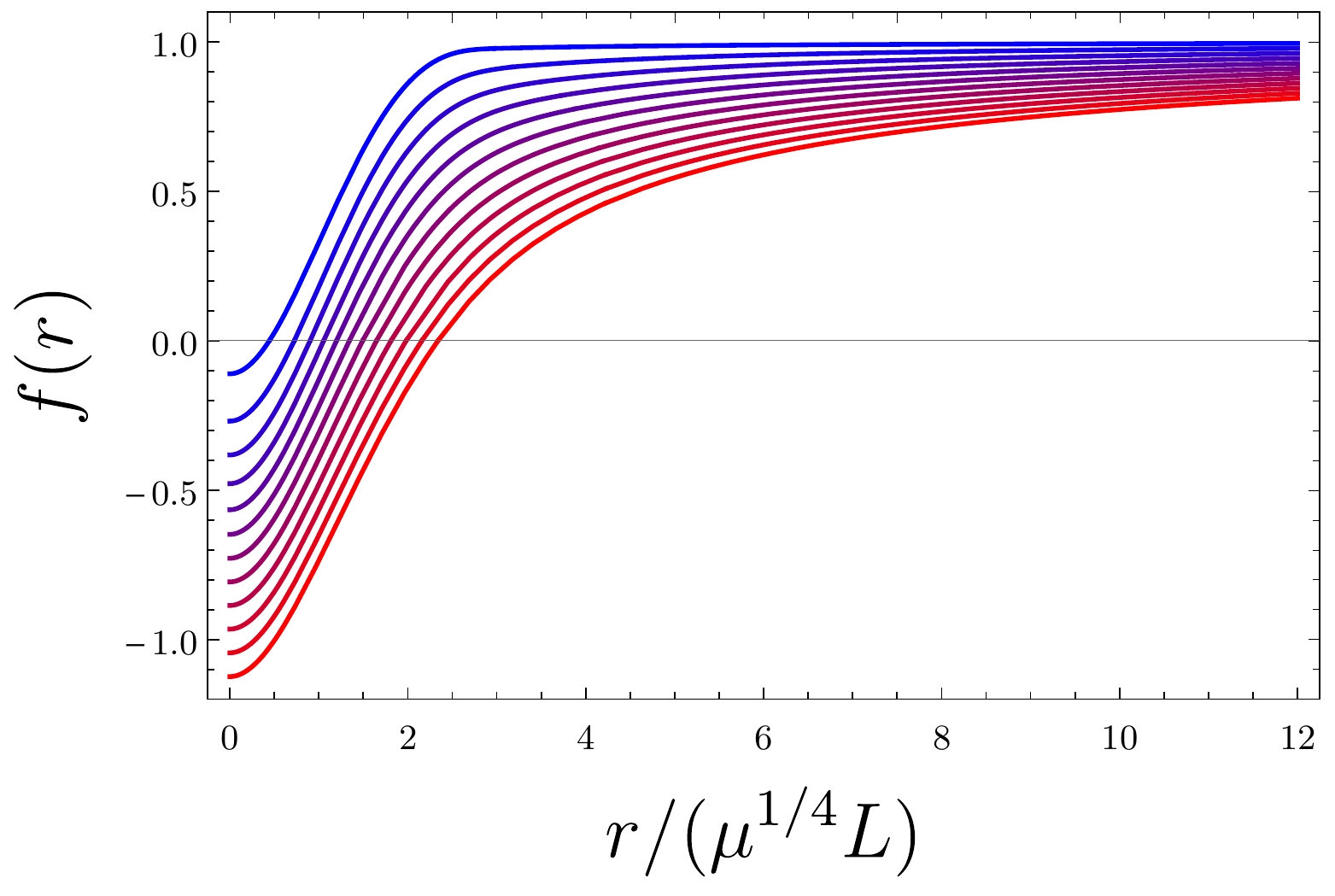}
\quad 
\includegraphics[scale=0.5]{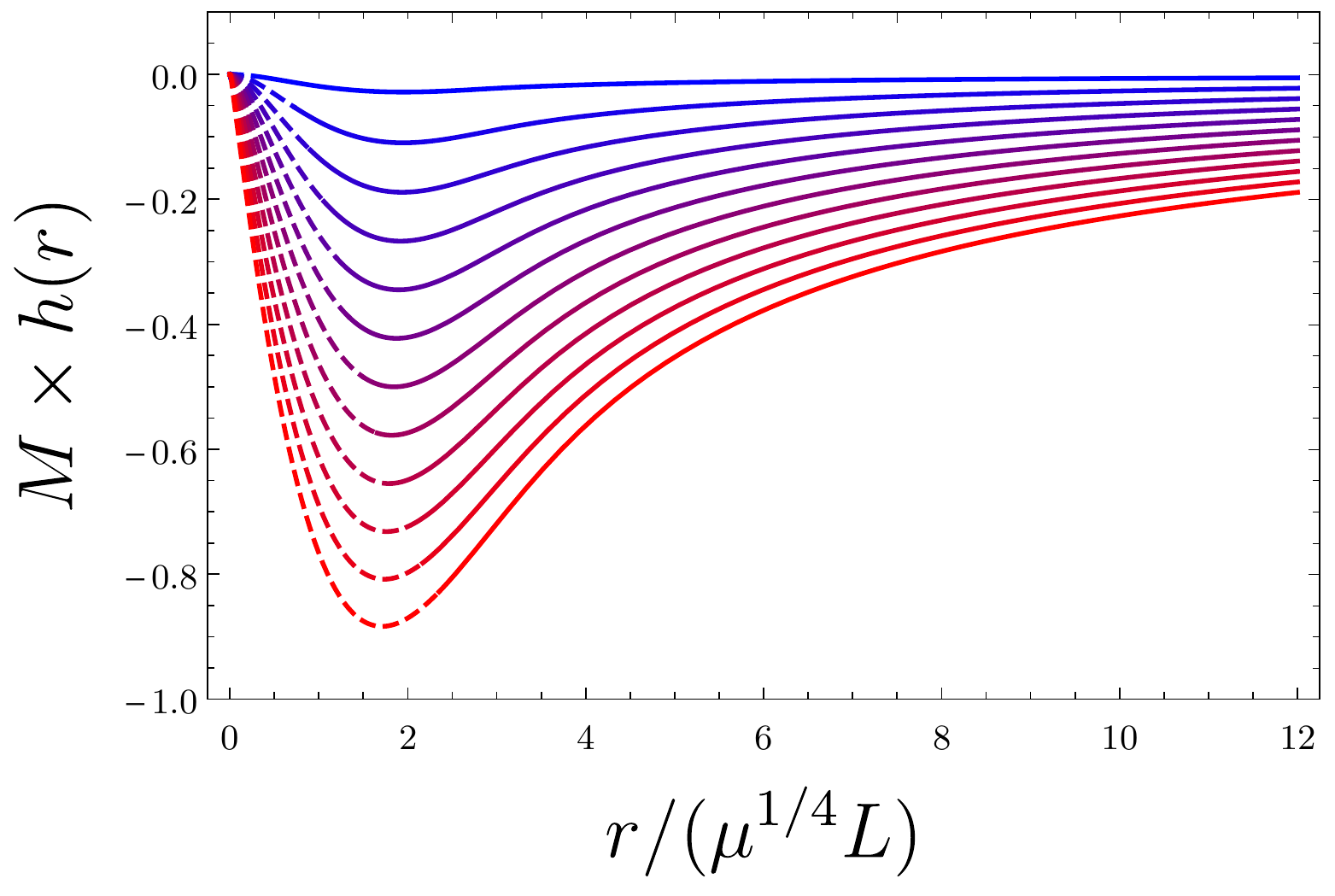}
\caption{Numerical profiles for the metric functions $f(r)$ (left) and $M h(r) = M r^2 p(r)$ (right) for fixed $\mu L^4$ and several values of the mass. From red to blue curves the mass ranges from $\tfrac{34}{30} \mu^{1/4}L$ to $\tfrac{1}{30}  \mu^{1/4}L$. For fixed $\chi=a/M$, rotation becomes irrelevant when $M\ll\mu^{1/4} L$, and the zero-mass limit seems to correspond to a non-rotating extremal black hole of vanishing area. The dashed lines in the right plot correspond to the part of the solution which is inside the horizon.}
\label{fhplotsMass}
\end{figure}

In Fig.~\ref{fgplots} we have plotted $f(r)$ and $h(r)$ for a fixed value of the mass and various values of the coupling $\mu L^4$, in order to explore how the solution is affected by the ECG correction. However, in practice we would have a fixed value of $\mu L^4$ and black holes of several masses. The effects of the corrections outside of the horizon only become relevant when the mass (or the horizon radius) is of the order of $\mu^{1/4} L$, and it is interesting to see how the profile of the solution changes with the mass. In Fig.~\ref{fhplotsMass} we show the profiles of $f(r)$ and $M h(r)$ for fixed $\mu L^4$ and various masses. As we mentioned, $M h(r)$ goes to 0 everywhere for small masses. We also observe a very interesting phenomenon in the case of $f(r)$ when $M\rightarrow 0$. While the radius $\rh$ vanishes in that limit, the solution develops a potential well of length $\sim \mu^{1/4} L$ that is present for arbitrarily small mass. Thus, the zero mass limit of these black holes is not flat space. Instead, it seems to a be a massless, non-rotating extremal black hole whose horizon is reduced to a point (and coincides with a curvature singularity). This intriguing behavior will prove to have remarkable observational consequences.

\section{Properties of the solution}\label{props}
In this section we study several physical properties of the solutions constructed above. Firstly, we evaluate the angular velocity of the horizon as a function of the mass. Then, we move on to the study of geodesics. We find the general equations and then we compute the photon sphere. Then, restricting the discussion to geodesics in the equatorial plane, we compute the innermost stable circular orbit for timelike geodesics,  photon rings and the Lyapunov exponents associated with their instability, and finally we study how the black hole shadow is modified with respect to the Einstein gravity case.

\subsection{Angular velocity of the horizon}

%\comment{words and plots to appear here}

The angular velocity of the horizon is defined as
\begin{equation}
\Omega=-\frac{g_{t\phi}}{g_{\phi\phi}}\bigg|_{r=\rh}\, .
\end{equation}
\begin{figure}[t]
\centering
\includegraphics[scale=0.65]{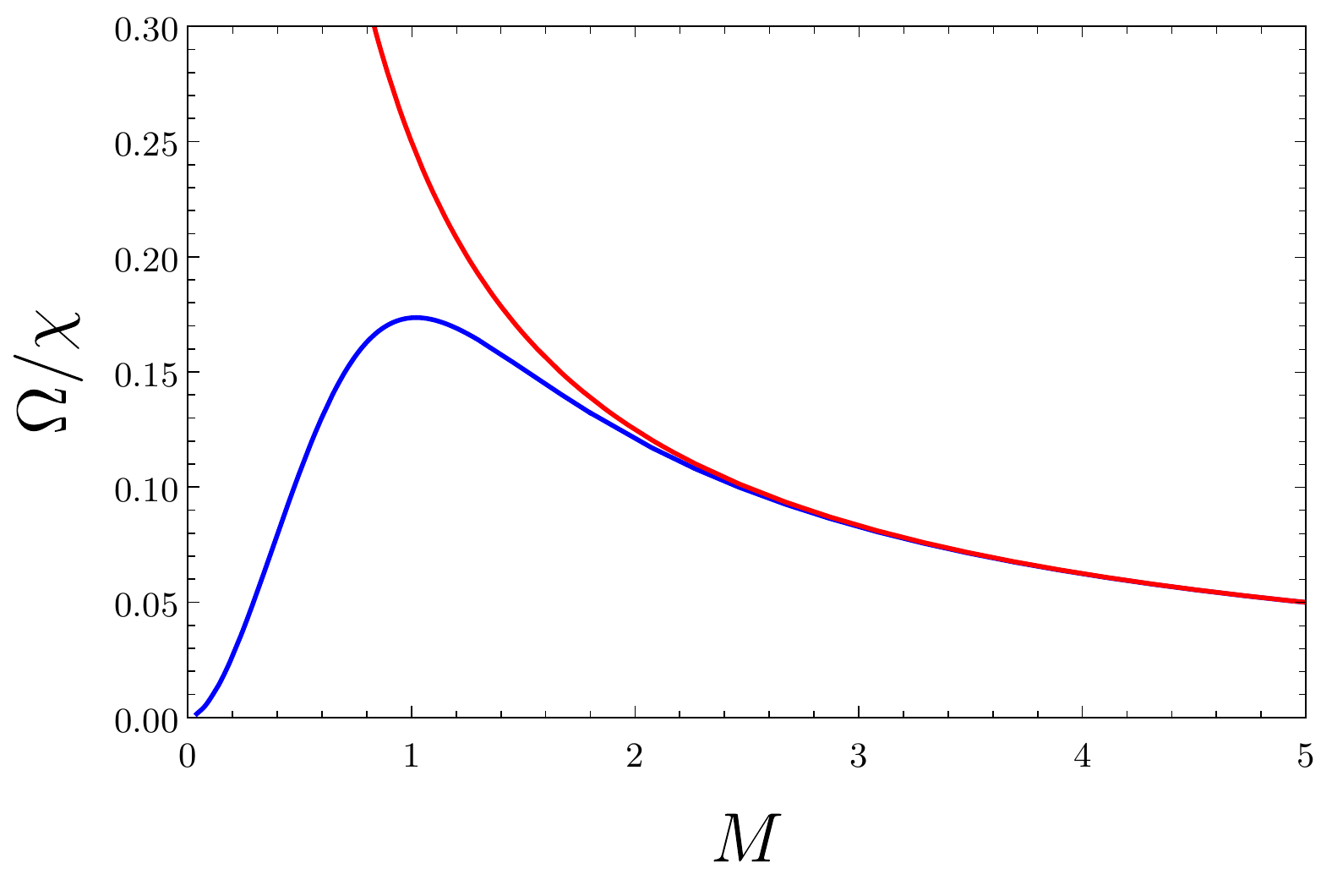}
\caption{Angular velocity of the horizon as a function of the mass for fixed $\chi$. In red we show the result in Einstein gravity $(\Omega=\chi/(4M)$) and in blue the case of ECG. The mass $M$ is expressed in units of $\mu^{1/4} L$ and $\Omega$ in the reciprocal units.}
\label{Omegaplot}
\end{figure}
In the perturbative regime, using Eqs.~\eqref{eq:fpert} and \eqref{eq:ppert}, we get 
\begin{equation}
\Omega=\frac{\chi}{4 M}\left[1-\frac{35 \mu  L^4}{64 M^4}+\frac{26199 \mu ^2 L^8}{53248 M^8}+\ldots\right]\, .
\end{equation}
This expression works nicely for $M \gtrsim 1.7 \mu^{1/4} L$. For smaller values of $M$ we need to use the numerical solution, and in that case we get the profile of $\Omega$ shown in Fig.~\ref{Omegaplot}. We observe that, unlike the Einstein gravity case, $\Omega$ no longer diverges for $M\rightarrow 0$. Instead, there is a maximum value $\Omega_{\rm max}$ that is reached at a mass $M(\Omega_{\rm max})$:
\begin{equation}
\Omega_{\rm max}\approx \frac{0.1736\chi}{\mu^{1/4} L}\, ,\quad M(\Omega_{\rm max})\approx 1.02 \mu^{1/4} L\, .
\end{equation}
When $M\rightarrow 0$ the angular velocity also vanishes. Let us note that if we take $\chi$ to be small enough, the angular velocity remains small for every value of the mass, and hence the slowly spinning approximation should work for all masses. Nevertheless, for larger $\chi$ this approximation will break down at some point, and we expect the relation shown in Fig.~\ref{Omegaplot} to be modified at small masses. An interesting question is whether a maximum value of $\Omega$ would persist in that case. In order to answer it we would need to find the solution at least at order $\mathcal{O}(a^2)$.% \comment{Perhaps estimate the relation $\Omega(M)$ when $M\rightarrow 0$}

%\rbm{More comments here?  Should we expect this maximum to persist once higher orders in $a$ are taken into account?}

\subsection{Geodesics}\label{sec:Geodesics}
Let us study the geodesics of the slowly rotating solution. We can do that in general for a metric of the form 
\be\label{eq:rotatingmetric}
ds^2 = -f(r)dt^2 + \frac{dr^2}{f(r)} + 2 a h(r) \sin^2\theta dt d\phi + r^2 \left(d\theta^2 + \sin^2\theta d\phi^2 \right) \,.
\ee
The geodesics are most conveniently studied by introducing the Lagrangian
\be
L=\frac{1}{2}g_{ab}\dot{x}^{a}\dot{x}^{b}\, ,
\ee
where $\dot{x}^{a}=dx^{a}/d\lambda$, and where $\lambda$ is an affine parameter. From the Lagrangian we may construct the conjugate momenta and the Hamiltonian, that read
\be
p_{a}=g_{ab}\dot{x}^{b}\, ,\quad H=\frac{1}{2}g^{ab}p_{a}p_{b}\, .
\ee
Then, we can write the Hamilton-Jacobi equation,
\be
\frac{\partial S}{\partial \lambda}=\frac{1}{2}g^{ab}\frac{\partial S}{\partial x^{a}}\frac{\partial S}{\partial x^{b}}\, .
\ee
If we manage to solve this equation for the function $S$, then we get the solutions of the equations of motion using
\be
p_{a}=\frac{\partial S}{\partial x^{a}}\, .
\ee

We  apply this method for the metric \req{eq:rotatingmetric}. Since the Hamiltonian does not depend explicitly on $t$, $\phi$ and $\lambda$, we can write
\be
S=-\frac{1}{2}\xi^2\lambda-E t+\ell_z \phi+\tilde S(r,\theta)\, ,
\ee
where $\xi^2$, $E$ and $\ell_{z}$ are constants. Let us further assume that the function $\tilde S(r,\theta)$ can be expressed as $\tilde S(r,\theta)=\tilde S_{r}(r)+\tilde S_{\theta}(\theta)$.  Inserting this into the Hamilton-Jacobi equation, we realize that it is indeed separable:\footnote{We only keep terms linear in the spin $a$.}
\be
\left(\frac{\partial S_{\theta}}{\partial\theta}\right)^2+\frac{\ell_z^2}{\sin^2\theta}=-r^2\xi^2+\frac{r^2E^2}{f(r)} + \frac{2ah(r)E\ell_{z}}{f(r)}-r^2 f(r) \left(\frac{\partial S_{r}}{\partial r}\right)^2\, .
\ee
Since the left-hand-side only depends on $\theta$ while the right-hand-side only depends on $r$, we conclude that both terms are equal to a constant, that we may call $j^2$. Therefore, we have determined all the derivatives of $S$,
\begin{eqnarray}
\frac{\partial S}{\partial t}&=&-E\, ,\\
\frac{\partial S}{\partial \phi}&=&\ell_{z}\, .\\
\frac{\partial S}{\partial \theta}&=&\pm\sqrt{j^2-\frac{\ell_{z}^2}{\sin^2\theta}}\, ,\\
\frac{\partial S}{\partial r}&=&\pm\sqrt{-\frac{j^2}{r^2 f(r)}-\frac{\xi^2}{f(r)}+\frac{E^2}{f(r)^{2}} + \frac{2ah(r)E\ell_{z}}{r^2f(r)^{2}}}\, .
\end{eqnarray}
Now we substitute $\partial_{a}S\rightarrow p_{a}$, where the momenta read
\begin{eqnarray}
p_{t}=-f(r)\dot{t}+ah(r)\sin^2\theta \dot{\phi}\, ,\quad 
p_{\phi}=a h(r)\sin^2\theta \dot{t}+r^2\sin^2\theta \dot{\phi}\, ,\quad
p_{\theta}=r^2\dot{\theta}\, ,\quad
p_{r}=\frac{\dot{r}}{f(r)}
\end{eqnarray}
yielding a system of first-order equations,
\begin{eqnarray}\label{geodesicEqns}
r^2\dot{t}&=&\frac{Er^2+a h(r)\ell_z}{f(r)}\, ,\\
r^2\dot{\phi}&=&\frac{\ell_{z}}{\sin^2\theta}-\frac{a h(r)E}{f(r)}\, ,\\
r^2\dot{\theta}&=&\pm\sqrt{j^2-\frac{\ell_{z}^2}{\sin^2\theta}}\, ,\\ \label{rEq}
\dot{r}^2&=&-f(r)\left(\xi^2+\frac{j^2}{r^2}\right)+E^2+\frac{2ah(r)E\ell_{z}}{r^2}\, ,
\end{eqnarray}
where again we are expanding linearly in $a$. It is clear that, asymptotically, $j^2$ represents the total angular momentum of the orbit, while $\ell_{z}$ is the component of the angular momentum in the $z$ axis (this is, $\theta=0,\pi$). On the other hand, $\xi^2$ is the norm of the tangent vector
\begin{equation}
\xi^2=-g_{ab}\dot{x}^{a}\dot{x}^{b}\, .
\end{equation}

\subsubsection{The photon sphere}
%\comment{Apparently, the only error beyond $V_{ph}$ was in the sign in front of $a$. I have fixed this here and in the section 3.4}
For null geodesics, we have $\xi^2=0$, and rescaling the affine parameter $\lambda$, we can always choose $E=1$. Let us then write the equation for the radial coordinate as
\begin{equation}
\dot{r}^2+V_{\rm ph}(r)=0\, ,\quad \text{where}\, \quad V_{\rm ph}(r)=\frac{j^2f(r)-2a\ell_{z}h(r)}{r^2}-1\, .
\end{equation}
The photon sphere is formed by constant-$r$ photon orbits, that appear when
\begin{equation}
V_{\rm ph}(r_{\rm ps})=0\, ,\quad  V'_{\rm ph}(r_{\rm ps})=0\, .
\end{equation}
These conditions give us the radius of these orbits $r_{\rm ps}$ as well as the value of $j_{\rm ps}^2$. Since we are working perturbatively in $a$, let us write these quantities as
\begin{equation}
r_{\rm ps}=r_{\rm ps}^{(0)}+a r_{\rm ps}^{(1)}\, ,\quad j_{\rm ps}^2=(j_{\rm ps}^{(0)})^2+a(j_{\rm ps}^{(1)})^2\, .
\end{equation}
We find that $r_{\rm ps}^{(0)}$ is determined by the equation
\begin{equation}
r_{\rm ps}^{(0)}f'(r_{\rm ps}^{(0)})-2 f(r_{\rm ps}^{(0)})=0\, ,
\end{equation}
and the rest of quantities read
\begin{eqnarray}
r_{\rm ps}&=&r_{\rm ps}^{(0)}+\frac{2a\ell_{z}f(rh'-2h)}{r(r^2f''-2f)}\bigg|_{r=r_{\rm ps}^{(0)}}\, ,\\
j_{\rm ps}^2&=&\frac{(r_{\rm ps}^{(0)})^2}{f(r_{\rm ps}^{(0)})}+\frac{2a\ell_{z}h(r_{\rm ps}^{(0)})}{f(r_{\rm ps}^{(0)})}\, .
\end{eqnarray}
Thus, every choice of $\ell_{z}$ produces a family of constant-$r$ geodesics. Observe that in the case of equatorial orbits we have $\ell_z=\pm |j_{\rm ps}|$ and this fixes two types of orbits, either prograde or retrograde. 

In the case at hand, when $\mu L^4/M^4\ll1$ we may use the perturbative solution and we get
\begin{eqnarray}
r_{\rm ps}&=&3 M+\frac{13 \mu  L^4}{81 M^3}+\frac{1295 \mu ^2 L^8}{59049 M^7}-\frac{a\ell_{z}}{M}\left[\frac{2}{9}-\frac{460 \mu  L^4}{6561 M^4}+\frac{764 \mu ^2 L^8}{1594323 M^8}\right] \, .\\
j_{\rm ps}^2&=&27 M^2+\frac{35 L^4 \mu }{27 M^2}-\frac{2806 L^8 \mu ^2}{19683 M^6}+\frac{a\ell_{z}}{M}\left[4 M-\frac{140 L^4 \mu }{243 M^3}-\frac{48790 L^8 \mu ^2}{2302911 M^7}\right]\, .
\end{eqnarray}
On the other hand, when $\mu L^4/M^4$ becomes of order 1 or larger we need to use the numerical solution --- see section~\ref{sec:shadow} below.

\subsection{Geodesics in the equatorial plane}

%\comment{Pasting some rough notes/plots in here. I will flesh this out more when I wake up.} \comment{1 year later: I have woken up.}

Here we consider geodesics confined to the equatorial plane, \ie $x=0$. For this purpose, we will specialize the results from the previous subsection to this situation, which amounts to setting $\theta = \pi/2$ and $\dot{\theta} = 0$. Note that, via  these two constraints it is enforced that $j^2 = \ell_z^2$. We can understand \req{rEq} %have then
% Lagrangian
%\begin{align} 
%L &=  \frac{1}{2} g_{\alpha\beta} \dot{x}^\alpha \dot{x}^\beta  =   \frac{1}{2} \left[-f(r) \dot{t}^2 + \frac{\dot{r}^2}{f(r)} + 2 a r^2 p(r) \dot{t} \dot{\phi} + r^2 \dot{\phi}^2 \right]  \, ,
%\end{align}
%where we have used the notation $\dot{x}^\alpha \equiv d x^\alpha/d\lambda$ with $\lambda$ parametrizing the geodesics. The equations must obey the additional constraint that $ L = -\epsilon/2$ with $\epsilon = 0$ for null geodesics and $\epsilon = 1$ for timelike geodesics. 
%
%Since $t$ and $\phi$ make no explicit appearance in the Lagrangian their conjugate momenta represent quantities conserved along geodesics,
%\begin{align}
%-p_t &= -\frac{\partial L}{\partial \dot{t}} =  f(r) \dot{t} - a r^2 p(r) \dot{\phi} =  E = \text{constant} \, , 
%	\nn
%p_\phi &= 	 \frac{\partial L}{\partial \dot{\phi}} =  a r^2 p(r) \dot{t} +  r^2 \dot{\phi} =  \ell = \text{constant} \, .
%\end{align}
%Solving these expressions for $\dot{t}$ and $\dot{\phi}$ gives
%\be 
%\dot{t} = \frac{E + a p(r) \ell }{f(r)} \, , \quad \dot{\phi} = \frac{f(r) \ell - a r^2 p(r) E}{r^2 f(r)}  \, .
%\ee
%Using these expressions, the Lagrangian for the geodesic motion can be re-written in terms of the constants of motion. After a bit of massaging the result reads
%\be 
%\dot{r}^2 = E^2 - \xi^2 f(r)  - \frac{j^2 f(r)}{r^2} + \frac{ 2 a h(r) E \ell_z}{r^2} \, ,
%\ee
%where we have neglected terms ${\cal O}(a^2)$ in the spin parameter. We can view this equation 
as analogous to that describing a particle moving in a potential,
\be \label{Veff}
\dot{r}^2 + V_{\rm eff}(r) = 0 \, , \quad \text{where } \quad V_{\rm eff}(r) \equiv  f(r) \left( \mu^2  + \frac{j^2 }{r^2} \right)  - \frac{2 a h(r) E j}{r^2} - E^2 \, .
\ee
In the following two sub-sections, we will consider the special case of circular orbits, determining the inner-most stable circular orbit (ISCO) for timelike geodesics, and the photon ring for null geodesics.

\subsubsection{Timelike geodesics: ISCO}

First let us consider the case of circular, timelike geodesics, \ie those with $\xi^2=1$ and $\dot{r} = 0$. The conditions for the existence of these geodesics are
\be 
V_{\rm eff}(r) = 0\, , \quad  V'_{\rm eff}(r)  = 0\, ,
\ee
and the stability of the circular orbit can be deduced by considering the sign of $V''_{\rm eff}(r)$, with $V''_{\rm eff}(r) > 0$ indicating stability and $V''_{\rm eff}(r) < 0$ indicating instability. We can determine the location of the inner-most stable circular orbit by searching for circular orbits that are also inflection points, \ie orbits for which $V''_{\rm eff}(r) = 0$, leading to three conditions on the parameters $r$, $E$, and $j$.

\begin{figure}
\centering
\includegraphics[scale=0.45]{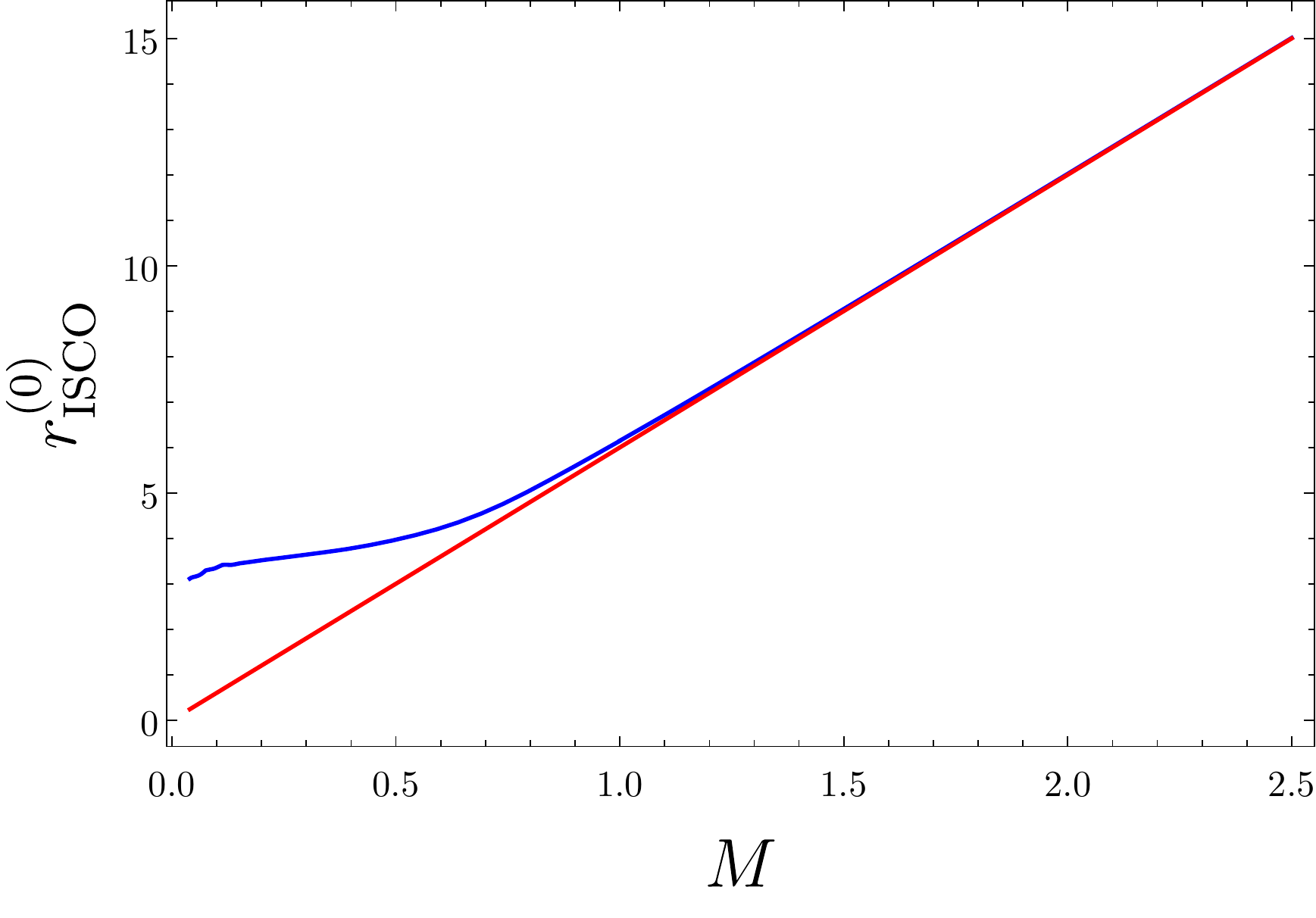}
\quad
\includegraphics[scale=0.45]{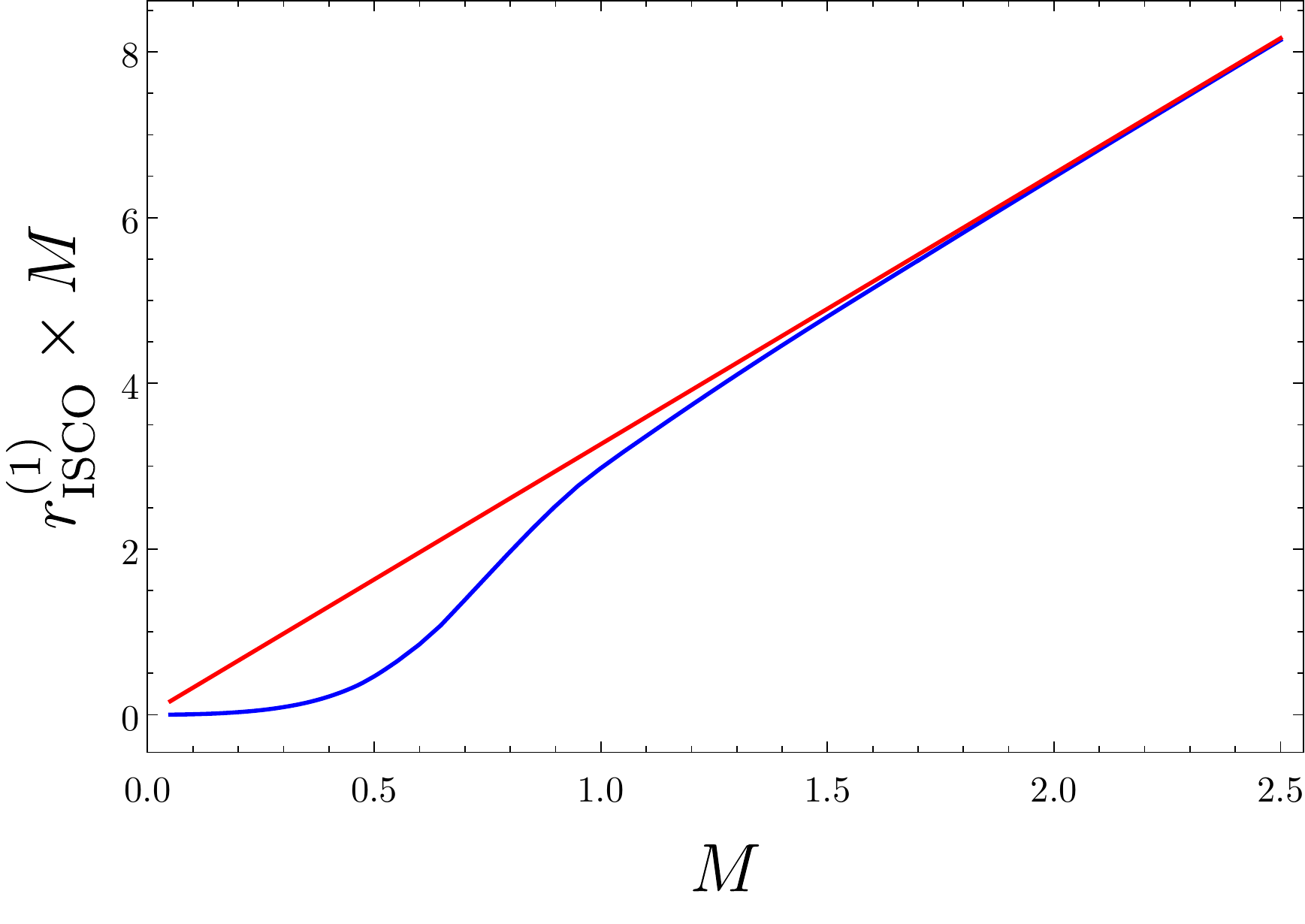}
\includegraphics[scale=0.45]{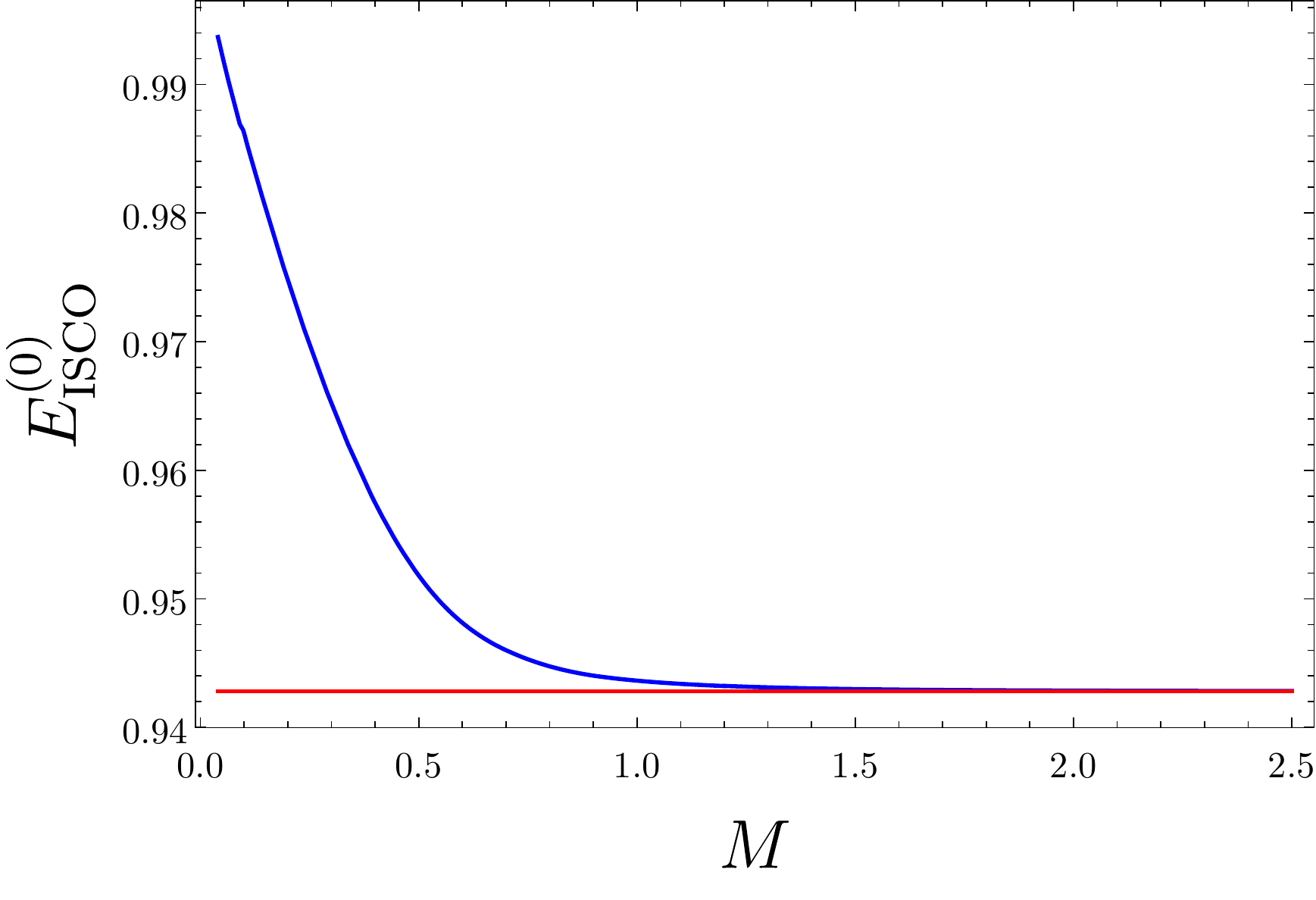}
\quad
\includegraphics[scale=0.45]{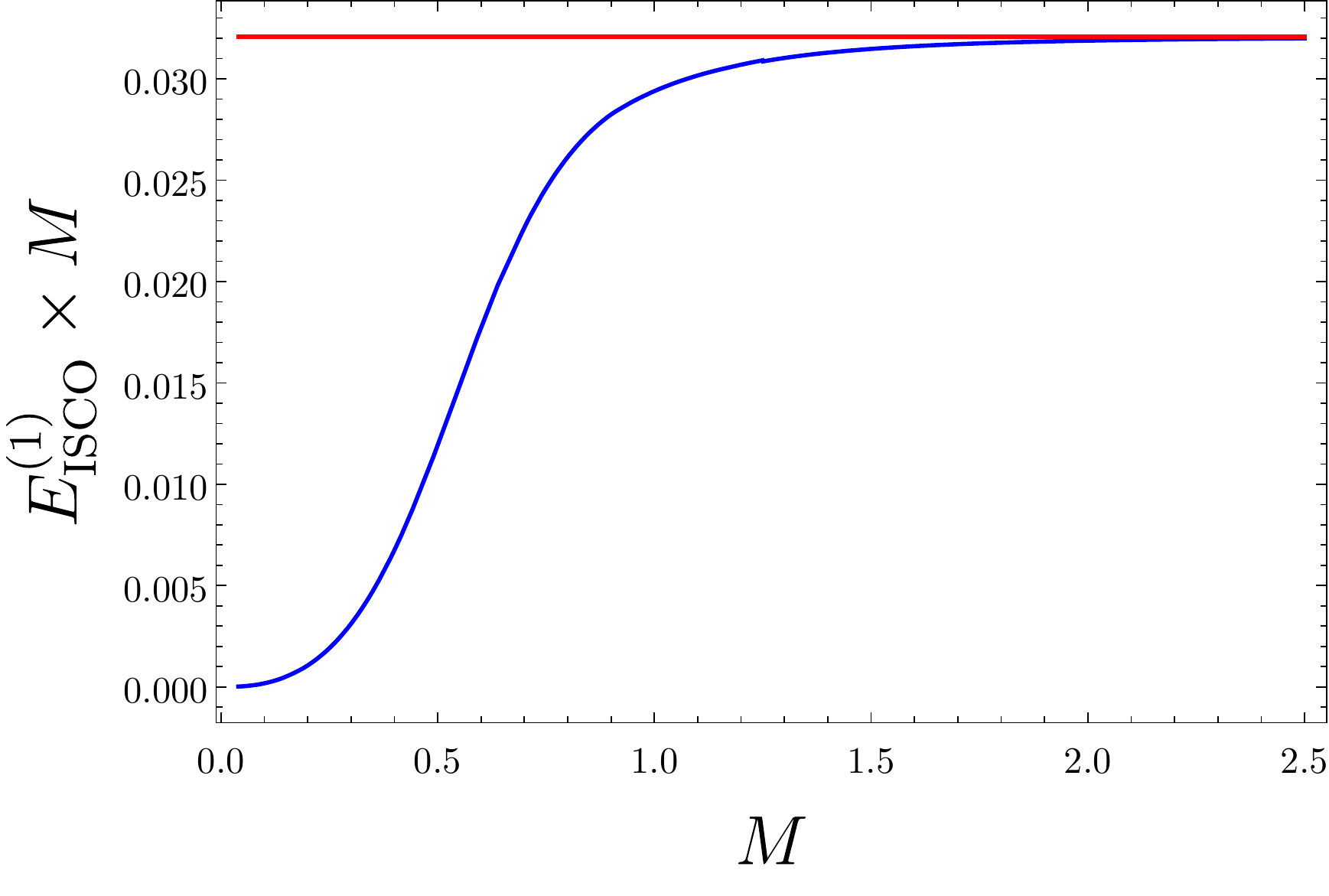}
\includegraphics[scale=0.45]{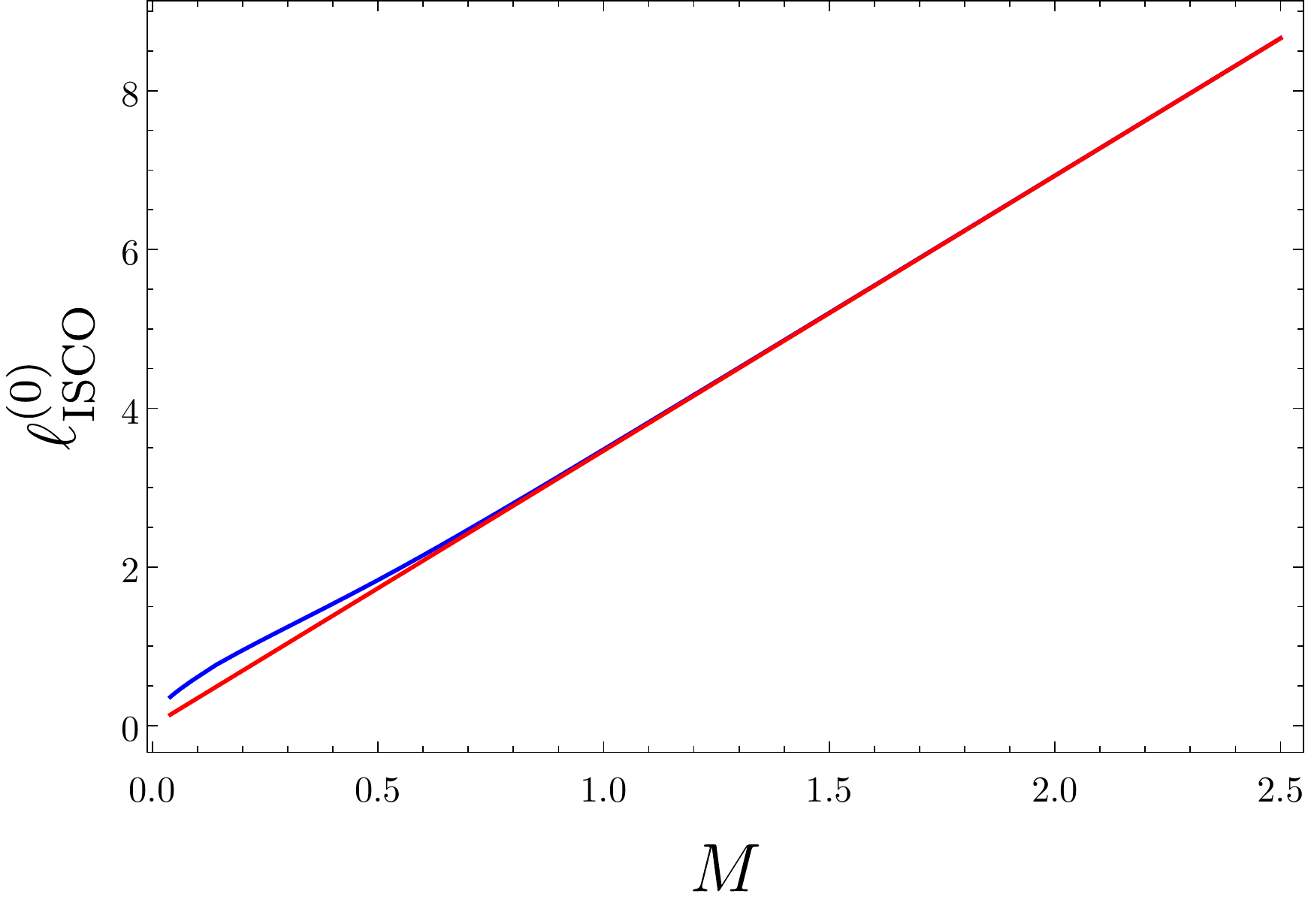}
\quad
\includegraphics[scale=0.45]{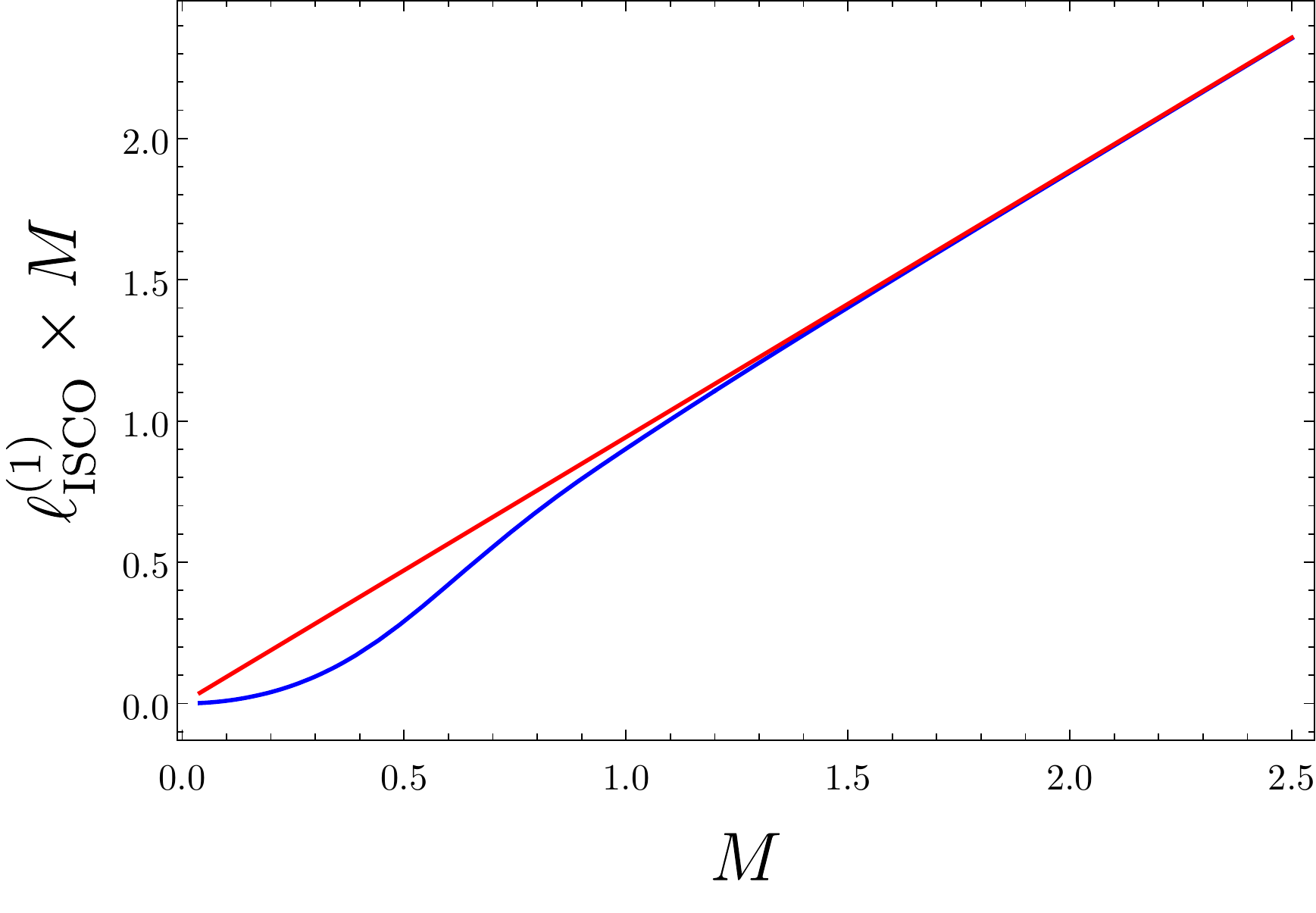}
\caption{Here we plot the parameters describing the ISCO. In the left column we present the zeroth-order terms, which are those corresponding to the static solution. The right column shows the leading-order correction due to rotation. In each case, the blue curves represent the ECG result, while the red curves represent the Einstein gravity result. In all cases the mass is expressed in units of $\mu^{1/4} L$. }
\label{iscoEquator}
\end{figure}

We wish to solve these equations, working to linear order in the rotation parameter. To this end, we make the following definitions:
\begin{align}
r_{\rm ISCO} = r_{\rm ISCO}^{(0)} + a r_{\rm ISCO}^{(1)} \, ,
	\quad
E_{\rm ISCO} = E_{\rm ISCO}^{(0)} + a E_{\rm ISCO}^{(1)} \, ,
	\quad
j_{\rm ISCO} = j_{\rm ISCO}^{(0)} + a j_{\rm ISCO}^{(1)} \, .		
\end{align}
Substituting these into the conditions $d^n V_{\rm eff}(r)/dr^n = 0$ for $n=0,1,2$ and expanding to linear order in $a$ yields a system of six equations that must be solved.  These equations themselves are not particularly illuminating, but let us note that within the perturbative regime they admit the following solution:
\begin{align}
r_{\rm ISCO} &=6 M  \left[1+ \frac{101 \mu L^4}{3888 M^4} - \frac{499601 \mu^2 L^8}{362797056 M^8} \right] \mp  4a \sqrt{\frac{2}{3}} \left[ 1 - \frac{10073 \mu L^4}{93312 M^4} + \frac{92651747  \mu^2 L^8}{17414258688 M^8} \right] \, ,
\\ \notag
E_{\rm ISCO} &=\frac{2 \sqrt{2}}{3} \left[ 1+ \frac{267 \mu L^4}{ 279936M^4} - \frac{782985 \mu^2 L^8}{17414258688  M^8} \right] 
 \mp \frac{a}{18 \sqrt{3} M}  \left[1 - \frac{499 \mu L^4}{5184  M^4} + \frac{20883769 \mu^2 L^8}{2358180864  M^8} \right] \, ,
\\ \notag
j_{\rm ISCO} &= \pm 2 \sqrt{3} M \left[1 + \frac{59 \mu L^4}{11664  M^4} - \frac{355163 \mu^2 L^8}{2176782336  M^8} \right]
-\frac{2 \sqrt{2} a }{3} \left[1 + \frac{4603 \mu L^4}{  93312  M^4}  - \frac{21098789 \mu^2 L^8}{ 5804752896  M^8} \right] \, .
\end{align}
Here taking the upper sign describes the ISCO for prograde orbits ($j_{\rm ISCO}^{(0)} > 0$), while the lower sign describes the ISCO for retrograde orbits ($j_{\rm ISCO}^{(0)} < 0$).  The perturbative solution provides an accurate description when $M/(\mu^{1/4} L) \gtrapprox 1.5$, but for smaller masses we must resort to the numerical solution.  The results for each of the parameters are plotted in Figure~\ref{iscoEquator}. The corrections due to ECG become  most significant for small masses, where they can either increase or decrease the relevant parameters.

\subsubsection{Null geodesics: photon rings}

%Let us start a discussion of geodesics by considering how the rotation deforms the photon rings of the black hole.  The photon rings are circular orbits (i.e. having $r = \text{constant}$) for null geodesics lying in the equatorial plane $x=0$. The geodesic equation reads
%\be 
%\ddot{x}^\mu + \Gamma_{\alpha\beta}^\mu \dot{x}^\alpha \dot{x}^\beta = 0 \, ,
%\ee 
%where $\dot{x}^\nu = dx^\nu/d\lambda$ with $\lambda$ parametrizing the worldline. As usual, we work to linear order in the rotation parameter $a$. Direct computation shows that the $\nu =x$ component of the geodesic equation is identically satisfied, while the $\nu  = t$ and $\nu = \phi$ components enforce $\dot{t} = \text{constant}$ and $\dot{\phi} = \text{constant}$. Taken together, these results imply that the angular velocity $\omega := d\phi/dt$ is constant.  The remaining $\mu = r$ component of the geodesic equation reads
%\be
%\label{phring} 
%0 = f(r) \left[ 2 \omega^2 r + 2 a \omega r \left(p'(r)  + 2 r p(r) \right) - f'(r) \right] \, .
%\ee
%This equation is satisfied trivially on the horizon where $f(r) = 0$, but we are interested in the case where it is the term in square brackets that vanishes. In this case, the above equation can be used to determine the values of $\omega$. To determine the radius of the photon orbit we must take into account the fact that we are interested in null geodesics which must satisfy $g_{\alpha\beta} \dot{x}^\alpha \dot{x}^\beta = 0$. This gives the following constraint:
%\be
%\label{nomass} 
%0 = \omega^2 r^2 + 2 a \omega r^2 p(r) - f(r) \, .
%\ee  

Let us now consider how the rotation deforms the photon rings of the black hole.  The photon rings are circular orbits (\ie  having $r = \text{constant}$) for null geodesics lying in the equatorial plane $x=0$. We therefore must seek determine the simultaneous zeros of the effective potential and its first derivative. In this case, rather than work with the conserved quantities $E$ and $j$, we will use the angular velocity $\omega \equiv d\phi/dt$, which is conserved along the photon trajectory. Written in terms of the angular velocity, the conditions determining the location of the photon rings read
\begin{align}
0 &= \omega^2 r^2 + 2 a \omega h(r) - f(r)  \, ,
\label{ps1}
\\
0 &= 2 \omega^2 r + 2 a \omega h'(r) - f'(r) \, .
\label{ps2}
\end{align}

To solve these equations to first order in $a$ it is useful to consider first the case where the rotation vanishes, corresponding to the photon sphere of the static, spherically symmetric solution. The equations reduce to 
\be 
\frac{\rps f'(\rps)}{2} - f(\rps) = 0 \, , \quad \omega_{\rm ps} = \frac{ \sqrt{ f(\rps)}}{ \rps } \, ,
\ee
where the first equation determines $\rps$ --- the radius of the photon sphere in the static solution ---  and we find that the solution is unique. Once $\rps$ is known, the second equation determines the angular velocity.  When the rotation is non-trivial, Eqs.~\eqref{ps1} and~\eqref{ps2} admit two distinct solutions for the radius of the photon rings and the angular velocity. We write the corrected radius of the photon ring, denoted ${\rpr}_\pm$, in the following way
\be 
{\rpr}_{\pm} = \rps \pm a \rpr^{(1)} \, ,
\ee  
where the leading order correction $\rpr^{(1)}$ is given by
\be 
\rpr^{(1)} =  \frac{2  \sqrt{f(\rps)} [\rps h'(\rps) - 2 h(\rps)] }{2 f(\rps) - \rps^2 f''(\rps)}  \, , 
\ee
and so it can be directly computed from the numeric/approximate solutions once the value of $\rps$ characterizing the static solution is known.   The results can then be plugged into the expression for the angular velocity and expanding to linear order in $a$, we obtain  two solutions
\be 
\omega_{{\rm pr} \pm} = \mp \omega^{(0)}_{\rm ps} + a \omega_{\rm pr}^{(1)}  = \mp \frac{ \sqrt{ f(\rps)}}{ \rps } - a \frac{h(\rps)}{\rps^2}  \, .
\ee
where the plus sign corresponds to the prograde photon ring and the minus sign corresponds to the retrograde photon ring.

%To first order in the spin parameter, we can write
%\be 
%\omega_{{\rm pr} \pm} = \mp \omega_{\rm ps} + a \omega_{\rm pr}^{(1)}  \, ,
%\ee
%where
%\be 
%\omega_{\rm pr}^{(1)} = - p(\rps) \, .
%\ee

Using the perturbative expansions for the metric functions, we can write
\begin{align}
r_{\rm pr} &= 3 M + \frac{13 \mu L^4}{81 M^3} + \frac{1295 \mu^2 L^8}{59049 M^7} \pm  a \left(\frac{2}{\sqrt{3}} - \frac{425 \mu L^4}{729 \sqrt{3} M^4} - \frac{13033 \mu^2 L^8}{2125764 \sqrt{3} M^8} \right)\, ,
\\ \notag
\omega_{\rm pr} &= \mp \left(\frac{1}{3 \sqrt{3} M} - \frac{35 \mu L^4}{4374 \sqrt{3} M^5} - \frac{7549 \mu^2 L^8}{12754584 \sqrt{3} M^9} \right) + a \left(\frac{2}{27 M^2} - \frac{280 \mu L^4}{19683 M^6}  - \frac{6247 \mu^2 L^8}{62178597 M^{10}} \right)
\end{align}

\begin{figure}[t]
\centering
\includegraphics[scale=0.45]{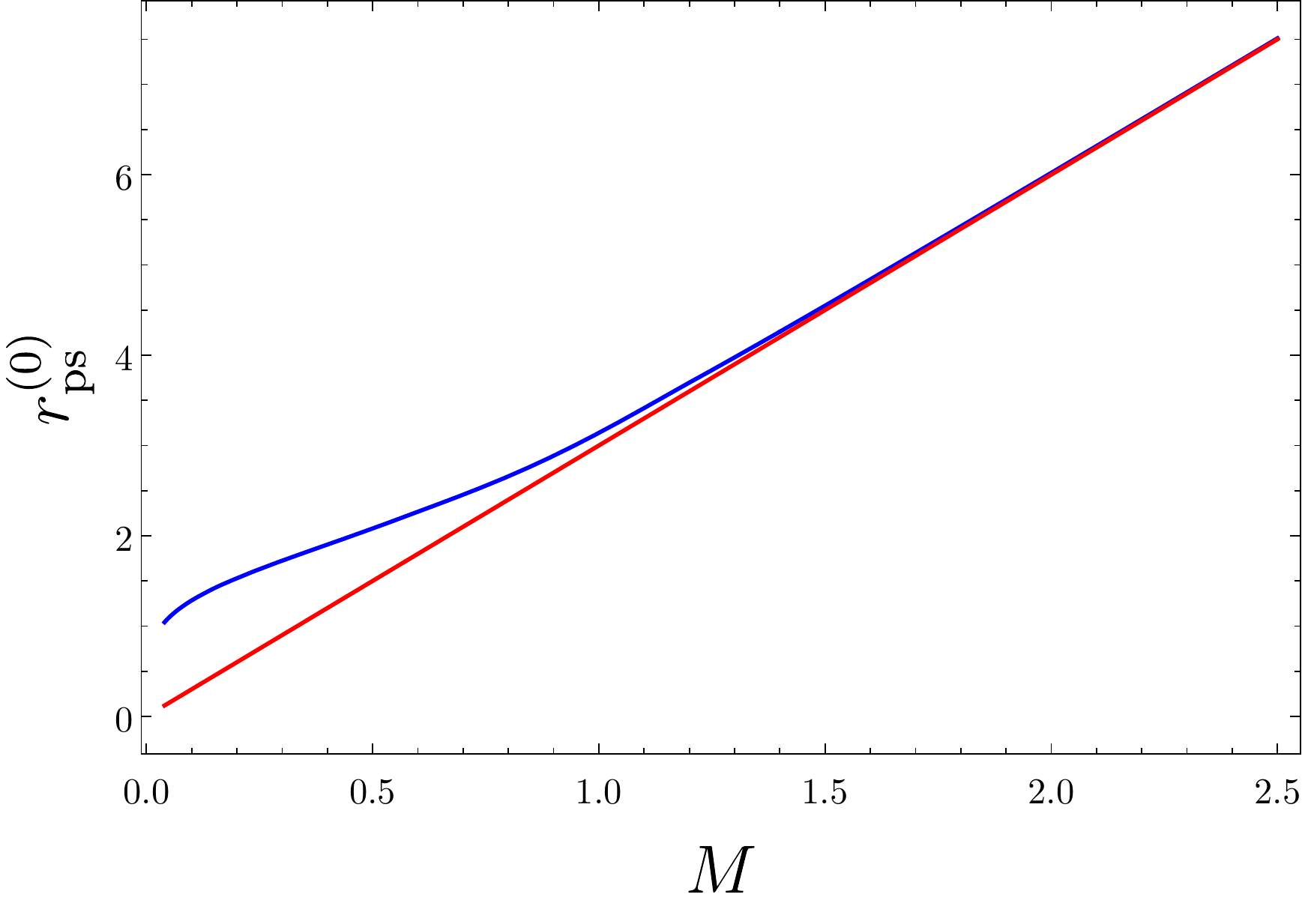}
\quad
\includegraphics[scale=0.45]{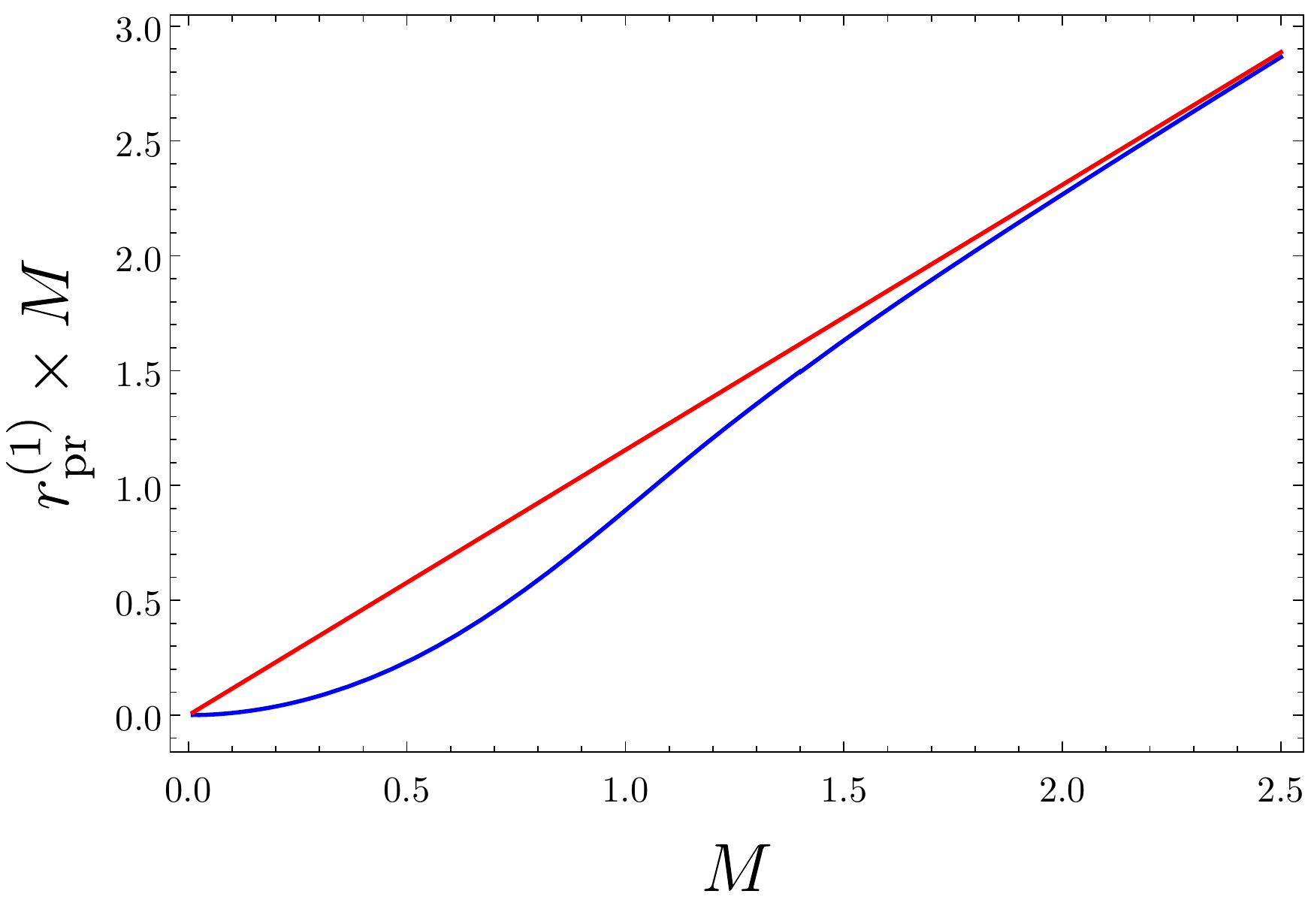}
\includegraphics[scale=0.45]{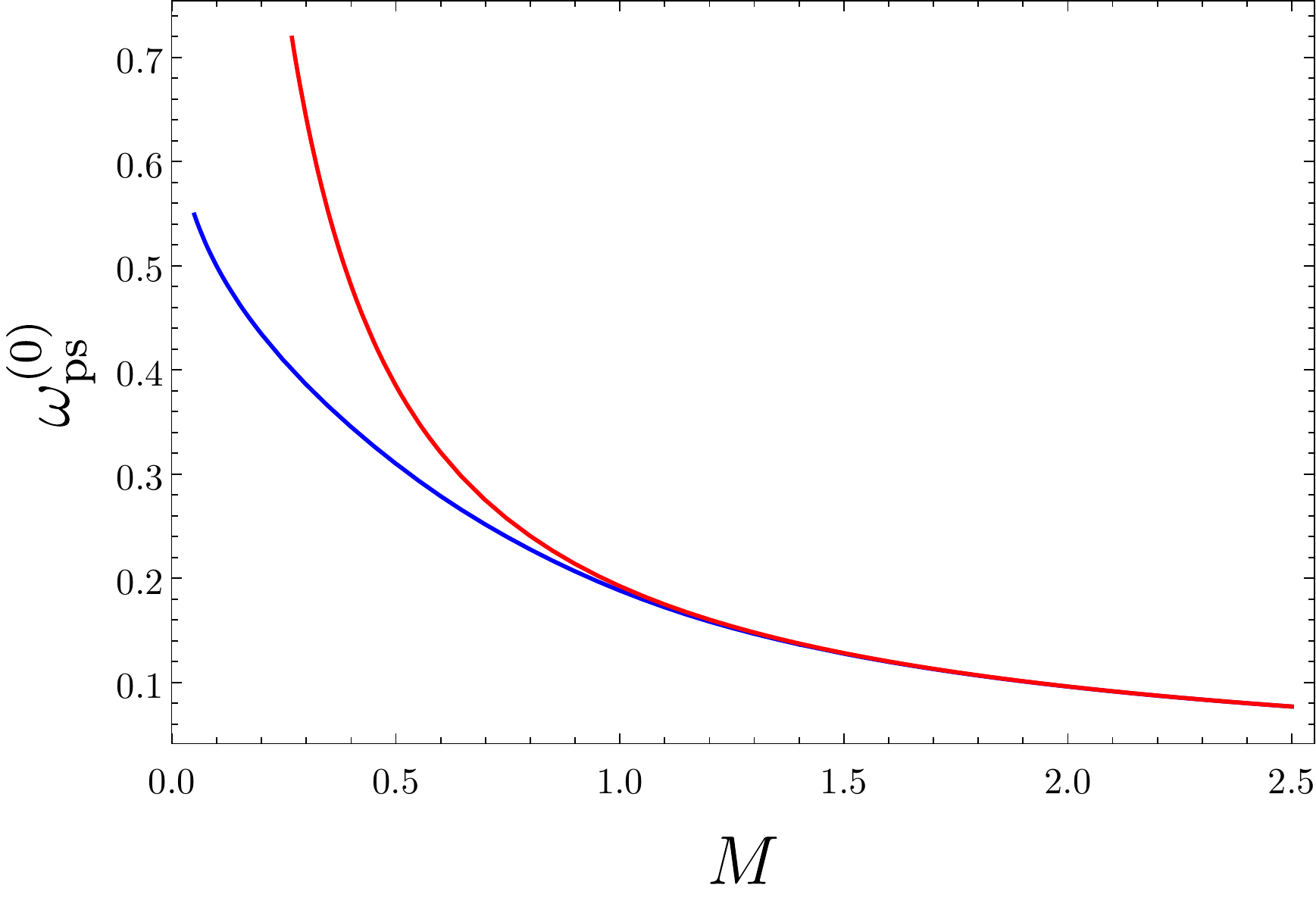}
\quad
\includegraphics[scale=0.45]{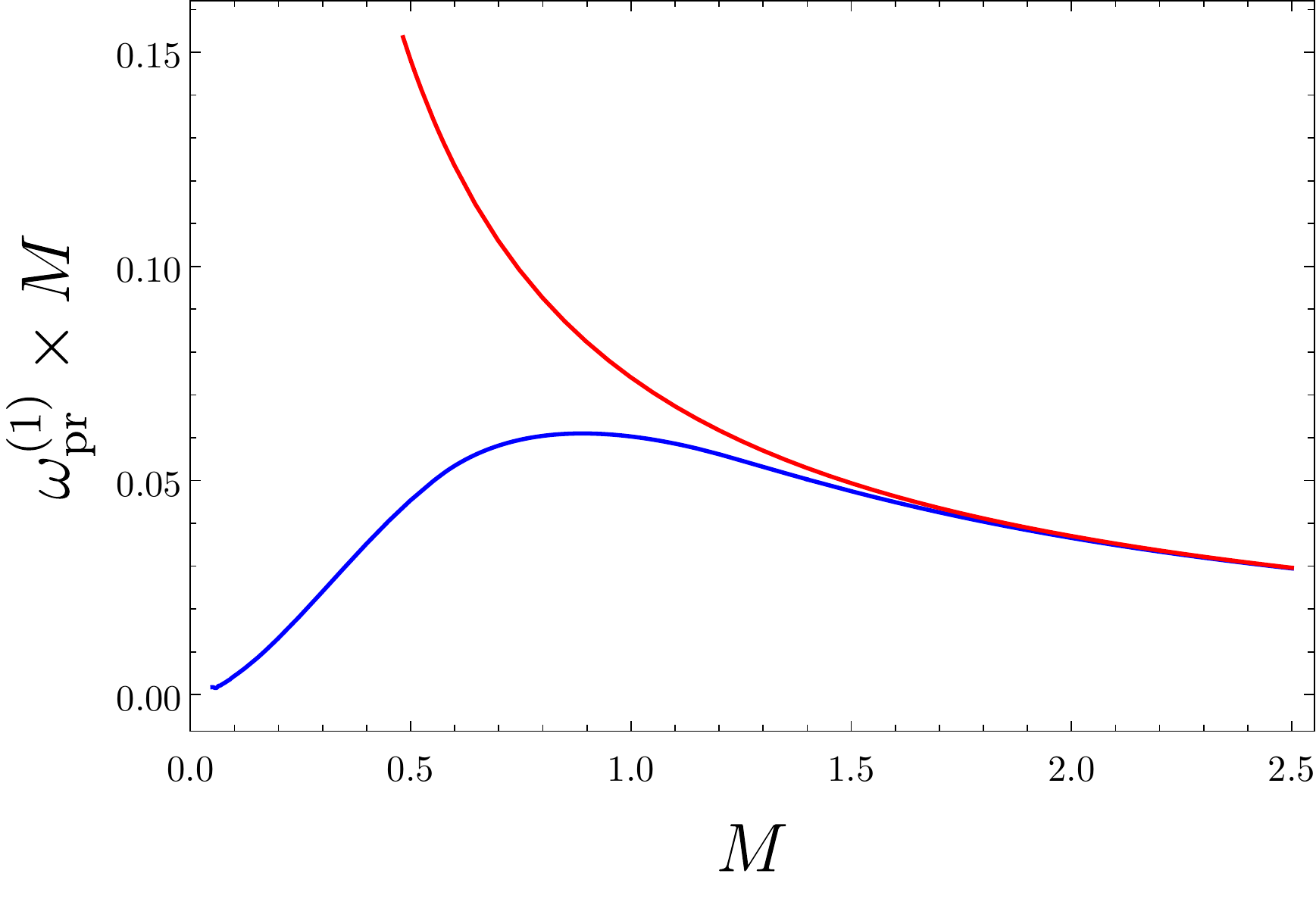}
\caption{Parameters describing the photon ring. The top row shows the radius of the photon ring, while the bottom row shows the angular velocity of the photon ring. The left column shows these parameters in  the static case, while the right column shows the leading-order corrections due to rotation. In each plot the mass is expressed in units of $\mu^{1/4} L$, and the red curves indicate the Einstein gravity result.}
\label{ptnSphereCorr}
\end{figure}

\begin{figure}[htp]
\centering
\includegraphics[scale=0.6]{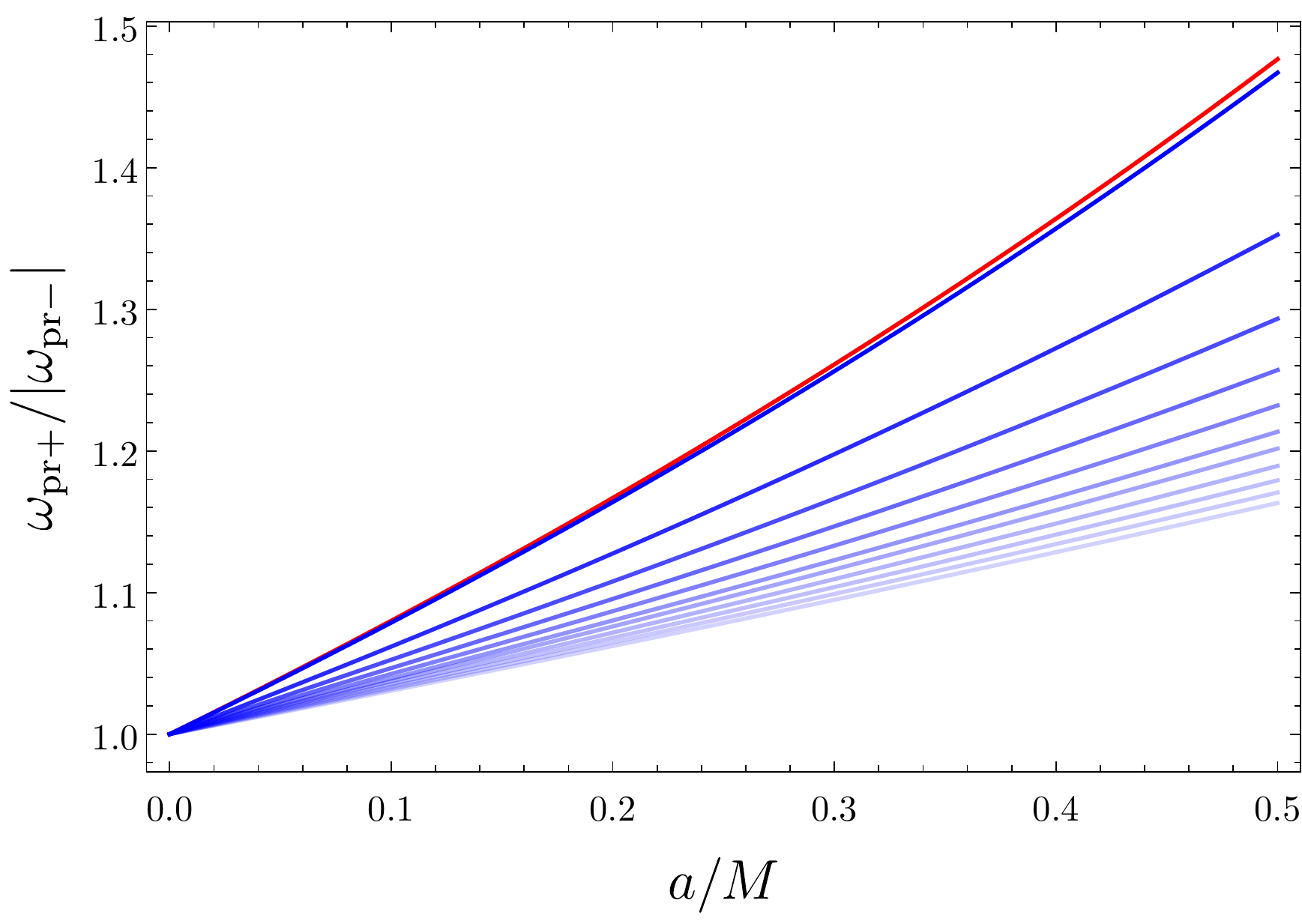}
\caption{Here we show the ratio of angular velocities for the photon rings as a function of the spin parameter $a/M$. The red curve corresponds to Einstein gravity, while the blue curves correspond to different values of $\mu$ ranging from $\mu = 1/10$ (darkest curve) to $\mu = 74/5$ (lightest curve), with the intermediate curves spaced by $\Delta \mu = 3/2$.  We have set $L = M$.}
\label{freqRat}
\end{figure}

In Figure~\ref{ptnSphereCorr} we plot the corrections to the radius and angular velocity of the photon sphere as a function of the black hole mass.  In the case of the zeroth-order terms, which correspond to the static solution, the ECG corrections are most prominent at small mass. 

In Figure~\ref{freqRat} we depict the ratio $\omega_+/|\omega_-$ for several values of the higher-order coupling. The idea here is the same as that in~\cite{Cano:2019ore} ---  in Einstein gravity this ratio is controlled only by the spin parameter, while here it depends also on the higher-order coupling. This feature could, in principle, be used to constrain the values of the ECG coupling, provided the spin parameter could be independently measured. From this plot we note that the effect of the ECG correction is to push this ratio below the corresponding curve for Einstein gravity. As the ECG coupling increases, the curves begin to ``bunch up'' --- in other words, the ratio is most sensitive to small differences in the coupling when the coupling is small.

\begin{figure}[t]
\centering
\includegraphics[scale=0.45]{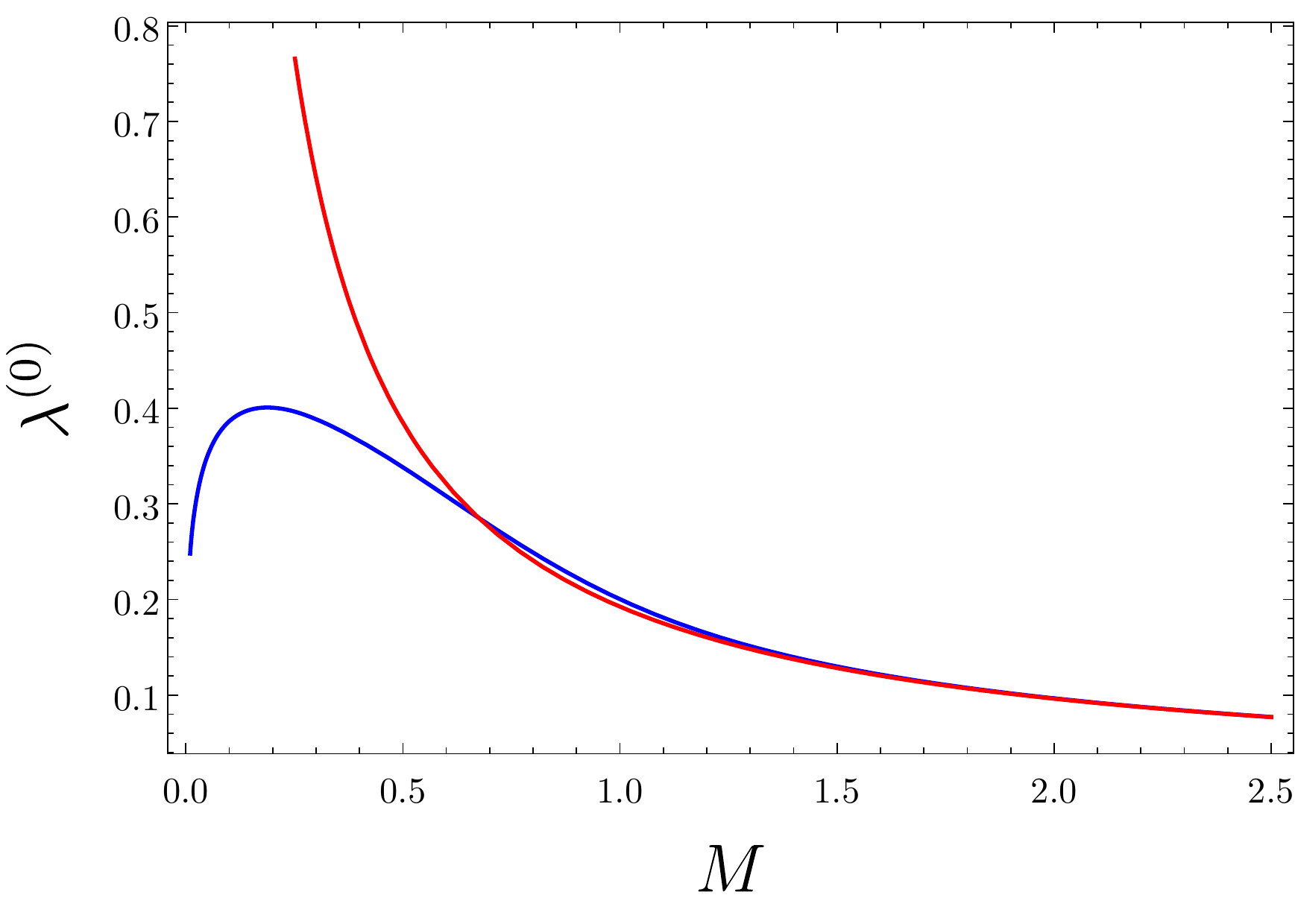}
\quad
\includegraphics[scale=0.45]{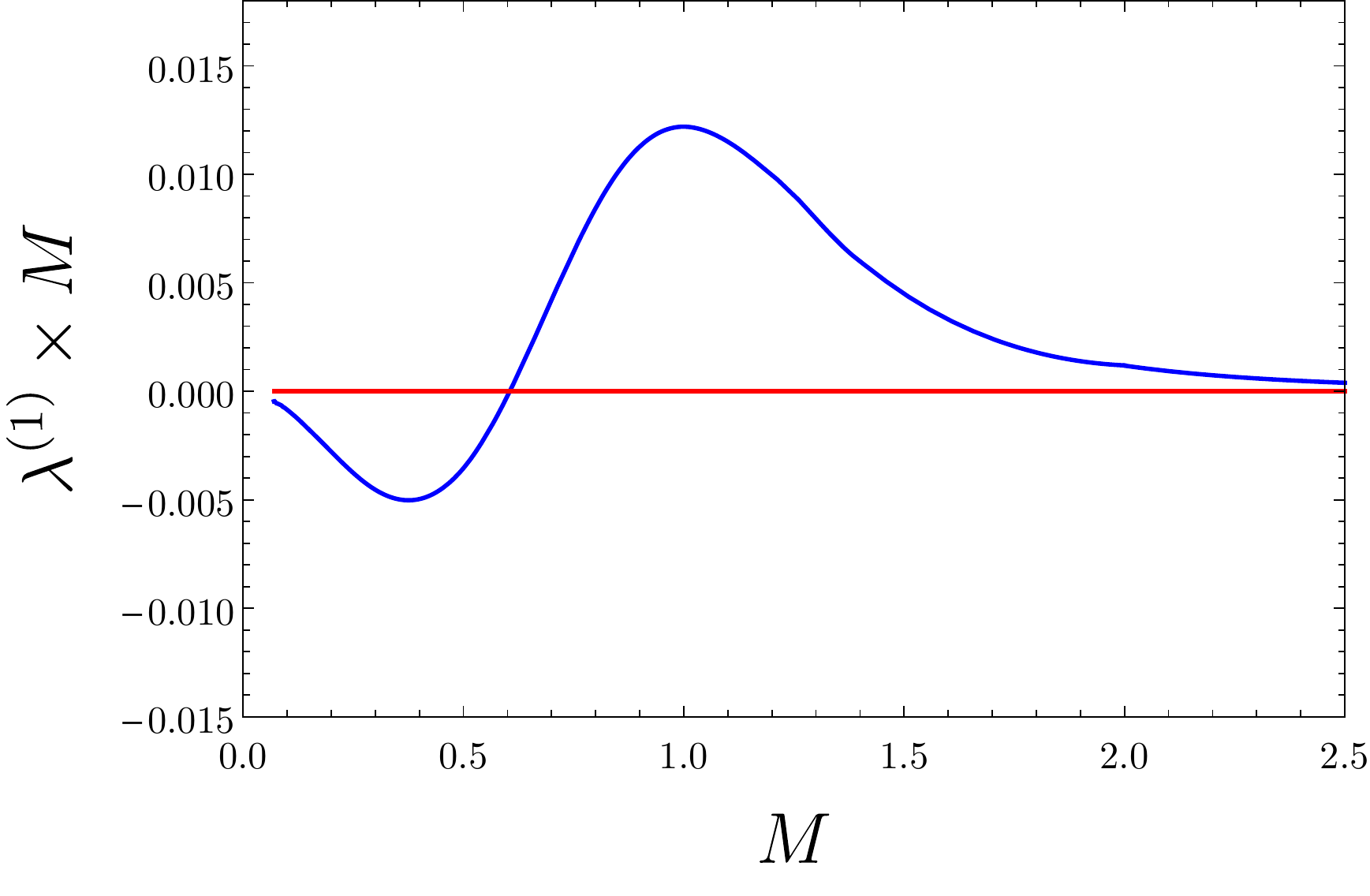}
\caption{Lyapunov exponent for photon ring. The left panel shows the zeroth-order term, corresponding to the static solution, while the right panel shows the leading correction due to rotation. In each case the red curve corresponds to the Einstein gravity result, while the blue curve is the ECG case. The mass is expressed in units of $\mu^{1/4} L$.}
\label{ptnLyap}
\end{figure}

We  turn next to the question of  stability of the photon ring orbits. Of course, it comes as no surprise that these orbits are unstable. Our goal is to compute the Lyapunov exponent associated with this instability. It has long been known~\cite{Mashhoon:1985cya} that the quasi-normal mode frequencies, in the Eikonal approximation, are related to the properties of unstable null geodesics. Specifically, the real part of the quasi-normal mode frequencies is related to the angular velocity of the unstable null orbit, while the imaginary part is related to the Lyapunov exponent. For any static, spherically symmetric spacetime one can prove this relationship to be~\cite{Cardoso:2008bp}:
\be 
\omega_{\rm QNM} = \omega_{\rm ps} \ell - i (n + 1/2) |\lambda|
\ee
where $\lambda$ is the Lyapunov exponent associated with the orbit.

To extract the Lyapunov exponent, we will follow~\cite{Mashhoon:1985cya}. We consider a solution of the equatorial geodesic equations corresponding to the photon ring. We then perturb those equations according to
\begin{align}
r(t) = \rps(1 + \epsilon F(t)) \, , \quad s(t) =  \frac{t}{\alpha} + \epsilon G(t) \, , \quad \phi(t) = |\omega_\pm| (1 + \epsilon H(t))
\end{align}
where $\epsilon$ is a small parameter controlling the perturbation and the system is subject to the boundary condition that the perturbation vanishes at $t = 0$. In the above, the constant $\alpha$ gives the relationship between the affine parameter $s$ and the coordinate time $t$ in the absence of the perturbation:
\be 
\alpha \equiv \dot{t}\big|_{r \to \rps} \, .
\ee
The only relevant correction for our purposes here is $F(t)$. This can be obtained via expanding the equation 
\eqref{Veff} involving the effective potential to first-order in the perturbation. The  differential equation reads
\be 
\frac{d^2 r}{ds^2} +  \frac{1}{2} \frac{d V_{\rm eff}}{dr} = 0 \, .
\ee
When we expand this to leading order in $\epsilon$, we obtain
\be 
2 \alpha^2 F''(t) + V_{\rm eff}''(\rps) F(t) = 0 \, .
\ee
Imposing the condition $F(0) = 0$ gives the solution
\be 
F(t) \propto \sinh \lambda t
\ee
where 
\be 
\lambda^2 \equiv - \frac{V_{\rm eff}''(\rps)}{2 \alpha^2} \, .
\ee

%We now turn to a discussion of the Lyapunov exponent in ECG. 

As with the other  quantities, we write the Lyapunov exponent as a zeroth-order term plus a correction linear in $a$:
\be 
\lambda = \lambda^{(0)} + a \lambda^{(1)} \, .
\ee
Although expressions for $\lambda^{(0)}$ and $\lambda^{(1)}$ can easily be obtained in terms of the metric functions, the resulting expressions are quite messy. When the mass is large we can obtain a perturbative solution for the Lyapunov exponent
\be 
\lambda = \frac{1}{2 \sqrt{3} M} + \frac{91 \mu L^4}{4374 \sqrt{3} M^5} + \frac{50987 \mu^2 L^8}{12754584 \sqrt{3} M^9} \mp a \left(\frac{28 \mu L^4}{729 M^6} - \frac{1531 \mu^2 L^8}{177147 M^{10}} \right)  
\ee
but when the mass, expressed in units of $\mu^{1/4} L$, becomes small we must, as before, resort to numerics. The results are shown in Figure~\ref{ptnLyap}. The zeroth-order term, which corresponds to the Lyapunov exponent for the static solutions, differs significantly from the Einstein gravity result at small mass. In particular, it reaches a maximum before turning rapidly toward zero in the small mass regime. This behavior is somewhat similar to that seen for rotating black holes in the extremal limit~\cite{Cardoso:2008bp}.  The leading-order correction due to rotation oscillates around the Einstein gravity value, with the ECG corrections becoming negligible at both small and large masses. The effect of this oscillation is to introduce a mass dependence to the slope of the Lynapunov exponent (as a function of $a$) in ECG.

\subsection{Black hole shadow}\label{sec:shadow}
Let us consider an observer that is placed far from the black hole, at a radius $r_0$, polar angle $\theta_{0}$ and, without loss of generality $\phi_{0}=0$. Then this observer receives a photon that moves in the direction $dr/dt>0$ and whose trajectory is defined by the angular momentum parameters $j^2$ and $\ell_{z}$. We want to determine the angle of incidence of this photon to the plane perpendicular to the $r$ direction at the position of the observer.  The spatial tangent vector at that point is 
\begin{equation}
u=-\dot{r}e_{r}+r_0\dot{\theta}e_{\theta}+r_0\sin\theta_0\dot{\phi}e_{\phi}\, ,
\end{equation}
where we have introduced the following orthonormal system for the observer, who is looking directly toward the black hole
\begin{equation}
e_{r}=-\partial_{r}\, ,\quad e_{\theta}=\frac{\partial_{\theta}}{r_0}\, ,\quad e_{\phi}=\frac{\partial_{\phi}}{r_0\sin\theta_0} \, .
\end{equation}
Let us define $\pi/2-\delta$ as the angle of incidence of the photon on the plane $r=r_0$ and $\alpha$ as the angle that the projected vector forms with the direction $e_{\phi}$. In other words, we parametrize the tangent vector as
\begin{equation}
u=-\dot{r}e_{r}+\sin\delta\left(e_{\theta}\sin\alpha + e_{\phi}\cos\alpha\right)\, ,
\end{equation}
\begin{equation}
\sin\delta=r_0\sqrt{\dot{\theta}^2+\sin^2\theta_0\dot{\phi}^2}\, ,\quad \cos\alpha=\frac{\sin\theta_0\dot{\phi}}{\sqrt{\dot{\theta}^2+\sin^2\theta_0\dot{\phi}^2}}\, .
\end{equation}

\begin{figure}[t]
\centering
\includegraphics[scale=0.48]{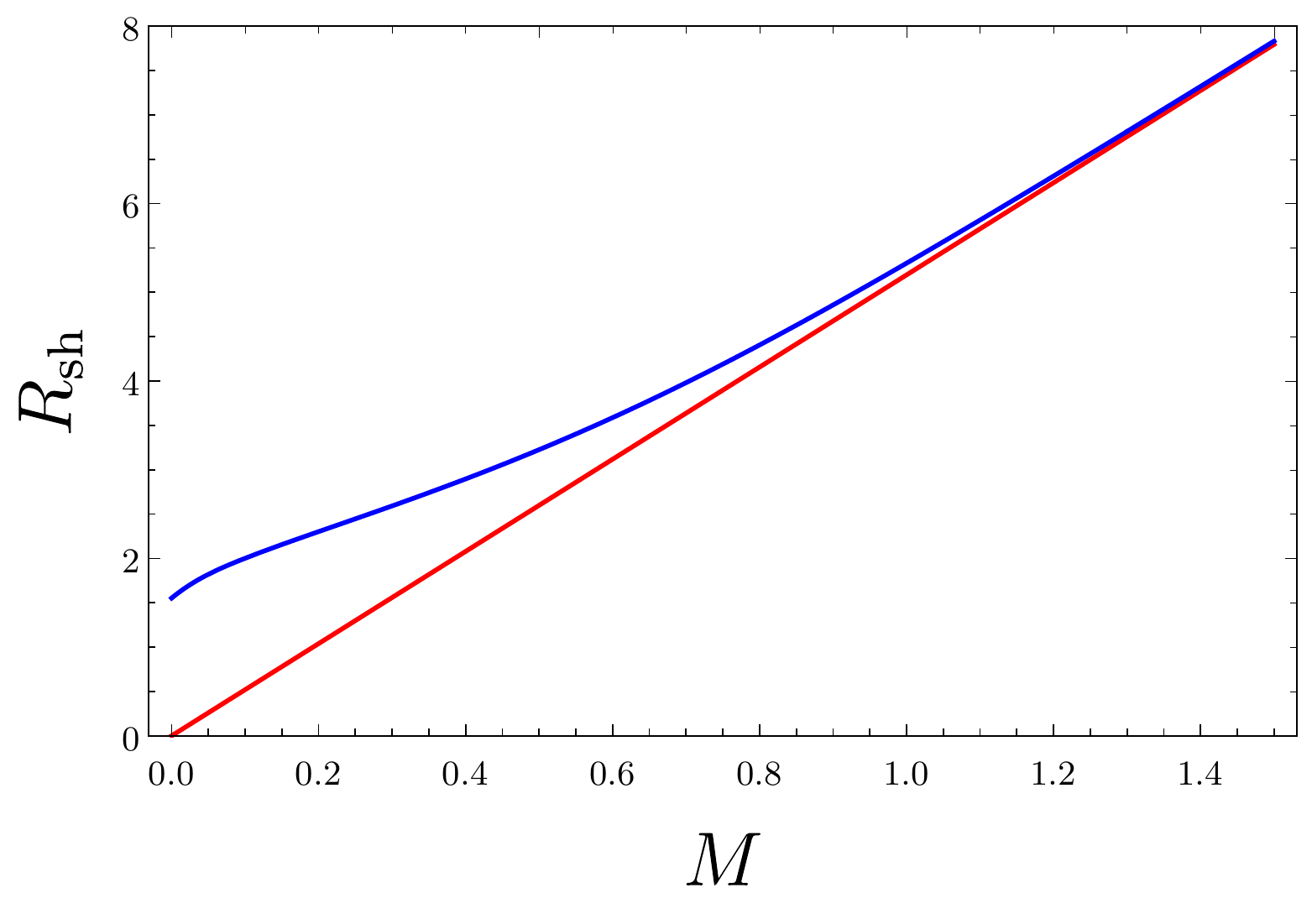}
\quad
\includegraphics[scale=0.5]{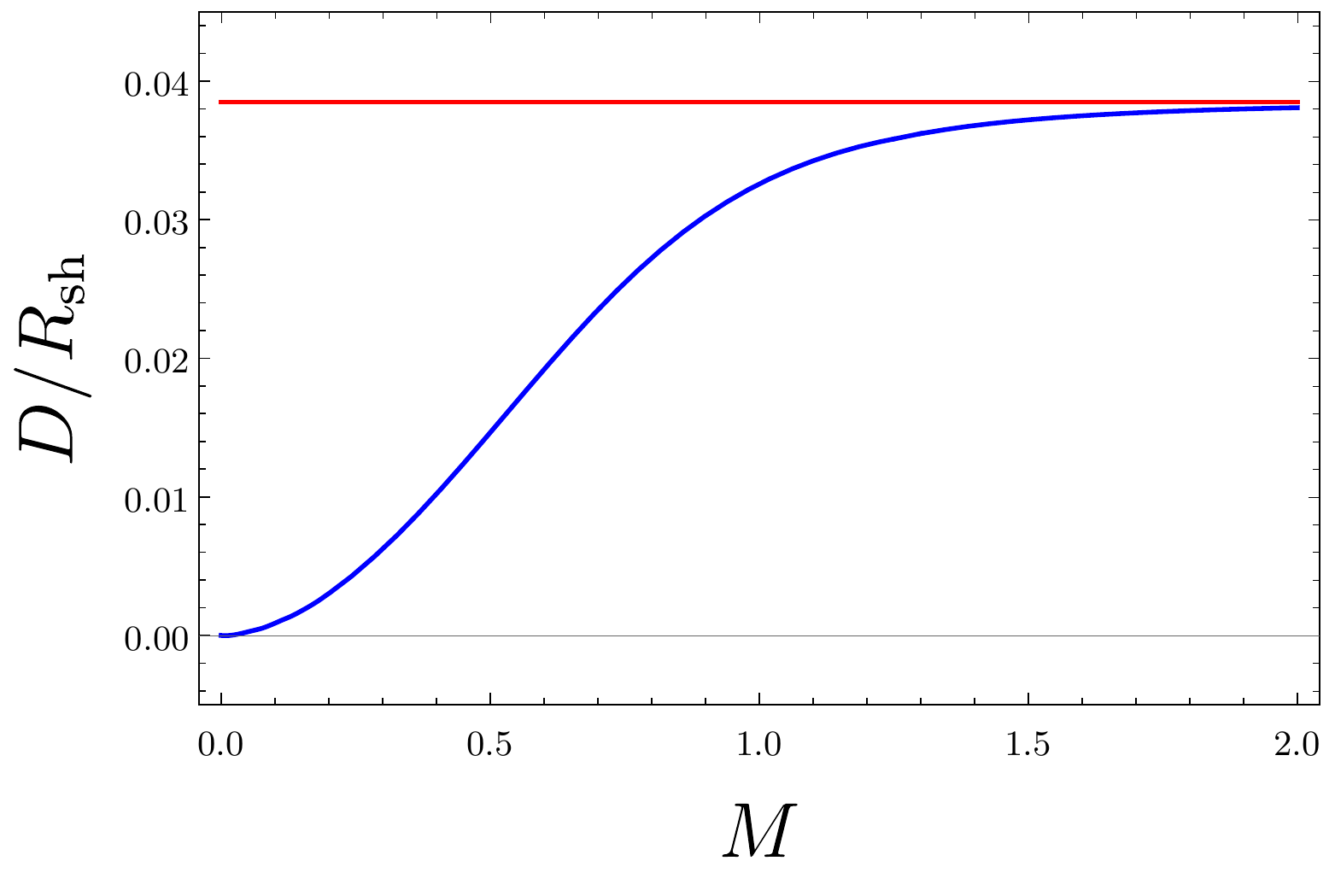}
\caption{Properties of the black hole shadow for Einstein gravity (red curve) and Einsteinian cubic gravity (blue curve). Left: we show the radius $R_{\rm sh}$ of the shadow as a function of the mass $M$. Right: we plot the ratio between the shift of the shadow $D$ and the radius $R_{\rm sh}$ as a function of the mass for $\chi=-0.1$. The mass is expressed in units of $\mu^{1/4} L/G$ and $R_{\rm sh}$ in the reciprocal length units.}
\label{shadow1}
\end{figure}

In these expressions we have already assumed that $r_0\sqrt{\dot{\theta}^2+\sin^2\theta_0\dot{\phi}^2}\ll1$ because we are taking $r_0\rightarrow\infty$. Finally, using the geodesic equations, we can write these angles in terms of the angular momentum; or conversely, we can express the angular momentum of the geodesic in terms of the angles:
\begin{equation}
j=r_0\sin\delta\, ,\quad \ell_z=r_0\sin\theta_0\cos\alpha\sin\delta\, .
\end{equation}
Now, the shadow of the black hole is determined by the photons that pass arbitrarily close to the photon sphere. Remember that for the photons in the photon sphere we have derived a relation between $j^2$ and $\ell_z$, and therefore the contour of the shadow corresponds precisely to photons with that value of the angular momentum.  Taking into account these points, we derive the following equation
\begin{equation}
r_0^2\sin^2\delta=\frac{(r_{\rm ps}^{(0)})^2}{f(r_{\rm ps}^{(0)})}+\frac{2a\sin\theta_0h(r_{\rm ps}^{(0)})}{f(r_{\rm ps}^{(0)})}\cos\alpha\;  r_0\sin\delta\, ,
\end{equation}
that determines the contour $\delta(\alpha)$ of the black hole shadow. Expanding linearly in $a$, the solution to this equation reads
\begin{equation}
r_0\sin\delta=\frac{r_{\rm ps}^{(0)}}{\sqrt{f(r_{\rm ps}^{(0)})}}+\frac{a\sin\theta_0h(r_{\rm ps}^{(0)})}{f(r_{\rm ps}^{(0)})}\cos\alpha\, ,
\end{equation}

Since $\delta\ll1$ when $r_0>r_{\rm ps}^{(0)}$, we can approximate $\sin\delta\approx\delta$, and we can see that the curve $\delta(\alpha)$ is approximately a circumference of radius $R_{\rm sh}$ centered at $\alpha=0$, $r_0\delta=D$, where
\begin{equation}
R_{\rm sh}=\frac{r_{\rm ps}^{(0)}}{\sqrt{f(r_{\rm ps}^{(0)})}}\, ,\quad D=-\frac{a\sin\theta_0h(r_{\rm ps}^{(0)})}{f(r_{\rm ps}^{(0)})}\, .
\end{equation}
Thus, at first order in $a$, the effect of rotation is to shift the shadow a distance $D$ from the radial direction. Let us compute $R_{\rm sh}$ and $D$ for the slowly rotating ECG black hole. First, if $\mu L^4/M^4\ll1$ we can perform a perturbative analysis and we find
\begin{equation}\label{RshD}
R_{\rm sh}=3 \sqrt{3} M+\frac{35 L^4 \mu }{162 \sqrt{3} M^3}+\frac{1111 L^8 \mu ^2}{52488 \sqrt{3}
   M^7}\, ,\quad D=-\frac{a\sin\theta_0}{M}\left(2 M-\frac{70 L^4 \mu }{243 M^3}-\frac{24395 L^8 \mu ^2}{2302911 M^7}\right)\, .
\end{equation}

\begin{figure}[t]
\centering
\includegraphics[scale=0.6]{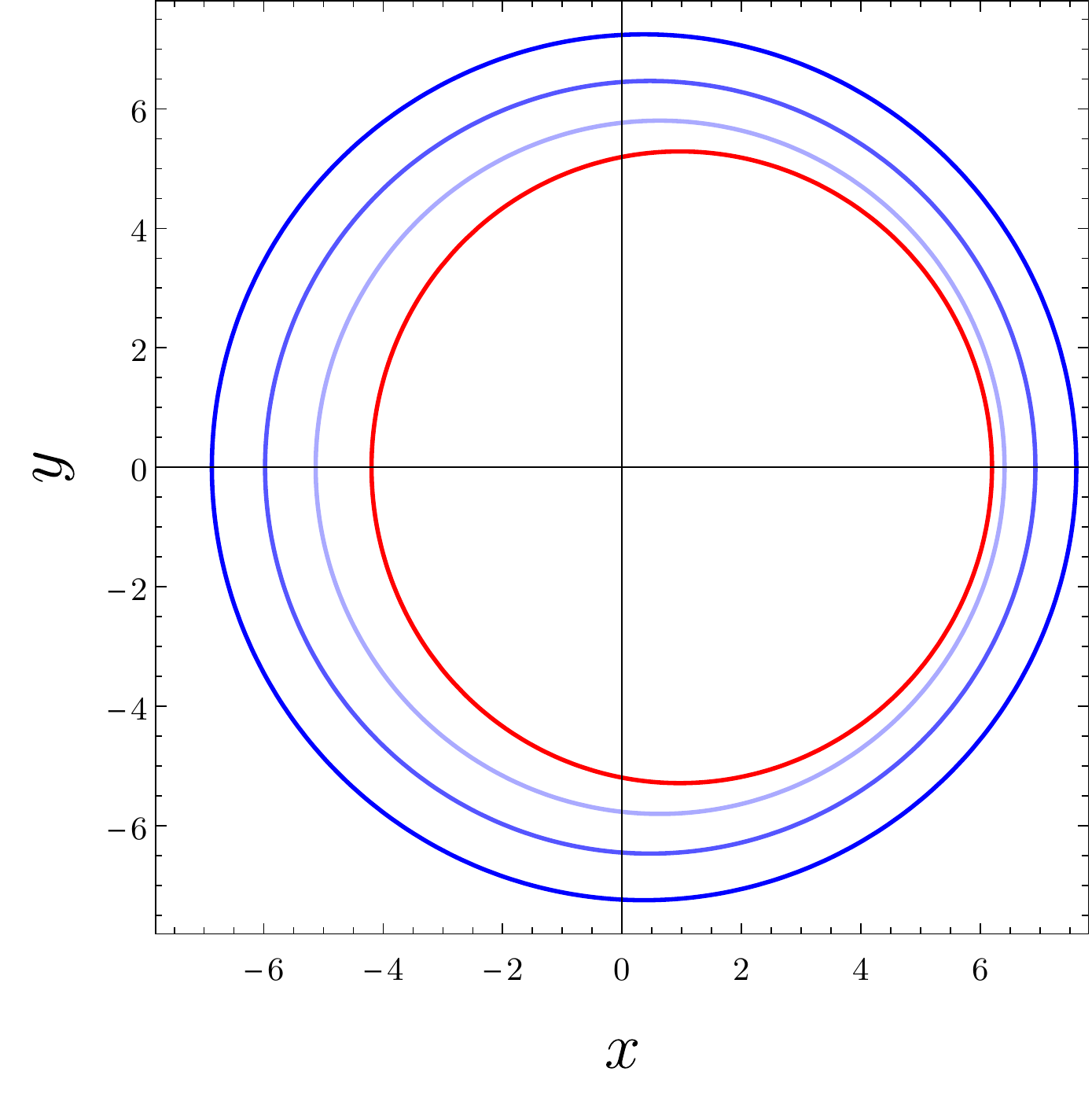}
\caption{Contour of the black hole shadow for a fixed mass and for $\chi=-0.5$. In red we show the Einstein gravity result and in blue the ECG one for several values of the higher-order coupling. From lighter to darker blue: $\mu^{1/4} L/M=1.5, 2, 2.5$.}
%$\mu L^4/M^4=2,4,\ldots 20$
\label{shadow2}
\end{figure}

For large $\mu L^4/M^4$ we need to use the numerical solution, and we find in that case some remarkable facts. First, we observe that when we decrease $M$ (keeping $a/M$ fixed), the ratio $D/M$ goes to zero, which is a manifestation of the fact that the effect of rotation becomes irrelevant for these black holes when the mass is small. %\rbm{I don't understand that last sentence for 2 reasons.  First it says that for fixed $a/M$
%$D/M \ to 0$ as $M$ increases.  But \eqref{RshD} indicates that $D/M \to -\frac{a\sin\theta_0}{M}\left(2 +\cdots\right)$, which doesn't go to zero for fixed $a/M$.  Second, the first part of the sentence says we are increasing $M$, but
%the last part says that $M$ is small.}\comment{Right, I meant ``decrease $M$''}

More strikingly, we find that when $M\rightarrow 0$, the radius of the shadow $R_{\rm sh}$ tends to a constant value. Performing a fit to the numerical values, we get that the behavior for $M\ll L \mu^{-1/4}$ is 
\begin{equation}
R_{\rm sh}\approx \frac{1.6 L}{\mu^{1/4}}+4.86 M\, \quad \text{when}\,\,\, M\rightarrow 0\, .
\end{equation}
%\rbm{How is the previous equation obtained?  Numerically?}
In Fig.~\ref{shadow1} we show the profile of $R_{\rm sh}$ and $D/R_{\rm sh}$ as functions of the mass and in Fig.~\ref{shadow2} we plot the contour of the shadow. We observe that the effect of the ECG term is to make the shadow larger and more centered with respect to the situation in Einstein gravity.

\section{Asymptotically AdS solutions}\label{adss}

In this section we study the case of slowly rotating black holes in ECG with a negative cosmological constant. The action reads
\be\label{ECG-act2} 
I = \frac{1}{16 \pi G} \int d^4x \sqrt{-g} \left[\frac{6}{L^2} + R - \frac{\mu L^4}{8}{\cal P} \right],
\ee
where now $L$ would coincide with the AdS scale if the corrections were not present. Instead, the AdS radius $\tilde L=L/\sqrt{f_{\infty}}$ is determined by the cubic equation
\be 
h(f_\infty) = 1- \fin + \mu \fin^3 = 0 \, .
\ee 
Here we have introduced the function $h(x)$ which we will refer to as the \textit{embedding function}.\footnote{This has been sometimes called \textit{characteristic polynomial}, and it has been shown to be useful in characterizing the thermodynamic properties of black holes \cite{Camanho:2010ru,Camanho:2011rj,Paulos:2011zu,Camanho:2015ysa,Bueno:2020odt}.}

The ansatz for the slowly rotating black holes takes the form,
\be \label{adsmetric}
ds^2 = -N(r)^2 f(r)dt^2 + \frac{dr^2}{f(r)} + 2 a r^2 p(r) (1-k x^2) dt d\phi + r^2 \left[\frac{dx^2}{1-k x^2} + (1-k x^2) d\phi^2 \right] \, ,
\ee
where again we work to linear order in the rotation parameter $a$. Unlike the asymptotically flat case, we can also consider planar $k=0$ or hyperbolic $k=-1$ transverse sections, besides spherical ones $k=1$. We will mainly focus on the latter case, but we will write the basic equations for general $k$.
The equations of motion can be reduced, again, to two second-order equations for $f$ and for $g=p'$, with two integration constants, $M$ and $C$:
\begin{align}
2 G M &= r^3/L^2 + r(k-f) + \frac{\mu L^4}{4 r^2} \left[6 \left(\frac{r f'}{2} + k- f \right) f'' f -  \left(r^2 f'^2 + 3 r k f' + 6 f (k-f) \right)f' \right] \, , \label{eqnf2}
\\
C &= r^4 g  -  \frac{3 \mu L^4}{2}  \bigg[  \left( \frac{r f'}{2} + k - f \right) r^2 f g'' +  \left(\frac{r^2 f}{2} f'' + \frac{r^2 f'^2}{2} + \frac{r(2k-f) f'}{2} + 2 f(k-f) \right) g'
	\nn
	&-5 \left( - \frac{3r^2 }{10} \left(-\frac{r f'}{3} + k + \frac{2 f}{3} \right)f'' - \frac{r^2 f'^2}{2} + \frac{r(k + 7 f) f'}{5} + (k-f)\left(k + \frac{6}{5} f \right) \right) g  \label{eqng2}
\bigg] \, .
\end{align}
On the other hand, $N$ is constant again. 

%\subsection{Boundary conditions }

\subsection{Asymptotic solution \label{sec:asympsol2}}
As we did before for the asymptotically flat case, we have to determine first the asymptotic behavior of the solution.  We assume that $f$ and $g$ can be expressed as a particular solution in the form of a $1/r$ expansion, plus the general solution of the corresponding homogeneous equation: 
\begin{equation}
f(r) = f_{1/r}(r) + f_{\rm h}(r)\, ,\quad g(r) = g_{1/r}(r) + g_{\rm h}(r)\, .
\end{equation}
The large-$r$ expansions read
\begin{align} 
f_{1/r} =& \fin \frac{r^2}{L^2} + k  + \frac{2 M}{h'(\fin) r} + \frac{42 M^2 L^2 \mu \fin}{ \left[h'(\fin) \right]^3  r^4} + \frac{27 k M^2 L^4 \mu}{\left[h'(\fin) \right]^3  r^6} + {\cal O}\left( r^{-7} \right)  \, ,\\
g_{1/r}(r) =& - \frac{C}{h'(\fin) r^4} - \frac{42 C M  L^2 \mu \fin}{\left[h'(\fin) \right]^3 r^7} + \frac{3 C M L^4 \mu (23 - 3786 \mu \fin^2 )}{\left[h'(\fin) \right]^{5} r^{10} } + {\cal O}\left(r^{-12} \right) \, . 
\end{align}
On the other hand, the linearized homogeneous equations satisfied by $f_{\rm h}(r)$ and $ g_{\rm h}(r)$ in the large-$r$ limit, read
\begin{align}
\frac{9 M \fin L^2 \mu}{2 h'(\fin)} f_{\rm h}''(r) + \frac{18 M L^2 \mu \fin}{[h'(\fin)] r} f_{\rm h}'(r) - h'(\fin) r f_{\rm h}(r) &= 0\, , \\
\frac{9 M \fin  L^2 \mu }{2 h'(\fin)}    g_{\rm h}''(r) + \frac{27 M  \fin L^2 \mu}{2 r h'(\fin)}    g_{\rm h}'(r) - h'(\fin) r  g_{\rm h}(r) &= 0 \,  
\end{align}
These equations are identical and have the following solutions
\begin{align}
f_{\rm h}(r)& \sim A_1 r^{-3/2} I_1\left(-\frac{2 \sqrt{2} r^{3/2} h'(\fin) }{9 L \sqrt{M \mu \fin} } \right) + B_1 r^{-3/2} K_1 \left(\frac{2 \sqrt{2} r^{3/2} h'(\fin) }{9 L \sqrt{M \mu \fin} } \right) \, ,\\
g_{\rm h}(r) &\sim \sim A_2 r^{-3/2} I_1\left(-\frac{2 \sqrt{2} r^{3/2} h'(\fin) }{9 L \sqrt{M \mu \fin} } \right) + B_2 r^{-3/2} K_1 \left(\frac{2 \sqrt{2} r^{3/2} h'(\fin) }{9 L \sqrt{M \mu \fin} } \right)   
\end{align}
where $K_\alpha(x)$ and $I_\alpha(x)$ are modified Bessel functions.
Thus, in order to guarantee a regular asymptotic limit we demand $\mu>0$, in which case by setting $A_1=A_2=0$ the homogeneous solutions decay faster than exponentially at infinity.\footnote{We always assume $h'(\fin)<0$.}

In the limit of large-$r$, the behavior of $p(r)$ is accurately described by integrating just the particular solution $g_{1/r}(r)$. This gives
\be 
p(r) = -\frac{\Omega_{\infty}}{a} + \frac{C}{3 \left[h'(\fin) \right] r^3} + \frac{7 C \mu M L^2 \fin}{\left[h'(\fin) \right]^3 r^6} +  \frac{C \mu M^2 L^4 (-23 + 3786 \mu \fin^2) }{3 \left[h'(\fin) \right]^5 r^9} + {\cal O} \left(r^{-11} \right) \, ,
\ee  
where $\Omega_{\infty}$ is a constant of integration which corresponds to the asymptotic angular velocity of the spacetime. From now on we will set $\Omega_{\infty}=0$.  To interpret the integration constant $C$, we will require that our solution asymptotically approaches the slow rotation limit of the Kerr-AdS solution with appropriately rescaled cosmological length scale and ADM charges \cite{Deser:2002jk,Arnowitt:1960es,Arnowitt:1960zzc,Arnowitt:1961zz,Abbott:1981ff}:
\begin{align} 
ds^2 \to& - \left(\fin \frac{r^2}{L^2} + k + \frac{2 M}{h'(\fin) r} \right) dt^2 + \frac{4 M a (1-k x^2)}{h'(\fin) r} dt d\phi + \left(\fin \frac{r^2}{L^2} + k + \frac{2 M}{h'(\fin) r} \right)^{-1} dr^2  
	\nn
	&+ \frac{r^2 dx^2}{1-k x^2} + r^2(1-k x^2) d\phi^2 \, .
\end{align}
Noting that $g_{t\phi}\sim -(1-kx^2) 2JG_{\rm eff}/r$, where $G_{\rm eff}=-G/h'(f_{\infty})$,
comparing this with our asymptotic expansion for $p(r)$ reveals
\be 
C = 6 M 
\ee
analogous  to the asymptotically flat case, with $J=aM$. 

\subsection{Near horizon solution \label{sec:nhsol2}}

The analysis of the near-horizon regime is completely analogous to the asymptotically flat case, so let us be brief. 
Near the horizon $r=\rh$, the functions $f$ and $g$ are required to admit a series expansion of the form
\be 
f(r) = 4 \pi T (r-\rh) + \sum_{n=2} a_n (r-\rh)^n \, ,\quad g(r) = \sum_{n=0} g_n (r-\rh)^n \, ,
\ee  
for certain coefficients $a_n$ and $g_n$. Together with an infinity number of equations for these parameters, the equation for $f$ \eqref{eqnf2} yields two constraints between the mass, the temperature and the horizon radius:\footnote{For a detailed analysis of the thermodynamic properties of static black holes, see \cite{HoloECG}.}
\begin{align}
2 G M &= k \rh +\frac{\rh^3}{L^2} - \frac{3 k (4 \pi T)^2 \mu L^4}{4 \rh} - \frac{(4 \pi T)^3 \mu L^4}{4}  \, ,
	\nn
0 &= -k + 4 \pi T \rh  -\frac{3 \rh^2 }{L^2}+ \frac{3 k (4 \pi T)^2 \mu L^4}{4 \rh^2}  \, .
\end{align}
On the other hand, we find that all the coefficients $a_{n\ge3}$ are determined by $a_2$. In the same way, the full sequence of coefficients $g_n$ contains only a free parameter which can be taken to be $g_0$. This is completely analogous to the asymptotically flat case. In particular, the few first relationships read
\begin{align}
6 M &= g_0 \left(\frac{3}{4} \mu  L^4 \left(2 k \rh \left[(4 \pi T)-3 a_2 \rh \right]+(4 \pi T) \rh^2 \left[2 a_2 \rh-5 (4 \pi T) \right]+10 k^2\right)+\rh^4\right)
	\nn
	&-\frac{3}{4} (4 \pi T) g_1 \mu  L^4 \rh \left[(4 \pi T) \rh+2 k\right] \, ,
\\
0 &= g_1 \left(\frac{3}{4} \mu  L^4 \left(-2 k \rh \left[5 a_2 \rh+3 (4 \pi T) \right] -(4 \pi T) \rh^2 \left[4 a_2 \rh+7 (4 \pi T) \right]+10 k^2\right)+\rh^4\right)
	\nn
	&+g_0 \bigg(\frac{3}{2} \mu  L^4 \left(k \left[2 (4 \pi T)-\rh
   (9 a_3 \rh+4 a_2) \right]+\rh \left[2 a_2^2 \rh^2+3 (4 \pi T) \rh (a_3 \rh-3 a_2)+2 (4 \pi T)^2\right]\right)
   \nn
   &+4 \rh^3\bigg) -3 (4 \pi T) g_2 \mu  L^4 \rh^2 \left[(4 \pi T) \rh+2 k\right] \, .
\end{align}
Thus, we need to fix the values of $a_2$ and $g_0$ in order to obtain a solution.  These are fixed by requiring that the solution has the correct behavior at infinity that we determined above.

\subsection{Numerical solution}
The space of solutions in the AdS case is richer than in the flat case since the black holes depend now on two parameters: the mass and the cosmological constant.  When the cosmological constant is zero, the corrections only depend on the dimensionless combination $\mu L^4/M^2$, and thus we only need to vary this parameter in order to sample the full space of solutions. In the AdS case, the corrections have two effects. On the one hand, they always become relevant for black holes of small masses, just like in the flat case --- in fact, for small black holes the effect of the cosmological constant is irrelevant. On the other hand, the corrections change the structure of the vacuum, so the modifications to black holes can be important even if the black holes are not small. In fact, there is a maximum value for the parameter $\mu$ for which an AdS vacuum exists. The maximum value is $\mu_{\rm cr}=\frac{4}{27}$, in whose case the AdS radius takes the value $\tilde L^2=\frac{2}{3}L^2$ and the theory is said to be at the critical point  \cite{Feng:2017tev}. The critical limit is characterized by the divergence of the effective Newton's constant, which in general reads
\begin{equation}
G_{\rm eff}=\frac{G}{1-3\mu \fin^2}\, ,
\end{equation}
or equivalently by the vanishing of the linearized equations on AdS. Let us now show our results for the rotating black hole solutions.

In the small coupling limit, $\mu\ll1$ and $\mu L^4/M^4\ll1$, we can obtain an approximate solution by performing a perturbative expansion of the functions $f$ and $p$. The result reads
\begin{align}
f(r)=&\frac{r^2}{L^2}+1-\frac{2 M}{r}+\mu  \left[\frac{r^2}{L^2}-\frac{6 M}{r}-\frac{42 L^2 M^2}{r^4}-\frac{27 L^4 M^2}{r^6}+\frac{46 L^4 M^3}{r^7}\right]\, ,\\
r^2p(r)=&-\frac{2 M}{r}+\mu  \left[-\frac{6 M}{r}-\frac{42 L^2 M^2}{r^4}+\frac{46 L^4 M^3}{r^7}\right]\, .
\end{align}
%\rbm{We are missing the analogue of this solution in the asymptotically flat case in earlier sections.}
Let us note that the $\mathcal{O}(r^2)$ and $\mathcal{O}{(1/r)}$ terms are related to the corrections to the AdS radius and to $G_{\rm eff}$. From these equations we can get for instance the mass and the angular velocity of the horizon as a function of the radius,
\begin{align}
M=&\frac{\rh \left(L^2+\rh^2\right)}{2 L^2}+\mu  \left[-\frac{27 \rh^3}{8 L^2}-\frac{27 \rh}{4}-\frac{27 L^2}{8 \rh}-\frac{L^4}{2 \rh^3}\right]\, ,\\
\Omega=&a\left[\frac{1}{L^2}+\frac{1}{\rh^2}+\mu  \left(\frac{1}{L^2}-\frac{27}{4 \rh^2}-\frac{27 L^2}{2 \rh^4}-\frac{27 L^4}{4 \rh^6}\right)\right]\, ,
\end{align}
perturbatively in $\mu$.

\begin{figure}[t]
\centering
\includegraphics[width=0.49\textwidth]{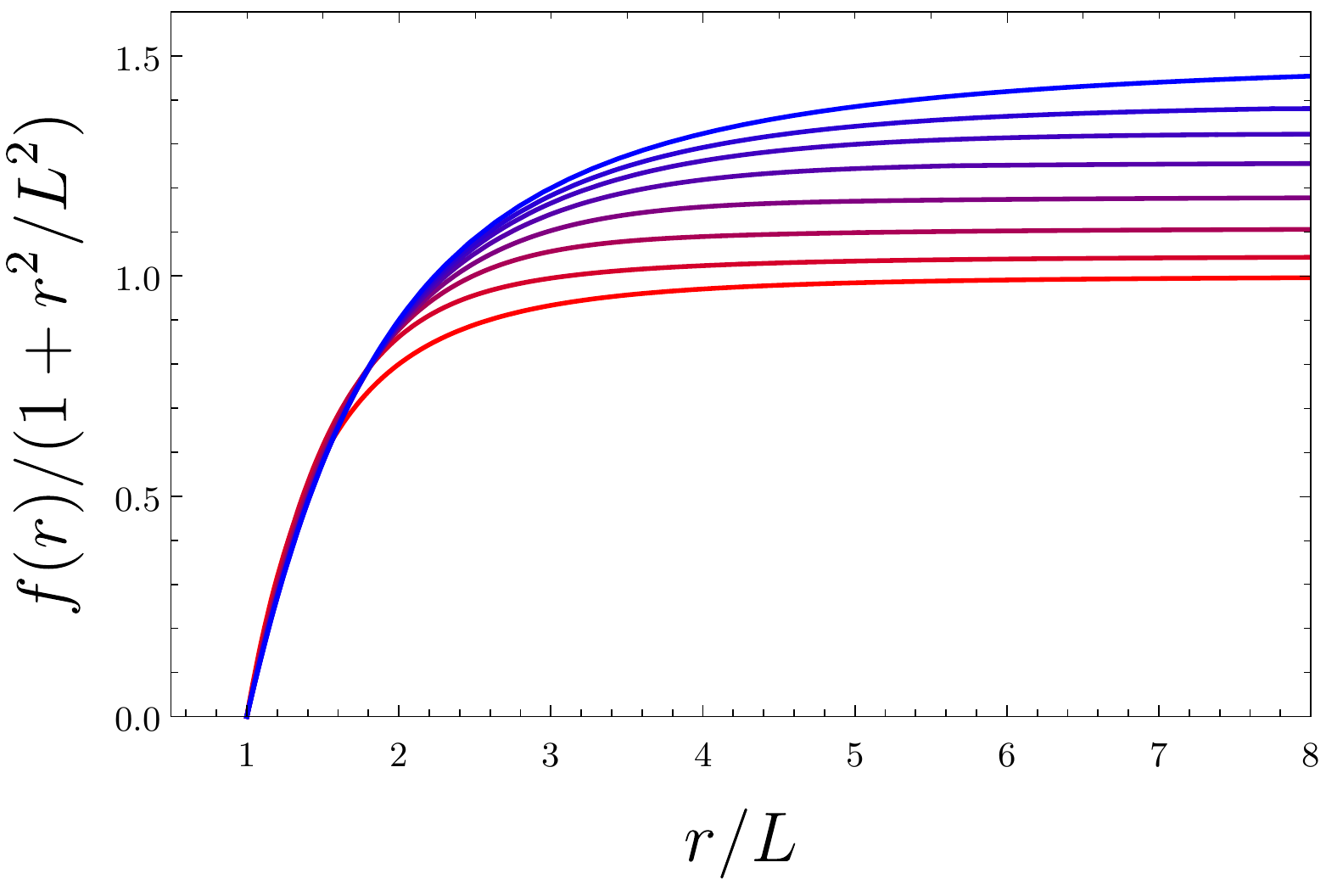}
\includegraphics[width=0.49\textwidth]{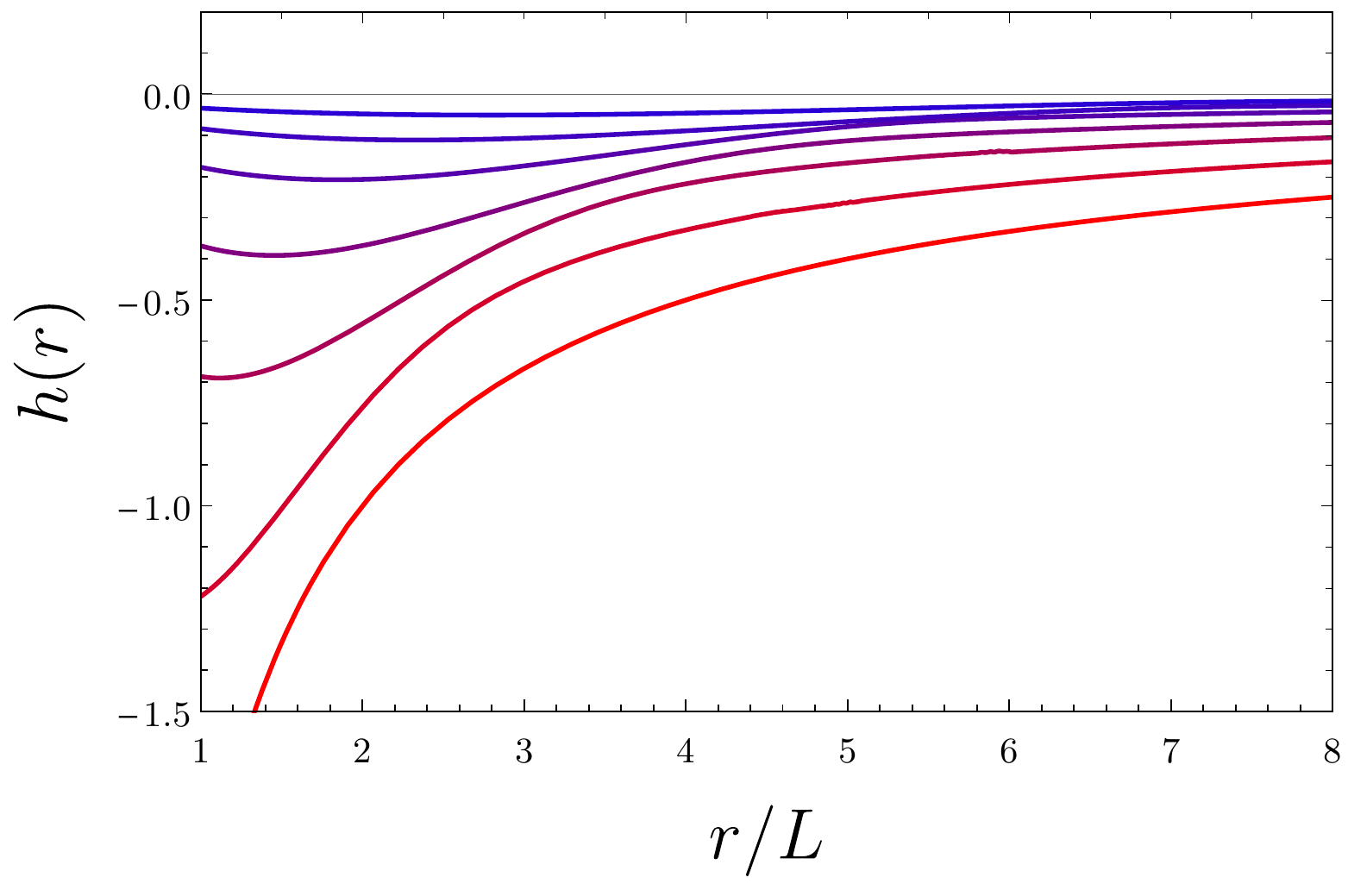}
\caption{Rotating AdS black holes in Einsteinian cubic gravity. We show solutions with $\rh=L$ and for $\mu=0$, $0.04$, $0.08$, $0.11$, $0.13$, $0.14$, $0.145$, $4/27$, with the $\mu=0$ Einstein solution the 
lowermost curves in red. Left: function $f(r)$ normalized by $1+r^2/L^2$. Right: function $h(r)=r^2p(r)$.}
\label{fig:AdSBH}
\end{figure}

If we wish to study the solution for higher values of the coupling or for small masses, we need to resort to numerical methods. 
Equations \eqref{eqnf2} and \eqref{eqng2} can be solved numerically by imposing the boundary conditions described in the previous subsection. For illustration purposes, let us study how the corrections modify the black hole solutions when we fix the size of the horizon while varying $\mu$ from zero to the critical value. 
In Fig.~\ref{fig:AdSBH} we show the numerical solution for black holes of radius $\rh=L$. The main effect of the corrections on the function $f$ is to change its asymptotic behavior due to the different value of the AdS radius. On the other hand, in the case of the function $h$, we see that its value gets smaller as we increase $\mu$. Let us recall that in the metric this function appears multiplied by $a=J/M$. Therefore, when we increase $\mu$ leaving $a$ and $\rh$ fixed, the effect of the angular momentum becomes less and less relevant. In the critical limit $\mu\rightarrow 4/27$ the solution seems to tend to $h\rightarrow 0$, so that no (regular) rotating solution exists in that case. Let us note that when we approach the critical limit leaving the radius fixed, the mass goes to zero, so it means that we can have slowly rotating black holes with $a\gg M$.

\begin{figure}[t]
\centering
\includegraphics[width=0.6\textwidth]{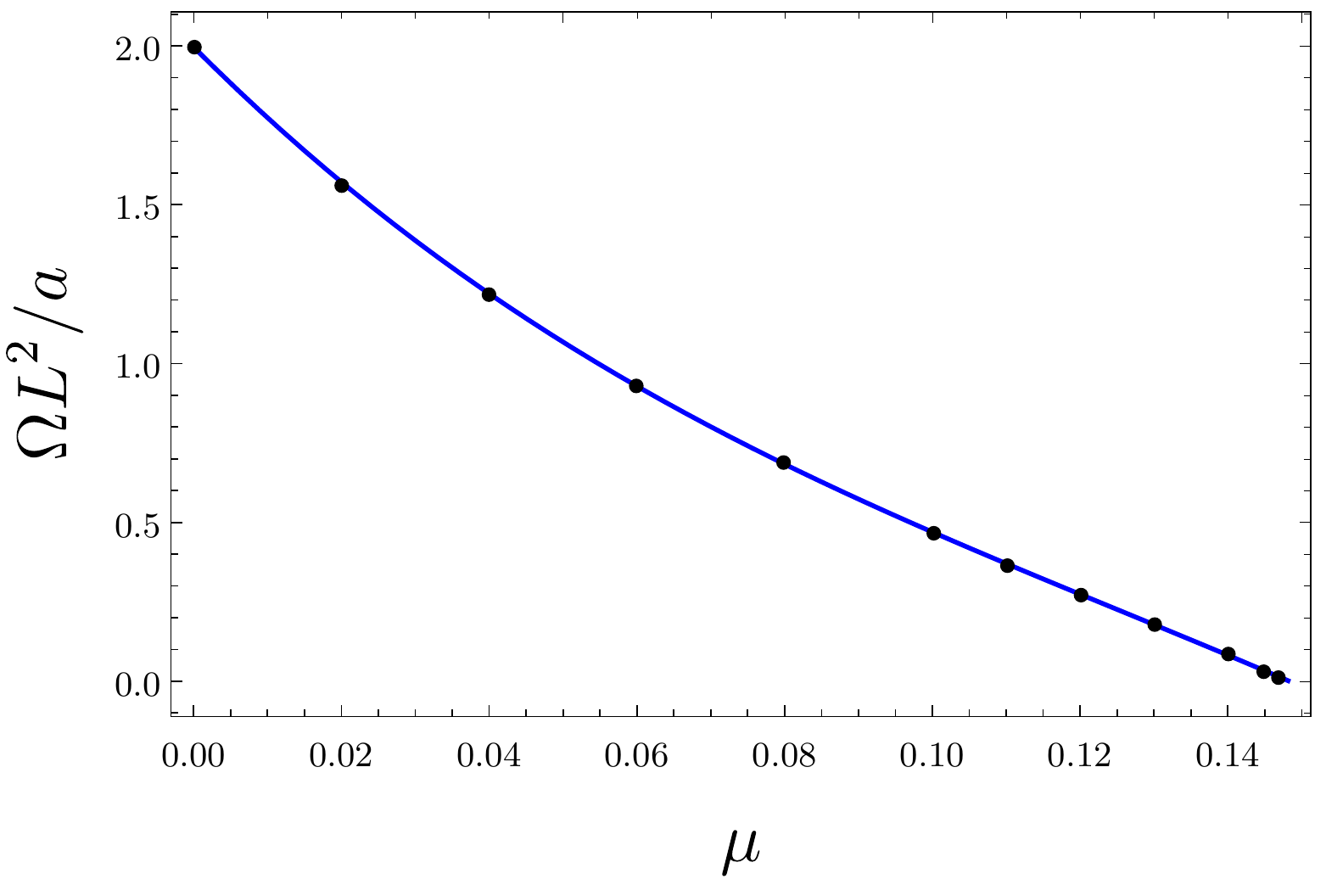}
\caption{Angular velocity of the horizon as a function of $\mu$ for black holes of radius $\rh=L$. The points represent numerical data while blue curve is a fitting polynomial of degree 4.}
\label{fig:AdSomega}
\end{figure}

The irrelevance of rotation as we increase $\mu$ is better illustrated by looking at the angular velocity, which we show in Fig.~\ref{fig:AdSomega}. As we can see, $\Omega L^2/a$ goes to zero as we approach the critical limit. This is a quite exotic behavior, since in the case of Einstein gravity this quantity is bounded from below according to $\Omega L^2/a>1$. Now, regarding the absolute value of the angular velocity $\Omega$ and not the ratio $\Omega/a$, if we considered black holes of fixed $J$ and took the critical limit, we would find that $\Omega$ diverges. However, it seems to make more sense to fix $a$, so that the total angular momentum $J=aM$ goes to zero in the same way as the mass. In that case, the angular velocity tends to zero in the critical limit and there are, in fact, no rotating solutions. 

\subsection{Spinning solutions in the critical limit}
As we have just seen, the critical theory is special, as it seems to allow only for massless and spinless black hole solutions. However, one can have solutions with non-vanishing charges if the condition of regularity is dropped. In particular, small-mass black holes in the critical theory were studied in Ref.~\cite{Feng:2017tev}, where they were found to develop a singularity at the horizon. Here, we are going to study rotating massless black holes.

\begin{figure}[t]
\centering
\includegraphics[width=0.6\textwidth]{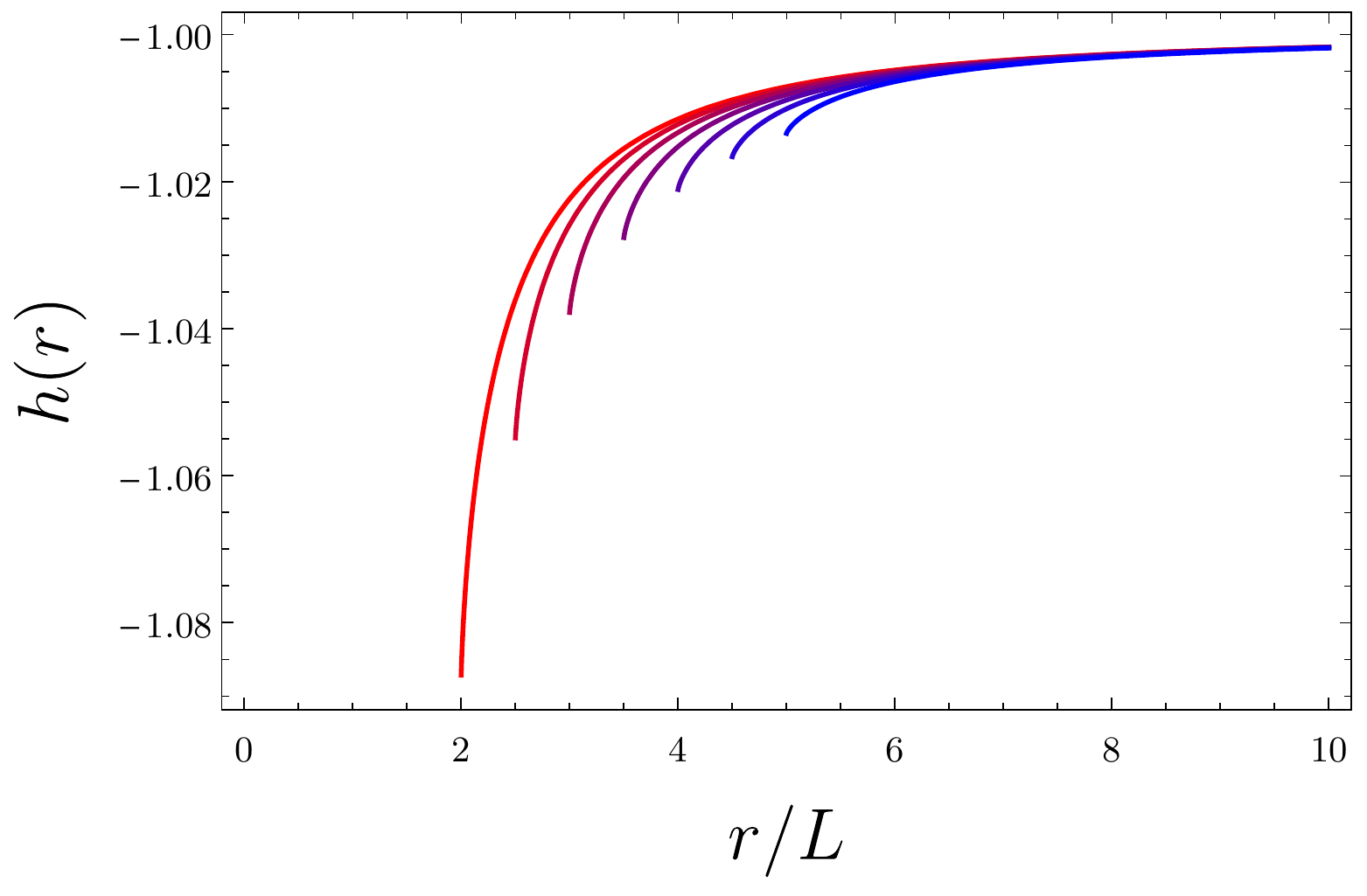}
\caption{Function $h(r)=r^2p(r)$ in the critical limit for several values of $\rh$, corresponding to the endpoint of each curve. From red to blue we have $\rh/L=2,$ 2.5, 3, 3.5, 4, 4.5, 5.}
\label{fig:hcrit}
\end{figure}

At the critical point, $\mu=4/27$, the equation \eqref{eqnf2} with $M=0$ is solved by \cite{Feng:2017tev}
\begin{equation}
f(r)=\frac{3(r^2-\rh^2)}{2L^2}\, .
\end{equation}
Thus, when $a=0$, the metric \eqref{adsmetric} represents that of a static black hole with a regular horizon placed at $r=\rh$. Let us then consider the effect of rotation. After some manipulations, equation \eqref{eqng2} can be written as
\begin{equation}
\tilde C+\left(\frac{10L^2}{3}-6\rh^2\right) g-\frac{d}{dr}\left[r^2(r^2-\rh^2)\frac{dg}{dr}\right]=0\, ,
\end{equation}
where $\tilde C$ is an integration constant related to $C$. The general solution to this equation can be expressed in several ways in terms of hypergeometric functions, but the most useful way is the following one: 
\begin{equation}
g(r)=-\frac{\tilde C}{\left(\frac{10L^2}{3}-6\rh^2\right)}+c_1\, _2F_1\left(1-\alpha ,\alpha -\frac{3}{2};-\frac{1}{2};\frac{\rh^2}{r^2}\right)+\frac{c_2}{r^3} \, _2F_1\left(\frac{5}{2}-\alpha ,\alpha ;\frac{5}{2};\frac{\rh^2}{r^2}\right)\, ,
\end{equation}
where 
\begin{equation}
\alpha=\frac{5}{4}\left(1+\sqrt{1-\frac{8L^2}{15\rh^2}}\right)\, .
\end{equation}
Now, the integration constants are fixed by analyzing the asymptotic behavior at infinity, which reads
\begin{equation}
g(r)=-\frac{\tilde C}{\left(\frac{10L^2}{3}-6\rh^2\right)}+c_1-\left(\frac{10L^2}{3}-6\rh^2\right)\frac{c_1}{2 r^2}+\frac{c_2}{r^3}+\ldots\, .
\end{equation}
Now, in terms of the function $h(r)=r^2p(r)$, the asymptotic expansion becomes
\begin{equation}
h(r)=r^3\left(-\frac{\tilde C}{\left(\frac{10L^2}{3}-6\rh^2\right)}+c_1\right)+\left(\frac{10L^2}{3}-6\rh^2\right)\frac{c_1 r}{2}-\frac{c_2}{2}+\ldots\, .
\end{equation}
Since the $r^3$ and $r$ divergent terms are not desired, we must set $c_1=\tilde C=0$. In that case, the behavior of $h$ at infinity is $h=-c_2/2+\mathcal{O}{(1/r)}$, so that it necessarily tends to a non-vanishing value. Although this might seem strange, this is similar to what happens to the function $f(r)$ in the critical limit, which instead of behaving as $f(r)=f_{\rm AdS}(r)+\mathcal{O}{(1/r)}$ it goes as $f(r)=f_{\rm AdS}(r)+\mathcal{O}{(1)}$. Thus, let us also set $c_2=2$, so that the metric component $g_{t\phi}$ takes the asymptotic value $g_{t\phi}\rightarrow -a \sin^2\theta$. In this case, $a$ is a parameter that controls the angular momentum, although it no longer can be interpreted as the angular momentum per mass. 
The final expression for $h(r)$ after integration of $g(r)$ reads
\begin{equation}
h(r)=-\frac{r^2}{\left(\rh^2-\frac{5L^2}{9}\right)}\left[1-\, _2F_1\left(\frac{3}{2}-\alpha ,\alpha -1;\frac{3}{2};\frac{\rh^2}{r^2}\right)\right]\, .
\end{equation}
It can be checked that this expression is real and regular for any value of $\rh$ --- in particular, the limit $\rh\rightarrow \sqrt{5}/3 L$ is finite. In Fig.~\ref{fig:hcrit} we show the profile of $h(r)$ for several values of $\rh$, where we can see that $h(r)$ is finite everywhere.  However, $h$ is not smooth at $r=\rh$, since its expansion near the horizon contains terms such as $(r-\rh)\log(1-r/\rh)$, so that its derivatives are divergent.

\begin{figure}[t]
\centering
\includegraphics[width=0.6\textwidth]{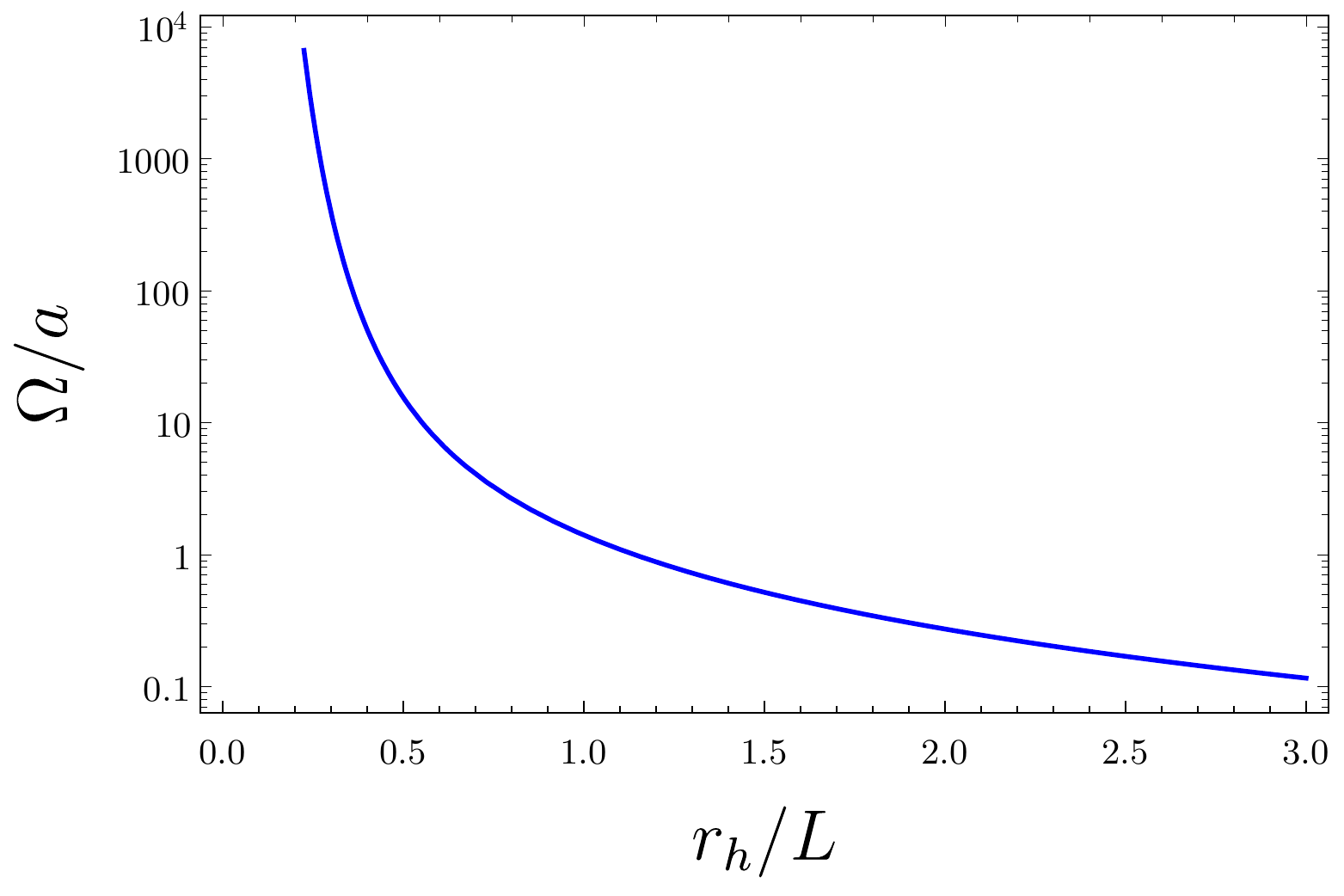}
\caption{Angular velocity for the critical rotating black holes as a function of the radius of the horizon.}
\label{fig:wcrit}
\end{figure}

The value of $h(\rh)$ is rapidly growing as we decrease $\rh$, and as a consequence, the angular velocity $\Omega=-a h(\rh)/\rh^2$, also becomes very high. The angular velocity is given by
\begin{equation}
\Omega=\frac{a}{\left( \rh^2-\frac{5L^2}{9}\right)}\left(1-\frac{\sqrt{\pi }}{2\Gamma \left(\frac{5}{2}-\alpha\right) \Gamma \left(\alpha\right)}\right)\, ,
\end{equation}
and we plot it in Fig.~\ref{fig:wcrit}, where we observe that it diverges exponentially for $\rh\rightarrow 0$. Obviously, the slowly rotating approximation is not valid when $\Omega$ is large, so we have to restrict to sufficiently large values of $\rh$.

\section{Final comments}\label{fincom}
In this paper we have constructed the slowly rotating black hole solutions of Einsteinian cubic gravity both with flat and AdS asymptotes and studied some of their properties. A summary of our results can be found in the introduction. Let us close with some final comments, mostly regarding possible future explorations.

In the static case, the order-reduction phenomenon observed in the equations of motion of ECG turns out to be a general property of GQT gravities in general dimensions and orders. Here we have observed a similar mechanism taking place in the slowly rotating case. It would be interesting to explore whether this is a common feature of this  general class of theories.

While the slowly rotating black holes of Lovelock theories have been already characterized --- see appendix \ref{love} --- no studies of the kind have been performed, to our knowledge, in the case of Quasi-topological gravities. It would be of course interesting to fill this gap. Moreover, as described in the appendix, the slowly rotating solutions in the Lovelock case are completely characterized by the metric for the static solution. This is reminiscent of the Newman-Janis algorithm and it could be interesting to better understand this feature. On the other hand, the slowly rotating solutions in ECG are not so obviously connected with the properties of the static solution.

In the present paper we have not given much attention to the thermodynamic properties of the solutions, the reason being that the effect of the angular momentum on the entropy and the temperature only appears at order $\mathcal{O}(a^2)$, and hence it is not captured by the leading-order solution. Nevertheless, it would be interesting to perform this kind of study, particularly from a holographic perspective. 

Finally, a study of the solutions including arbitrary values of the rotation parameter would be perhaps worth pursuing, but a priori considerably more challenging.

%\comment{Perhaps mention Kubiznak/Frolov comment about the solution not really being valid all the way to the event horizon}

\section*{Acknowledgements}
We thank  Jos\'e Edelstein and Julio Oliva for useful discussions.  
The work of PB was supported by the Simons foundation through the It From Qubit Simons collaboration. PAC was supported by the KU Leuven grant
``Bijzonder Onderzoeksfonds C16/16/005 --- Horizons in hoge-energie fysica''. The work of RBM and
RAH is supported by the Natural Sciences and Engineering Research Council of Canada, the latter 
through the Banting Postdoctoral Fellowship program. At least some of us were also supported physically by planet Earth through the electromagnetic and gravitational interactions.

\appendix
\section{Slowly rotating black holes in Lovelock gravities}\label{love}
In this appendix we review the single-axis slowly rotating black holes of Lovelock theories in general dimensions.
%\comment{\rd{slowly rotating for general Lovelock constructed in Camanho's thesis, so this section to be moved to appendix or something like that}}
\subsection{Einstein gravity}
Let us start considering $D$-dimensional Einstein gravity with a negative cosmological constant,
\begin{equation}
I=\frac{1}{16\pi G}\int d^D x \sqrt{|g|} \left[\frac{(D-1)(D-2)}{L^2}+R  \right]\, .
\end{equation}
Assuming a single axis of rotation, the slowly rotating version of the Kerr-AdS solution --- also known as Myers-Perry black hole in its more general form for $D\geq 5$ --- can be written as
\begin{align}\label{rotissKA}
ds^2=&- \left[k+\frac{r^2}{L^2}g(r) \right]dt^2+\frac{dr^2}{ \left[k+\frac{r^2}{L^2}g(r) \right]}+\frac{2 a r^2}{L^2} g(r)(1-k x^2) dt d\phi \\ \notag &+r^2 \left[\frac{dx^2}{(1-kx^2)}+(1-kx^2)d\phi^2+x^2 d\sigma^2_{(D-4)} \right]\, ,
\end{align}
where: 
\begin{align}
x\equiv \cos\theta  \quad  \text{for} \quad k=1 \, ;\quad
x \equiv \cosh \theta \quad  \text{for} \quad k=-1 \, .
\end{align}
%$x\equiv 1$ for $k=0$, $x\equiv \cos\theta$ for $k=1$ and $x\equiv \cosh \theta$ for $k=-1$. 
In the second line of \req{rotissKA}, $r^2$ multiplies the metric of $\mathbb{S}^{(D-2)}$, $\mathbb{R}^{(D-2)}$ and $\mathbb{H}^{(D-2)}$ for $k=0,1,-1$ respectively, which in the coordinates used above means that $d\sigma^2_{(D-4)}$ is the metric of a round $\mathbb{S}^{(D-4)}$ for $k=\pm 1$, and $d\sigma^2_{(D-4)}\equiv d \vec{y}_{(D-4)}^2/L^2$,  with $d \vec{y}_{(D-4)}^2$ the metric of $(D-4)$-dimensional Euclidean space for $k=0$. %Setting $a=0$, we recover the usual Schwarzschild-AdS solution with the usual three possible horizon geometries.

Observe that the same function $g(r)$ which characterizes the static limit of the solution appears in the  only new components arising at order $\mathcal{O}(a)$ --- namely, $g_{t\phi}=g_{\phi t}$. The function $g(r)$ is given by
\begin{equation}\label{fr}
g(r)=\left[1 -\frac{16\pi G M L^2}{(D-2)\Omega_{(D-2)}r^{(D-1)}}\right]\,  \Rightarrow \,  -g_{tt}=g_{rr}^{-1}=\left[k-\frac{16\pi G M }{(D-2)\Omega_{(D-2)}r^{(D-3)}}+\frac{r^2}{L^2}\right]\, .
\end{equation}
In the above expression, $M$ is the ADM mass of the solution, and $\Omega_{(D-2)}\equiv 2\pi^{\frac{(D-1)}{2}}/\Gamma\left[\frac{D-1}{2} \right]$ is the area of $\mathbb{S}^{(D-2)}$. Naturally, when $a\rightarrow 0$, \req{rotissKA} simply reduces to the usual Schwarzschild-AdS solution with various horizon geometries. %At order $\mathcal{O}(a)$, the only new components in the metric are $g_{t\phi}=g_{\phi t}$, which are also controlled by $g(r)$. 

Another convenient set of coordinates is found by setting  $\phi \rightarrow \phi -at/L^2$, which replaces
\begin{equation}
2g_{t\phi} \rightarrow 2g_{t\phi} - \frac{2a r^2}{L^2}(1-k x^2)= \frac{2a r^2}{L^2} \left[g(r)-1 \right](1-k x^2)=\frac{-32\pi G M (1-k x^2)}{(D-2)\Omega_{(D-2)}r^{(D-3)}}\, .
\end{equation}
The advantage of these coordinates is that the asymptotic angular speed of the solution vanishes, 
\begin{equation}
 \Omega_{\infty}\equiv \lim_{r \rightarrow \infty }\frac{-g_{t\phi}}{g_{\phi\phi}}=\lim_{r\rightarrow \infty} \frac{16\pi G M }{(D-2)\Omega_{(D-2)}r^{(D-1)}}=0\, ,
\end{equation}
 for all $D$.

A way to understand the appearance of the Schwarzschild-AdS blackening factor $g(r)$ in the $g_{t\phi}=g_{\phi t}$ components which will turn out to be useful when we turn on the higher-order Lovelock couplings is the following. First, note that the metric \req{rotissKA} in the $k=0$ case is actually related to the $a=0$ metric by a coordinate transformation. In particular, applying  $t \rightarrow t-a \phi$ to the Schwarzschild-AdS black brane produces \req{rotissKA} at order $\mathcal{O}(a)$. While this is no longer the case in the $k=\pm 1$ cases, if we consider an ansatz of the form  \req{rotissKA} with $g(r)$ replaced by some other function $p(r)$ in the $g_{t\phi}=g_{\phi t}$ components, once $g(r)$ is determined using the $\mathcal{O}(a^0)$ equations, the only components of Einstein's equations which are modified at order $\mathcal{O}(a^1)$  --- namely, $\mathcal{E}_{t\phi}=\mathcal{E}_{\phi t}=0$ --- do not depend on $k$ explicitly.
% once $f(r)$ is determined using the $\mathcal{O}(a^0)$ equations, the only components of Einstein's equations which are modified at order $\mathcal{O}(a)$, namely $\mathcal{E}_{t\phi}=\mathcal{E}_{\phi t}=0$, do not depend on $k$ explicitly when an ansatz of the form \req{rotissKA} is considered with $f(r)$ replaced by some other function $p(r)$ in $g_{t\phi}=g_{\phi t}$. 
Therefore, the fact that $p(r)=g(r)$ holds for $k=0$ --- which is just a consequence of the existence of the coordinate transformation relating the rotating and static black brane solutions --- implies that it also holds for the less trivial cases $k=\pm 1$. A similar phenomenon occurs for Lovelock gravities, as we explain now.

\subsection{Lovelock gravities}
Let us now consider a generic $D$-dimensional Lovelock gravity \cite{Lovelock1,Lovelock2}, whose action can be written as
\begin{equation}\label{loV}
I=\frac{1}{16\pi G}\int d^D x \sqrt{|g|} \left[\frac{(D-1)(D-2)}{L^2}+R+ \sum_{n=2}^{\lfloor D/2 \rfloor} \frac{\lambda_{2n} L^{2(n-1)} (D-2)!}{(D-2n)!}  \mathcal{X}_{2n}  \right]\, ,
\end{equation}
where $\lambda_{2n}$ are dimensionless couplings and $ \mathcal{X}_{2n} $ are the dimensionally-extended Euler densities of $2n$-dimensional manifolds --- \eg $\mathcal{X}_4\equiv R^2-4 R_{ab}R^{ab}+R_{abcd}R^{abcd}$ is the usual Gauss-Bonnet density.

The static black hole solutions of \req{loV} have been extensively studied in the literature --- see \eg \cite{Wheeler:1985nh,Wheeler:1985qd,Boulware:1985wk,Myers:1988ze,Cai:2001dz,Dehghani:2009zzb,deBoer:2009gx,Camanho:2011rj,Garraffo:2008hu}. Generalizations of the slowly rotating Kerr-AdS solution appearing in eqs. (\ref{rotissKA}) and (\ref{fr}) exist for this class of theories. Just like in the Einstein gravity case, these solutions are characterized by the static-solution blackening-factor, namely, they also take the form \req{rotissKA}, where now $g(r)$ is determined by the algebraic equation
\begin{equation}\label{frL}
1-g(r)+\sum_{n=2}^{\lfloor D/2 \rfloor} \lambda_{2n} g(r)^n=\frac{16\pi G M L^2}{(D-2)\Omega_{(D-2)}r^{(D-1)}}\, .
\end{equation}
For example, including the Gauss-Bonnet density alone, one finds
\begin{equation}
g(r)=\frac{1}{2\lambda_{4}}\left[1\mp \sqrt{1-4\lambda_{4}+\frac{64\pi G M \lambda_{4}L^2}{(D-2)\Omega_{(D-2)}r^{(D-1)}}} \right]\, .
\end{equation}
%\begin{equation}
%x \equiv \begin{cases} 1 \quad & \text{for  } k=0 \\ \cos \theta \quad  &\text{for  } k=1 \\ \cosh \theta  \quad &\text{for  } k=-1\end{cases}\,
%\end{equation}
The slowly rotating solutions including the Gauss-Bonnet and cubic Lovelock densities were constructed for generic $k$ in \cite{Kim:2007iw} and \cite{Yue:2011et} respectively. For $k=0$, the full boosted black branes (with arbitrarily large values of $a$) were obtained in \cite{Dehghani:2002wn,Dehghani:2006dh}. The general slowly rotating case was studied in \cite{Camanho:2015ysa}.

The mechanism explained at the end of the previous subsection regarding the $k$-independence of the equations of motion and the role played by the function $g(r)$ in the slowly rotating solutions  holds for the general Lovelock theory in \req{loV}, which therefore possesses solutions of the form \req{rotissKA} with $g(r)$ determined by \req{frL}.

%\section{Slowly rotating black holes in Quasi-topological gravities}
%Beyond Lovelock theories, another natural higher-curvature modification of Einstein gravity corresponds to Quasi-topological gravity. In $D=5$, the action reads
%\begin{equation}\label{QTG}
%I=\frac{1}{16\pi G}\int d^5 x \sqrt{|g|} \left[\frac{12}{L^2}+R +\frac{\lambda_{\rm GB}L^2}{2} \mathcal{X}_4 +\frac{ 7\mu L^4}{4}  \mathcal{Z}_5 \right]\, .
%\end{equation}
%\comment{hope that same mechanism as for Lovelock theories works also for Quasi-topological gravities. To be verified, in progress}

\renewcommand{\leftmark}{\MakeUppercase{Bibliography}}
\phantomsection
\bibliographystyle{JHEP}
\bibliography{Gravities}
\label{biblio}
\end{document}